\documentclass[useAMS,usenatbib,twocolumn]{mnras}
\usepackage{hyperref}
\usepackage{color}

\usepackage{amsmath}

\usepackage{multicol,lipsum}

\usepackage{graphicx, cuted}
\usepackage{amssymb}
\usepackage{pifont}
\usepackage{etoolbox}
\usepackage{multicol}
\newcommand{\mincir}{\raise
-2.truept\hbox{\rlap{\hbox{$\sim$}}\raise5.truept
\hbox{$<$}\ }}
\newcommand{\magcir}{\raise
-2.truept\hbox{\rlap{\hbox{$\sim$}}\raise5.truept
\hbox{$>$}\ }}
\newcommand{\minmag}{\raise-2.truept\hbox{\rlap{\hbox{$<$}}\raise
6.truept\hbox{$>$}\ }}

\newcommand{\hagn}{\mbox{{\sc \small Horizon-AGN}}}
\newcommand{\hnoagn}{\mbox{{\sc \small Horizon-noAGN}}}

\newcommand{\hagnn}{\mbox{{\sc \small Horizon-AGN\,\,}}}
\newcommand{\hnoagnn}{\mbox{{\sc \small Horizon-noAGN\,\,}}}

\newcommand{\lyaa}{Ly-$\alpha$\,}
\newcommand{\lya}{Ly-$\alpha$}
\newcommand{\lymasw}{LyMAS2\,}

\definecolor{grey}{rgb}{0.75,0.75,0.75}
\definecolor{Orange}{rgb}{1.0,0.5,0.15}
\definecolor{brown}{rgb}{0.7,0.25,0.0}
\definecolor{Pink}{rgb}{1.0,0.5,0.5}
\definecolor{darkerred}{rgb}{0.8,0,0}
\definecolor{darkerblue}{rgb}{0,0,0.8}
\definecolor{Blue}{rgb}{0,0.08,0.65}
\definecolor{Red}{rgb}{0.65,0.08,0.05}
\definecolor{Green}{rgb}{0.15,0.45,0.25}

\let\vec\mathbf
\let\mat\mathbf
\let\dd\delta

\let\tth\theta
\let\oo\omega

\def\simlt{\lower.5ex\hbox{$\; \buildrel < \over \sim \;$}}
\def\simgt{\lower.5ex\hbox{$\; \buildrel > \over \sim \;$}}
\def\simpropto{\lower.2ex\hbox{$\; \buildrel \propto \over \sim \;$}}

\title[LyMAS reloaded]
{LyMAS reloaded: improving the predictions of the  large-scale Lyman-$\alpha$ forest statistics from  dark matter density and velocity fields}

\author[S.~Peirani et al.]
 {S. Peirani$^{1,2}$\thanks{E-mail: sebastien.peirani@oca.eu}, 
 S. Prunet$^{1,2}$, S. Colombi$^2$, C. Pichon$^{2,3}$, D.H. Weinberg$^4$, \newauthor
 C. Laigle$^2$,
 G. Lavaux$^2$, Y. Dubois$^2$ and J. Devriendt$^{5,6}$\\
$^{1}$ Universit\'e C\^ote d'Azur, Observatoire de la C\^ote d'Azur, CNRS, Laboratoire Lagrange, Bd de l'Observatoire,\\
 CS 34229, 06304 Nice Cedex 4, France \\
$^{2}$ Sorbonne Universit\'e, CNRS, UMR7095, Institut d'Astrophysique de Paris, 98 bis Boulevard Arago, 75014 Paris, France \\
$^{3}$ IPhT, DRF-INP, UMR 3680, CEA, Orme des Merisiers Bat 774, F-91191 Gif-sur-Yvette, France\\
$^{4}$ Department of Astronomy, The Ohio State University, Columbus, OH \\
$^{5}$ Sub-department of Astrophysics, University of Oxford, Keble Road, Oxford OX1 3RH, UK \\
$^{6}$ Universit\'e de Lyon, Universit\'e Lyon 1, ENS de Lyon, CNRS, Centre de Recherche Astrophysique de Lyon UMR5574, \\ F-69230 Saint-Genis-Laval, France \\
}

\begin{document}

\maketitle

\begin{abstract}
We present LyMAS2, an improved version of the ``Lyman-$\alpha$ Mass Association Scheme'' aiming at predicting the large-scale 3d clustering statistics of the Lyman-$\alpha$
forest (\lya) from  moderate resolution simulations of the dark matter (DM) distribution, with prior calibrations
from high resolution hydrodynamical simulations of smaller volumes.
In this study, calibrations are derived 
from the Horizon-AGN suite simulations, (100 Mpc/h)$^3$ comoving volume, using 
Wiener filtering, combining information 
from dark matter density and velocity fields (i.e. velocity
dispersion, vorticity,  line of sight 1d-divergence and  3d-divergence).
All new predictions have been done at $z=2.5$ in redshift-space, while considering
the spectral resolution of the SDSS-III BOSS Survey
and different dark matter smoothing (0.3, 0.5 and 1.0 Mpc/h comoving). 
We have tried different combinations of dark matter
fields and found that LyMAS2, applied to the Horizon-noAGN dark matter fields,
significantly improves the predictions of the  \lyaa 3d clustering statistics,
especially when the DM overdensity is associated with the velocity dispersion or the vorticity fields.
Compared to the hydrodynamical simulation trends, the 2-point correlation functions of pseudo-spectra 
generated with LyMAS2 can be recovered with relative differences of $\sim$5\% even for high angles, 
the flux 1d power spectrum (along the light of sight)
with  $\sim$2\% and the flux 1d probability distribution function exactly. 
Finally, we have produced several large mock BOSS spectra (1.0 and 1.5 Gpc/h) expected to lead to much more reliable and accurate theoretical predictions.
\end{abstract}



\section{Introduction}

Distant quasars emit light that crosses a large part of the Universe before being  
observed with instruments on Earth. In particular, the spectrum of each
quasar presents fluctuating absorption that corresponds to the Lyman-$\alpha$ forest
(\lya, \cite{lynds71,sargent+80}).
The study of the Lyman-$\alpha$ forest has become a major focus of modern cosmology, 
as it is supposed to trace the neutral hydrogen density
that fills most of the Universe in a way that
 approximately corresponds to the underlying dark matter density 
\citep{croft+99,peeples+10}.
Since a single background source only provides one dimensional information
along the corresponding line-of-sight (or ``skewer''), characterizing the 3d density of the
high redshift Universe with the \lyaa forest requires large samples of quasar spectra.
Successful surveys such as the (extended) Baryon Oscillation Spectroscopic Survey
(BOSS/eBOSS, \cite{dawson+13,dawson+16}), of the Sloan Digital Sky Survey (SDSS-III and SDSS-IV, \cite{blanton+17,eisenstein+11}
have measured the Lyman-$\alpha$ forest spectra of 160,000 quasars at
redshifts $2.2< z < 3$. Thanks to this large sample,
the study of the Lyman-$\alpha$ forest has proved to be a complementary probe 
to low redshift galaxy surveys. For instance, the large sample of quasar spectra
have permitted accurate measurements of 
3d flux auto-correlation functions \citep{slosar+11} as well as
the cross-correlation between the \lyaa Forest
and specific tracers, namely damped-\lyaa systems (DLAs) and quasars \citep{font-ribera+12, font-ribera+13,font-ribera+14}.
Such 3d measurements also enable measurements of
the distance-redshift relation and the Hubble expansion via baryon accoustic oscillations
(BAO)\citep{busca+13,slosar+13,delubac+15,bautista+17,dumasdesbourboux}.
Moreover, BOSS spectra also permit accurate measurements of
the line-of-sight power spectrum \citep{palanque+13} and flux probability distribution
function (PDF) \citep{lee+15}.
In the near future, the {\it Dark Energy Spectroscopic Instrument} (DESI, \cite{desi}),
the {\it William Herschel Telescope Enhanced Area Velocity Explorer} (WEAVE-QSO, \cite{pieri+16,dalton+16,dalton+20})
and the {\it Subaru Prime Focus Spectrograph} (PFS, \cite{takada+14})
will go well beyond the present surveys and
will open new perspectives on the high redshift intergalactic medium  probed
by the \lyaa forest.

In parallel to the development of these large quasar surveys, theoretical modeling needs
to reach the high level of complexity and accuracy to interpret the observational data.
Nowadays, hydrodynamical cosmological simulations represent an ideal tool as 
they manage to model the intergalactic medium with a high degree of realism
with appropriate resolution \citep[e.g.][]{dubois14,illustris,eagle,bolton+17}. 
However, to properly model the 3d correlations
of the \lyaa forest, one needs to resolve the pressure-support scale (Jeans scale)
of the diffuse intergalactic medium (IGM), which is typically $\sim$0.25 Mpc/h
comoving for matter overdensity of $\sim$10 \citep{peeples+10},
while considering
$\sim$Gpc$^3$ simulation volumes to exploit the statistical precision achieved
by the different observational surveys while avoiding box size effects. Combining such resolution and simulation
volume is currently not feasible mainly because of computational limits.
To tackle such an issue, several methods exist in the literature.
One of the most popular is to use the so-called ``Fluctuating Gunn-Peterson Approximation''
\citep[FGPA][]{weinberg+97,croft+98}
which links the \lyaa optical depth to the local dark matter density. This approach
is relatively straightforward as it assumes a deterministic relation and only information
on the density field (extracted from N-body simulations or log-normal density fields) is required. However, the FGPA approach is expected to be accurate enough 
only on very large scales, e.g. those of the BAO features ($\sim$ 100 Mpc/h, e.g. Fig. 5 of \cite{sinigaglia+21}). But 3d \lyaa forest surveys also enable precise
measurements of flux correlations at much smaller scales where the FGPA might not be adequate.

Another approach is to apply relevant calibrations, derived first from small hydrodynamical simulations,
to large-scale dark matter distributions extracted from pure dark matter
simulations, which are much cheaper to perform. 
In particular, \cite{peirani14} (hereafter, P14) have developed
the Lyman-$\alpha$ Mass Association Scheme (LyMAS) which follows such a philosophy.
The main idea is that flux correlations on small and large scales are mainly driven
by the correlations of the dark matter density field. More specifically,
the flux statistics can be estimated by combining the DM density field
with the conditional probability distribution $P(F|\rho)$ of the transmitted flux
$F$ on the DM density contrast $\rho$. In its most sophisticated form, LyMAS
creates ensemble of coherent pseudo-spectra at the BOSS resolution using 
the Gaussianized percentile distribution of the conditional flux, while re-scaling 
the line-of-sight power spectrum and PDF at the last step.
One of the main results of LyMAS is to improve the predictions of flux 3d correlations especially
with respect to deterministic mapping  (e.g. FGPA) which tends to significantly overestimate
them especially when the DM density is smoothed at scale greater than 0.3 Mpc/h.
Similarly, \cite{sorini+16} have developed ``{\it Iterative Matched Statistics}" (IMS)
in which the PDF and the power spectrum of the real-space 
\lyaa flux are derived from small hydrodynamical simulations.
Then, these two statistics are 1d (1D-IMS) or 3d (3D-IMS) iteratively mapped onto a pseudo-flux field of an N-body simulation from which the matter density is first Gaussian smoothed.
 In  3D-IMS,  smoothing is followed by matching the 3D power spectrum and PDF of the flux in real space to the reference hydrodynamic simulation.  With 1D-IMS, 
 the 1d power spectrum and PDF of the flux are additionally matched. 
 Both methods have proved to be again more accurate than the FGPA approach (which strongly
relies on the DM smoothing scale)
when reproducing line-of-sight observables, such as the PDF and power spectrum as
well as the 3d flux power spectrum (5-20\%). 
Finally, Machine Learning based methods start to be considered and lead to 
promising results \citep{harrington+21, sinigaglia+21, chopitan+prep}.

Although the LyMAS full scheme is able to model the BOSS 3d clustering quite accurately and
has been already used in different analysis related to
the quasar-\lyaa forest cross-correlation \citep{lochhaas16}, 
the three-point correlation functions \citep{suk} and the 
correlations  between  the  \lyaa transmitted  flux  and  the mass overdensity \citep{cai+16},
 we aim at  investigating whether other sets of calibrations could still improve the
theoretical predictions. To this regard,
we consider a new approach based on Wiener Filtering, which has been used
for 3d map reconstruction from an ensemble of 1d lines-of-sight \citep[e.g.][]{pichon+01,caucci+08,ozbek+16,lee+18,japelj+19,ravoux+20}.
The underlying philosophy in LyMAS2 remains unchanged. We still find that the flux correlations
are driven mainly by the correlations of the DM density field, but with potential 
refinements from the correlations of the DM velocity field.

The paper is organised as follows. In section~\ref{sec:wiener}, we describe 
the Wiener equations as multivariate normal conditional probabilities, and 
we explain their application to hydrodynamical simulations. Section~\ref{sec:simu} briefly describes
how we extract the flux and all relevant dark matter fields from the \hnoagnn simulation.
We also present the potential correlations that arise between these different fields.
Then in section~\ref{sec:pseudospectra} we present
the statistics in the line of sight power spectrum, the PDF and
the 2-point correlation function of pseudo-spectra
produced when LyMAS2 is applied to different associations of dark matter fields
of \hnoagn. Such trends are compared to the flux statistics from the hydrodynamical 
simulation (``hydro flux"). 
In section \ref{sec:largemock} we apply LyMAS2
to N-body simulations of  1.0 and 1.5 Gpc/h comoving volumes.
We summarize our results and conclusions in section \ref{sec:conclusions}.
We also add three appendices. In appendix~\ref{appendix1}, we compute the mean 
two point correlations functions from five different hydrodynamical simulations of
lower resolution to check the robustness of the results presented 
in section~\ref{sec:pseudospectra}. Appendix~\ref{appendix2} presents
the performance of specific deterministic samplings. Appendix~\ref{appendix3} provides details on how estimates of
density and velocities are performed on the dark matter particle
distribution, relying on adaptive softening.

\section{LyMAS VS LyMAS2}
\label{sec:lymasvslymas2}

{
We briefly describe the fundamental  differences between the first version of LyMAS, detailed
in \cite{peirani14} (hereafter P14) and the new scheme, LyMAS2, presented in this work.
The two versions basically share the same philosophy: specific calibrations are first derived from hydrodynamical simulations of small volume and then applied 
to large DM simulations, assuming that the correlations of the \lyaa at small
and large scales
are mainly driven by the correlations of the underlying DM density and (eventually) velocity fields.  The main differences
essentially lie in 1) the derivation and characterisation
of the cross-correlation between the different fields (namely the hydro spectra and the DM fields) and 2) the way we apply such calibrations
to the DM distributions.

More specifically, in the first version a hydrodynamical simulation was used to calibrate the conditional
probability distribution  $P(F|\rho)$ to have a transmitted flux
value $F$ given the value of the DM density contrast $\rho$ at the same location. 
In its simplest form, LyMAS randomly and independently draws transmitted flux values according to $P(F|\rho)$ and the value of $\rho$ at each pixel of a 
regular grid used to sample the DM overdensity field. 
Although the 3d clustering statistics of the pseudo-spectra generated by this approach
is quite close to that of the hydro flux, the main drawback
of this procedure is to create very noisy 
spectra as any coherence along each line-of-sight is lost.
To solve this issue and make the pseudo spectra more realistic, the most sophisticated form
of LyMAS uses the fact that neighboring pixels along a given line-of-sight are supposed to have close
 probability distribution $P(F|\rho)$. Hence, one can introduce a coherence 
 by defining percentile spectra i.e. the fractional
position of the flux value in the cumulative distribution of $P(F|\rho)$.
Then these percentile spectra derived from every lines-of-sight of the grid can be gaussianized and one can derive a characteristic power spectrum from these
1d Gaussian fields. Thus, from this new input parameter 
 LyMAS generates first a 1d Gaussian field, degaussiannizes it to get a realization of a 
percentile spectrum, and finally derive a coherent spectrum
using the different values of the DM density contrast $\rho$
and the percentile value in $P(F|\rho)$ along the considered los. Here again,
the predictions of the 3d clustering from such coherent spectra is proved
be very accurate.

LyMAS2 does not derive and consider $P(F|\rho)$ as well as
percentile spectra. Instead, as we explain in detail
in section~\ref{sec:wiener}, LyMAS2 makes good use of Wiener Filtering 
to characterize the correlations between the transmitted flux and the DM density contrast. The statistics are directly made los by los 
which naturally introduces a coherence in the pseudo-spectra.
Furthermore,  this 
approach has the advantage to naturally take into account not only
the DM density field (like in LyMAS) but other fields such as specific
DM velocity fields (e.g. velocity dispersion, vorticity, divergence) that potentially
bring new information to improve the predictions.

In the very last step, both LyMAS and LyMAS2 end similarly 
by rescaling the flux line-of-sight power spectrum and PDF. 
These transformations are useful to slightly correct the line-of-sight
1d power spectrum and PDF of pseudo spectra to make them
identical or quasi identical to those of the hydro spectra.
This step, however, does not significantly modify the 3d clustering statistics.

In the beginning of section~\ref{sec:pseudospectra}, We summarize all the steps
of LyMAS and LyMAS2 to create a pseudo spectrum.

}

\section{Wiener equations}
\label{sec:wiener}

\subsection{Multivariate  conditional probabilities}
\label{sec:general}

Let us assume a complex Gaussian random (vector) variable $\vec{x}$ 
that can be separated into two sub-vectors $\vec{x}=(\vec{x}_1,\vec{x}_2)$, whose mean and covariance can be written as 
\begin{align*}
\vec{\mu} &= (\vec{\mu}_1,\vec{\mu}_2),\\
\mat{\Sigma} &= 
\begin{pmatrix}
\mat{\Sigma}_{11} & \mat{\Sigma}_{12} \\
\mat{\Sigma}_{12}^H & \mat{\Sigma}_{22}
\end{pmatrix} \,,
\end{align*}
where the $H$ superscript denotes hermitian conjugate.
By using formulae for block inverses, it is possible to derive from the joint distribution of $\vec{x}_1$ and $\vec{x}_2$ the formula for the conditional distribution of $\vec{x}_1$ given $\vec{x}_2$. As expected, it is a gaussian multivariate distribution, of mean and covariance:
\begin{align}
    \bar{\vec{\mu}}_1 &= \vec{\mu}_1 + \mat{\Sigma}_{12}\mat{\Sigma}_{22}^{-1}(\vec{x}_2-\vec{\mu}_2)\,,
    \label{eq:conditional-mean}\\ 
    \bar{\mat{\Sigma}} &= \mat{\Sigma}_{11} - \mat{\Sigma}_{12}\mat{\Sigma}_{22}^{-1}\mat{\Sigma}_{12}^H\,.
    \label{eq:conditional-variance}
\end{align}

Now, consider the joint gaussian distribution of the (complex) spatial Fourier modes $(f_k,\delta_k,\theta_k,\omega_k)$, where, by definition for field $a$, we have 
$a_k = \int a(x) \mathrm{e}^{-ik\cdot x}\mathrm{d}^3x$. We assume they are centered fields (zero mean), and of covariance (we drop the subscript $k$ for lisibility):
\begin{equation*}
    \Sigma = 
    \left(\begin{array}{c|ccc}
    P_{f\!f}     & P_{f\dd}     & P_{f\tth}     & P_{f\oo} \\ \hline
    P_{f\dd}^* & P_{\dd\dd}   & P_{\dd\tth}   & P_{\dd\oo}\\
    P_{f\tth}^* & P_{\dd\tth}^* & P_{\tth\tth}   & P_{\tth\oo} \\
    P_{f\oo}^* & P_{\dd\oo}^* & P_{\tth\oo}^* & P_{\oo\oo} 
    \end{array}\right) =
    \left(\begin{array}{c|c}
    \mat{\Sigma}_{11} & \mat{\Sigma}_{12} \\ \hline
    \mat{\Sigma}_{12}^H & \mat{\Sigma}_{22}
    \end{array}\right)\,,
\end{equation*}
where by definition $P_{ab}(k) \equiv \langle a^*_k b_k \rangle$ is the cross-spectrum of fields $a$ and $b$ at wavenumber $k$, and we have partitioned the fields according to $\vec{x}_1 \equiv f_k$ and $\vec{x}_2 \equiv (\delta_k,\theta_k,\omega_k)^T$.
Applying the equations~\eqref{eq:conditional-mean} and \eqref{eq:conditional-variance} we get the conditional mean and variance of the field $f_k$ given $(\dd_k,\tth_k,\oo_k)$:
\begin{align}
    \bar{f_k} &= 
    \begin{pmatrix}
    P_{f\dd},\!\!\! & P_{f\tth},\!\!\! & P_{f\oo}
    \end{pmatrix}
    \cdot
    \mat{\Sigma}_{22}^{-1}
    \cdot
    \begin{pmatrix}
    {\dd_k} \\
    {\tth_k} \\
    {\oo_k}
    \end{pmatrix}\,,\label{eq:fk} \\
    \bar{\mat{\Sigma}} &= P_{f\!f} -
    \begin{pmatrix}
        P_{f\dd},\!\!\! & P_{f\tth}, \!\!\! & P_{f\oo}
    \end{pmatrix}
    \cdot
    \mat{\Sigma}_{22}^{-1}
    \cdot
    \begin{pmatrix}
    P_{f\dd}^* \\
    P_{f\tth}^* \\
    P_{f\oo}^*
    \label{eq:sigma}
    \end{pmatrix}\,.
 \end{align}
Computing the inverse $\mat{\Sigma}_{22}^{-1}$ using the cofactor matrix formula, we get
\begin{equation*}
    \mat{\Sigma}_{22}^{-1} = \frac{1}{|\mat{\Sigma}_{22}|}
    \begin{pmatrix}
    A_{11} & A_{12} & A_{13}  \\
    A_{21} & A_{22} & A_{23}  \\
    A_{31} & A_{32} & A_{33} 
    \end{pmatrix}\,,
\end{equation*}
with
\begin{align*}
     A_{11} &= P_{\tth\tth}P_{\oo\oo} - P_{\tth\oo}P_{\tth\oo}^*\,, &
     A_{12} &= -P_{\dd\tth}^*P_{\oo\oo} + P_{\dd\oo}^*P_{\tth\oo}\,,\\
     A_{13} &= P_{\dd\tth}^*P_{\tth\oo}^* - P_{\dd\oo}^*P_{\tth\tth}\,,&
     A_{21} &= -P_{\dd\tth}P_{\oo\oo} + P_{\dd\oo}P_{\tth\oo}^*\,,\\
     A_{22} &= P_{\dd\dd}P_{\oo\oo} - P_{\dd\oo}P_{\dd\oo}^*\,,&
     A_{23} &= -P_{\dd\dd}P_{\tth\oo}^* + P_{\dd\tth}P_{\dd\oo}^*\,,\\
     A_{31} &= P_{\dd\tth}P_{\tth\oo} - P_{\tth\tth}P_{\dd\oo}\,,&
     A_{32} &= -P_{\dd\dd}P_{\tth\oo} + P_{\dd\tth}^*P_{\dd\oo}\,, \\
     A_{33} &= P_{\dd\dd}P_{\tth\tth} - P_{\dd\tth}P_{\dd\tth}^*\,,
\end{align*}
and
\begin{align*}
    |\mat{\Sigma}_{22}| = P_{\dd\dd}(P_{\tth\tth}P_{\oo\oo}-P_{\tth\oo}P_{\tth\oo}^*) + (P_{\dd\tth}P_{\tth\oo}P_{\dd\oo}^* + \mathrm{c.c.})\\ -P_{\tth\tth}P_{\dd\oo}P_{\dd\oo}^* - P_{\oo\oo}P_{\dd\tth}P_{\dd\tth}^*\,,
\end{align*}
where ``c.c" denotes the conjugate complex. 

Note that we limit our study to a maximum of three different input fields (i.e. $\delta$, $\theta$ and $\omega$) to construct the field $f$. 
But obviously, this formalism can be extended to a higher number of fields, leading to more
and more complex analytical solution. However, we will see that the statistical trends derived when considering  two and three input fields are quite similar (when judiciously chosen), suggesting that  
adding more than two fields
will not noticeably improve the results anymore.

\subsection{Application to hydro simulations}

Let us call $F$, $\rho$, $v_1$ and $v_2$ respectively the local \lyaa transmitted flux,
the local dark matter density, velocity divergence and vorticity amplitude, extracted from a given hydrodynamical simulation.
Let us call now 
$G$ the cumulative distribution of a standard normal $\mathcal{N}(0;1)$
distribution ($G[x]=\int^x \exp(-u^{2}/2)/\sqrt{2\pi}du$),  and 
\begin{align}
G_{\rho}&=\int_{-\infty}^{\rho}{\rm PDF}(\rho')d\rho',\\
G_{v_i}&=\int_{-\infty}^{v_i}{\rm PDF}(v')dv',\\
G_{F}&=\int_{0}^{F}{\rm PDF}(F')dF'\,,
\end{align}
the cumulative distributions of the measured
DM density and velocity fields  (in the hydrodynamical simulation), and $G_{f}$ the cumulative distribution
of the measured flux, $F$. 
We can then define new fields, namely
\begin{align}
f&=G^{-1}(G_{F}(F),&
\delta&=G^{-1}(G_{\rho}(\rho)),\\
\theta&=G^{-1}(G_{v_1}(v_1)),&
\omega&=G^{-1}(G_{v_2}(v_2))\,.
\end{align}
which should be normally distributed by construction (or ``gaussianized"). 
Let us compute these different fields from the hydro simulations and extract from them 
the relevant cross-spectra, using Fourier space:
\begin{align*}
P_{f\!f} &=\langle f_k^* f_k\rangle, &
P_{\delta\delta}&=\langle\delta_k^*\delta_k\rangle, &
P_{\theta\theta}&=\langle\theta_k^*\theta_k\rangle, \\
P_{\omega\omega}&=\langle\omega_k^*\omega_k\rangle, &
P_{f\delta}&=\langle f_k^* \delta_k\rangle, &
P_{f\theta}&=\langle f_k^* \theta_k\rangle, \\
P_{\delta\theta}&=\langle\delta_k^*\theta_k\rangle, &
P_{\theta\omega}&=\langle\theta_k^*\omega_k\rangle, &
\mathrm{etc} ...
\end{align*}

 
\noindent{}
which are going to depend typically on a transverse and a longitudinal
separation radius. If we assume that the fields $f$, $\delta$, $\theta$
and $\omega$ are Gaussian random fields (GRF, not just its one point statistics
is now required to be normal) then for a given measured set of $\rho$, $v_1$ and
$v_2$
(correspondingly a set of $\delta_k$, $\theta_k$ and $\omega_k$) say along a set of lines of sight (LOS),
one can estimate the most likely field $\bar{f}$
following equations~\eqref{eq:fk}:
\begin{equation}
\bar{f_k}= T_1\cdot \delta_k + T_2\cdot \theta_k + T_3\cdot \omega_k\,,
\label{eq:fluxmean}
\end{equation}
where $T_1$, $T_2$ and $T_3$ are functions of cross-spectrum $P_{ab}(k)$.
This approach can be done along a given LOS or in volume.
However, in the present study we will only analyse LOS individually and independently, ignoring transverse
correlations between different LOS for now.
This allows us to use 1D Fast Fourier Transforms (FFT), and assuming stationarity along the LOS, 
the multiplication is simply done frequency by frequency since in Fourier space the covariance sub-blocks $P_{ab}$ are
diagonal.
To illustrate,  if only one field is considered (i.e. the DM overdensity field in our study)
we simply get:
\begin{equation}
\bar{f_k}= P_{f\delta}\cdot P_{\delta\delta}^{-1}\cdot\delta_k\,.
\label{eq:fluxmean1d}
\end{equation}

If we add additional information from a specific velocity field (e.g. velocity dispersion), the expression of $\bar{f}$
becomes:
\begin{equation}
\bar{f}_k= \frac{ P_{f\delta}(P_{\theta\theta}-P_{\delta\theta})}{P_{\delta\delta} P_{\theta\theta} - P_{\delta\theta} P_{\delta\theta}^*}\cdot \delta_k + \frac{  P_{f\theta}(P_{\delta\delta}-P_{\delta\theta}^*)}{P_{\delta\delta} P_{\theta\theta} - P_{\delta\theta} P_{\delta\theta}^*} \cdot \theta_k   \,.
\label{eq:fluxmean2d}
\end{equation}
%
Adding a second velocity field leads to a more complex expression of  $\bar{f}_k$.

We can then draw samples as $\tilde{f}_k\equiv\bar{f}_k+\Delta f_k$
where $\Delta f_k$ obeys a GRF of mean zero and variance as equation~\eqref{eq:sigma}:
\begin{equation*}
\Delta f_k\sim {\cal G}(0,P_{f\!f}-P_{f\delta}\cdot P_{\delta\delta}^{-1}\cdot P_{f\delta}^*)\,,
\end{equation*}
when only the dark matter density is considered and
\begin{equation*}
\Delta f_k\sim {\cal G}(0,P_{f\!f}-\left(\!\begin{array}{c}
P_{f\delta}\\
P_{f\theta}
\end{array}\!\right)^{T}\!\!\cdot\left[\begin{array}{cc}
P_{\delta\delta} & P_{\delta\theta}\\
P_{\delta\theta}^{*} & P_{\theta\theta}
\end{array}\right]^{-1}\!\!\cdot\left(\!\begin{array}{c}
P_{f\delta}^{*}\\
P_{f\theta}^{*}
\end{array}\!\right)\!),\label{eq:deltaf}
\end{equation*}
in the case of two dark matter fields. After computing the inverse Fourier transform
of $\tilde{f_k}$ to get $\tilde{f}$, the corresponding flux obeys $\tilde{F}=G_{f}[G^{-1}(\tilde{f})]$.
By construction the one point statistics of $\tilde{f}$ will be random normal, so that
the one point PDF of its de-gaussianized transform will be that of the original field.
The power spectrum $P_{ff}$ of $\tilde{f}$ will be the same as that of $f$.
Indeed, let us consider the
case with only one input file for simplicity. We have:
\begin{equation*}
P_{\tilde{f}_k\!\tilde{f}_k}\equiv\langle|\bar{f}_k+\Delta f_k|^{2}\rangle=\big\langle \left|\frac{1}{P_{\delta\delta}}P_{f\delta}\delta_k\right|^{2}+|\Delta f_k|^{2}\big\rangle \,,
\end{equation*}
because the expectation and the fluctuations are uncorrelated. Therefore
\begin{eqnarray*}
P_{\tilde{f}_k\!\tilde{f}_k} & = & \frac{P_{f\delta}^2}{P_{\delta\delta}^2}\left\langle |\delta_k|^{2}\right\rangle +P_{f\!f}-\frac{P_{f\delta}^2}{P_{\delta\delta}}
  =  P_{f\!f}\,,
\end{eqnarray*}
since $\left\langle |\delta_k|^{2}\right\rangle \equiv P_{\delta\delta}$.

Recall that all equations above are valid independently for each Fourier mode $k$, and for each mode, all $P_{ab}$ terms are scalars for a given
pair of fields $(a,b)$.

\begin{figure}
\begin{center}
\rotatebox{0}{\includegraphics[width=\columnwidth]{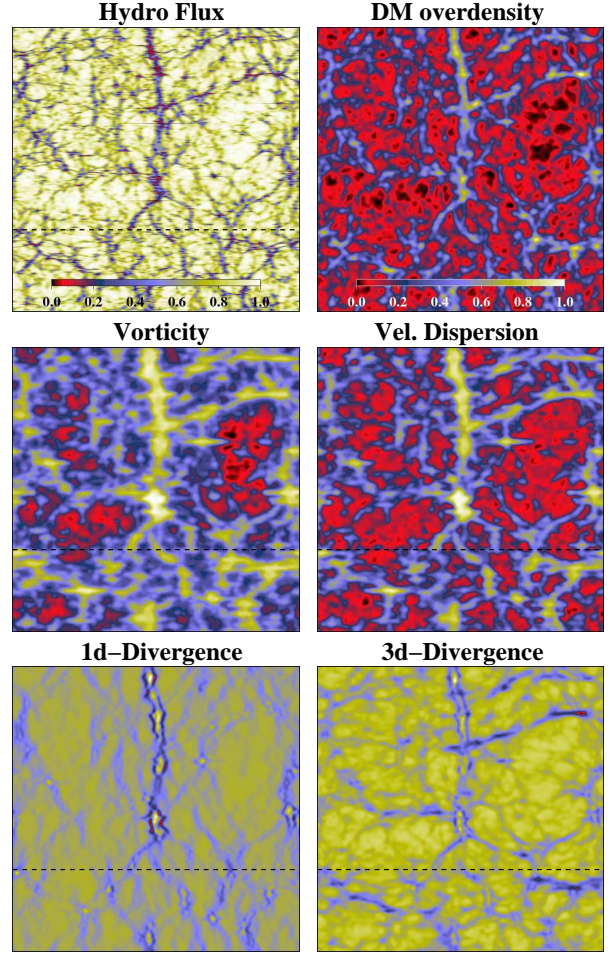}}
\caption{Slices through the Flux (1-dimensionally smoothed at the BOSS resolution) 
as well as corresponding DM overdensity and velocity fields (3-dimensionally smoothed at 0.5 Mpc/h) in redshift space (horizontal direction). 
Each field has been extracted from the \hnoagnn simulation at $z=2.5$ and
has been normalized to help the visual comparison. The dashed lines correspond to the same line of sight (see~Figure \ref{los}).}
\label{slices} 
\end{center}
 \end{figure}

\begin{figure}
\begin{center}
\rotatebox{0}{\includegraphics[width=\columnwidth]{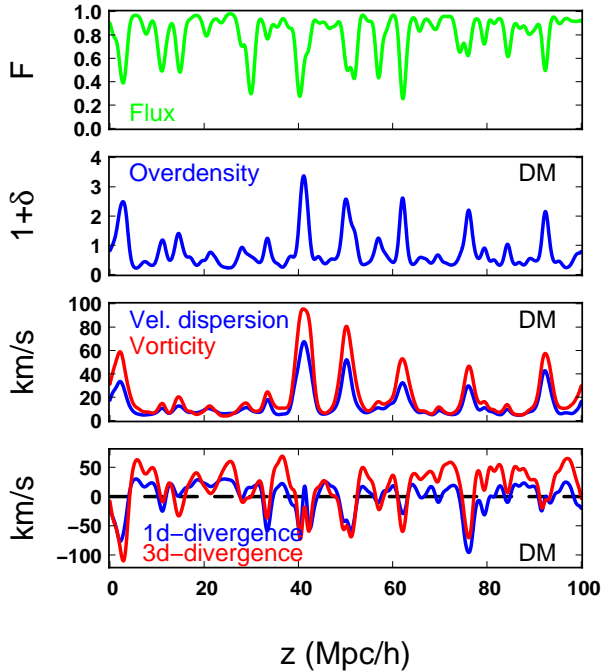}}
\caption{An example showing clear correlations (or anti-correlations) 
between the evolution of the hydro flux (green line) and DM density and velocity fields
(blue and red lines) along the same skewer extracted from the \hnoagnn at $z=2.5$. These skewers
are extracted from the slices studied in Fig.~\ref{slices}.}
\label{los}
\end{center}
 \end{figure}

\section{Flux and Dark Matter fields}
\label{sec:simu}

Throughout the present analysis, we have used the \hnoagnn simulation \citep{peirani17}
to characterize any relevant cross correlations between the transmitted flux and the
different dark matter fields. 
\hnoagnn is a hydrodynamical simulation of 100 h $^{-1}$Mpc comoving box side
run with the RAMSES code \citep{ramses}. It evolves 1024$^3$ dark matter particles with a mass resolution of 8.27$\times$10$^7$M$_\odot$ while the initially uniform grid is
refined in an adaptive way down to $\Delta x=1$ proper kpc at all times.
The simulation adopts a standard  $\Lambda$CDM
cosmology compatible with WMAP-7 results \citep{wmap7}, namely a total matter density
$\Omega_m$=0.272, a dark energy density $\Omega_\Lambda$=0.728, an amplitude of the matter power spectrum $\sigma_8$=0.81, a baryon density $\Omega_b$=0.045, a Hubble constant
$H_0=70.4 \,\rm km\, s^{-1}\,Mpc^{-1}$ and n$_s$=0.967.
\hnoagnn is the twin simulation of \hagnn \citep{dubois14}.
It contains all relevant physical processes such as metal-dependent cooling, photoionization  and  heating  from  a  UV  background,  supernova  feedback  and  metal  enrichment,
but does not include black hole growth and therefore
AGN feedback.

The choice of using \hnoagnn instead of \hagnn was mainly motivated by
the fact that we have performed five additional but slightly lower resolution
hydrodynamical simulations to estimate the accuracy and robustness of the LyMAS2 predictions. Thus, turning out the AGN feedback 
processes in the simulations has permitted us to limit the computational time. 
These results are presented in appendix~\ref{appendix1}.
Furthermore, we have tuned in the present study
the UV background intensity in the process of generating the ``noAGN" flux grid (see below)
in order to get the same mean transmitted Ly-$\alpha$ forest flux $\overline{F}$ derived from \hagn.
By doing this, the flux statistical predictions from the two simulations in the 3d \lyaa clustering tend to be almost the same. This 
has been already noticed in \cite{lochhaas16}
when studying the cross-correlations between dark matter haloes and transmitted flux in the 
\lyaa forest.
Note however that AGN feedback is expected to have non negligible effect on the \lyaa
3d clustering such as, for instance, the 1d power spectrum of the \lyaa forest 
 \citep[e.g.][]{viel+13,chabanier+20}.

In the following, we describe briefly how we derived the hydro spectra field and
the different DM density and velocity fields from the \hnoagnn simulation. Similarly
to P14, we analyse the simulation outputs at redshift $2.5$.
Each field is sampled in a regular grid of 1024$^3$ pixels and the size of a single pixel
is therefore $\sim$0.1 comoving Mpc/h or 0.04 physical Mpc.

\subsection{Transmitted flux}
From the \hnoagn,  we follow the method to generate the hydro spectra that is fully described
in \citep{peirani14}. 
The optical depth of Ly-$\alpha$ absorption is calculated based on the neutral hydrogen density along each line of  sight. 
Basically, the opacity at observer-frame frequency  $\nu_{\rm obs}$ is
$\tau(\nu_{\rm obs})=\sum_{\rm cells}n_\mathrm{HI}\sigma({\nu_{\rm obs}})dl$
where the sum extends over all cells traversed by the line-of-sight,
$n_\mathrm{HI}$ is the numerical density of neutral H atoms
in each cell, 
$\sigma({\nu_\mathrm{obs}})$ is the cross section of
Hydrogen to Ly-$\alpha$ photons, and
$dl$ is the physical size of the cell.
Then each
spectrum is smoothed with  a  1d  Gaussian of dispersion 0.696 h$^{-1}$ Mpc, equivalent to BOSS spectral resolution at $z\approx 2.5$.
The optical depth along each spectrum is converted to \lyaa forest flux by $F=e^{-\tau}$.
Following common practice in Ly$\alpha$ forest modelling, 
the UV  background  intensity  is  chosen  to  give  a  mean  transmitted
Ly-$\alpha$ forest flux $\overline{F}=\langle e^{-\tau}\rangle=$0.795, matching the metal-corrected
$z= 2.5$ value measured from high-resolution spectra by \cite{faucher}.

\subsection{Density, velocity and mean square velocity}
DM skewers that correspond to the ``hydro" spectra are also extracted from the hydrodynamical simulation.
We use the same 3-step algorithm introduced in P14 to derive both the overdensity, the velocity field and the velocity dispersion fields:
\begin{enumerate}
\item adaptive interpolation of the DM particle distribution on a regular
grid \citep{dens_smooth}, as detailed in Appendix~\ref{appendix3};
\item smoothing with a Gaussian window in Fourier space;
\item  Extraction of the skewers from a grid of LOS aligned along the
  $z$-axis.
\end{enumerate}

In step (ii), DM field is three-dimensionally smoothed using different choices of smoothing scales.
In P14, we found that a smoothing scale of 0.3 Mpc/h has proved to be optimal leading
to the most accurate predictions.
However, we prefer a value of 0.5 Mpc/h in the present study since the predictions
are very similar to those obtained with 0.3 Mpc/h (see Appendix~\ref{appendix1}). Furthermore,
we anticipate with the fact that it is computationally much easier to smooth a DM field to 0.5 Mpc/h than 0.3 Mpc/h
when considering large volumes namely with box sides at least greater or equal to 1 Gpc/h.

\subsection{Dark matter vorticity}
The velocity field is projected (using Cloud-in-Cell interpolations) on a regular grid of resolution 1024 and smoothed over 
0.5 Mpc/h with a gaussian filter. The vorticity $\Omega$ is then computed as being the curl of the
velocity field using FFT. Slightly smoothing the input velocity field allows us
 to avoid Gibbs artefacts.  

\subsection{Dark matter velocity divergence}
We have considered both the 3d velocity divergence and the 1d velocity divergence along
the line of sight direction.

For the 3d case, we employed two different methods to see whether this could affect
our results and trends. The first one is based on a centered
finite difference approximation, namely the divergence
$\nabla_{3d}^i$ at a pixel $i$ is given by
\begin{equation}
\nabla_{3d}^i \simeq \frac{V_x^{i+1}\! -\! V_x^{i-1}}{2h} +    \frac{V_y^{i+1}\! -\! V_y^{i-1}}{2h} +  \frac{V_z^{i+1}\! -\! V_z^{i-1}}{2h},
\end{equation}
%
where $V_x^i$, $V_y^i$ and $V_z^i$ are the velocity components at pixel $i$ 
and $h$ the size of a pixel (i.e. $100/1024$ Mpc/h here).
The second method uses the exact expression of the divergence in Fourier space. 
However, we found that the two methods lead to very similar results so we will 
only show results from the Fourier space method for the 3d case.

For the 1d case, we simply use the finite difference approach and the divergence
$\nabla_{1d}^i$ at a pixel $i$ becomes:
\begin{equation}
\nabla_{1d}^i \simeq \frac{V_z^{i+1} - V_z^{i-1}}{2h}\,,
\end{equation}
since we define the $z$-axis as the direction of the LOS.

\medskip
We summarize in Table~\ref{tab1} the different DM fields used in this
study. It is worth mentioning that we changed some of the notations that can be 
found in P14. 
We first replaced the definition of the smoothed flux at the BOSS
resolution $F_s$ in P14 to $F$ here. We also changed the definition of
the 3d smoothed DM overdensity, $\rho = 1+\delta$ instead of  $\Delta_s = 1+\delta_s$ in
P14.
\begin{table}
\caption{Summary of fields and corresponding notation used in the text.}
\label{tab1}
\begin{tabular}{lcc}
Hydro spectra & & \\
\hline
Flux (smoothed at BOSS res.) & $F$\\
Optical depth & $\tau$=-ln $F$\\
\hline
 Dark Matter Fields & & \\
\hline
Smoothed density & $\rho_s$ \\
Overdensity & $\rho = 1+\delta$ = ${\rho_s}/{\langle\rho_s\rangle}$ \\
Vorticity&  $\Omega$ \\
Velocity dispersion & $\sigma$  \\
1d Vel. divergence & $\nabla_{1d}$\\
3d Vel. divergence & $\nabla_{3d}$\\
\hline
\end{tabular}
\end{table}

In Fig.~\ref{slices}, we show a slice of the hydro flux (smoothed at BOSS resolution) and the corresponding
DM density and velocity fields (smoothed at 0.5 Mpc/h) extracted from the \hnoagnn simulation
at redshift 2.5. As expected, clear correlations are noticeable between the transmitted flux 
and the different DM fields. This trend can also be seen when studying the evolution
of each field along the same LOS, and a typical example is
given in Fig.~\ref{los}. We note that high absorptions in 
the flux correspond  to high density regions or high values in 
 the vorticity or the velocity dispersion. But the relative amplitudes of peaks in the 
 density contrast may differ from those of the the velocity dispersion/vorticity. Indeed,
 the density contrast and the velocicy dispersion/vorticity do not necessary put the emphasis of the same structures (e.g. walls, filaments) as suggested by Fig.~\ref{slices} or, for instance, by Fig.~2 of \cite{buehlmann&hahn19}. 
 On the contrary, these
high absorptions rather coincide with high negative values in 
the 3d or 1d velocity divergence. 
This is consistent since high density regions are associated with 
DM haloes in which matter tends to sink toward the center of the
objects. Note also that the variations of the modulus of the vorticity and velocity dispersion are very similar.

\begin{figure}
\begin{center}
\rotatebox{0}{\includegraphics[width=\columnwidth]{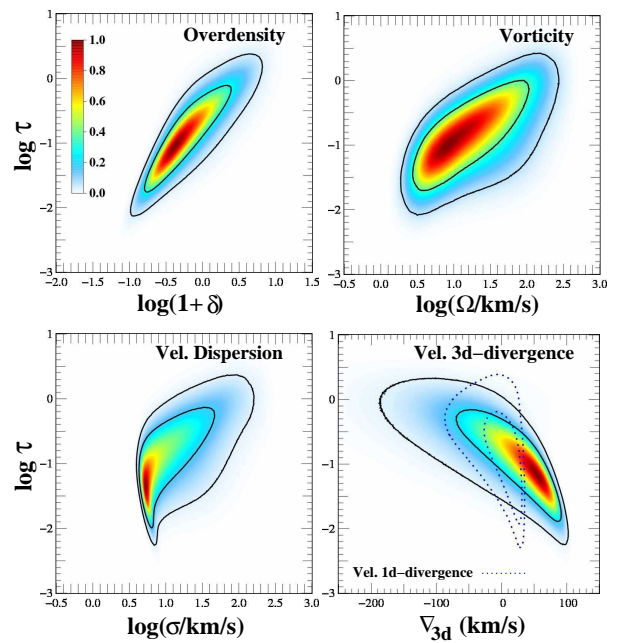}}
\caption{Correlations between the optical depth $\tau =$ -ln$F$ in the hydro spectra
(smoothed at the BOSS resolution) and dark matter quantities smoothed at 0.5 Mpc/h namely 
the overdensity ($1+\delta$), the vorticity ($\Omega$), the velocity dispersion ($\sigma$),
the 3d-velocity divergence and the 1d-velocity divergence (along the line of sight) at
$z=2.5$. 
Colors show the density of pixel using normalized values and contour line mark areas enclosing 68.27\% and 95.45\%. Hydro spectra and DM fields have been computed in redshift space and
extracted from the \hnoagnn simulation.}
\label{scatter1}
\end{center}
 \end{figure}

%

\begin{figure}
\begin{center}
\rotatebox{0}{\includegraphics[width=\columnwidth]{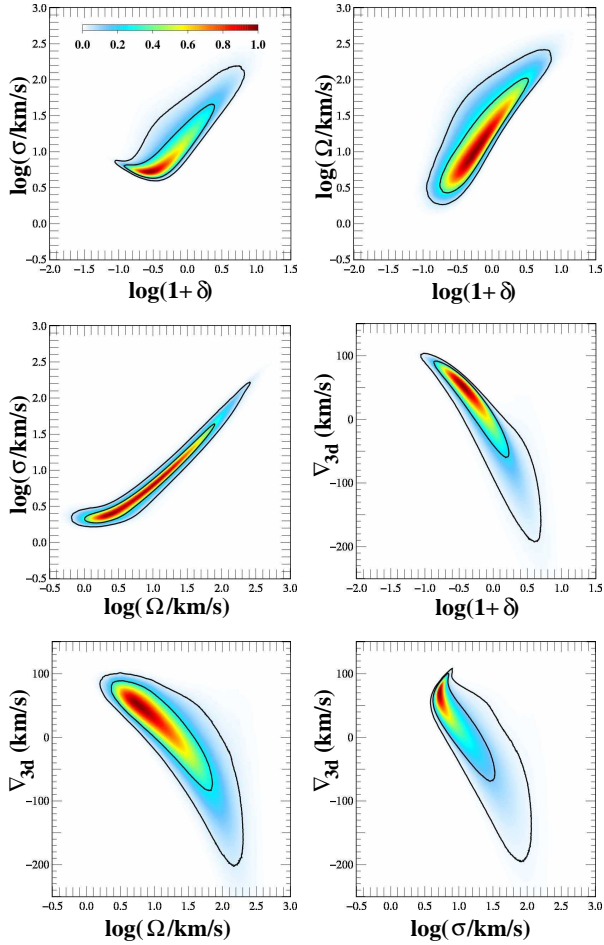}}
\caption{Some correlations between different dark matter quantities from the
Horizon-noAGN simulation at $z=2.5$. All field have been smoothed to 0.5 Mpc/h and have been computed in redshift space. Colors show again
the density of pixel using normalized values and contour line mark areas
enclosing 68.27\% and 95.45\%.}
\label{scatter2}
\end{center}
 \end{figure}

\begin{figure}
\begin{center}
\rotatebox{0}{\includegraphics[width=\columnwidth]{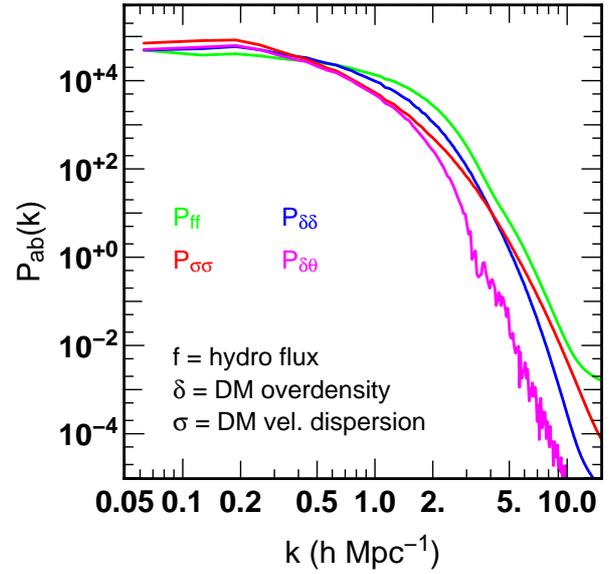}}
\caption{Examples of cross-spectrum $P_{ab}(k)= \langle a^*_k b_k \rangle$
where $a$ and $b$ refer either to the hydro flux or a DM field, 
 derived from the \hnoagnn simulation  at $z=2.5$. The flux field is 1d smoothed at the BOSS resolution
 while the DM field are smoothed at 0.5 Mpc/h.}
\label{fig:cc}
\end{center}
 \end{figure}

\subsection{Field cross correlations}
In order to characterize the correlations that emerge from Figures~\ref{slices}
and \ref{los},
we first plot in Fig.~\ref{scatter1} some relevant scatter plots  between the
optical depth $\tau$ and the DM overdensity and velocity fields.
The correlations between the optical depth in the hydro spectra
and the smoothed DM overdensity (1+$\delta$) is quite similar 
to the trend found in P14 using the 50 $h^{-1}\,\rm Mpc$ ``Horizon MareNostrum" simulation.
In Fig.~\ref{scatter2}, we additionally show the correlations between the different
DM fields.
As noticed in Fig.~\ref{los}, the velocity dispersion and vorticity field are highly
correlated. 
We do not show the correlations using the 1d velocity divergence as there are quite
similar with trends found using the 3d velocity divergence.

All these plots suggest that there are more or less pronounced correlations between
the different input DM fields. It is however tricky to anticipate which
combinations of fields through the LyMAS2 scheme would lead to the most accurate theoretical predictions.
As specified in section~\ref{sec:wiener}, we consider combinations
with up to three different DM fields
which offers 85 different possibilities (5, 20 and 60 respectively for
one, two and three input fields).
However, as the main philosophy of LyMAS is to trace the \lyaa flux from the underlying DM distribution with potential corrections from the DM veloctiy field,
we will always consider the DM overdensity field in each combination reducing this number to 17. 
Moreover, since the velocity dispersion and the vorticity fields are highly correlated, 
we will also always use the velocity dispersion in the 3d case.
Consequently, we limit our study to 8 different combinations  presented in Table~\ref{tab2}.
Nevertheless, we have checked that combinations
using only DM velocity fields do not lead to satisfactory theoretical predictions.

\begin{figure*}
\begin{center}
\rotatebox{0}{\includegraphics[width=16cm]{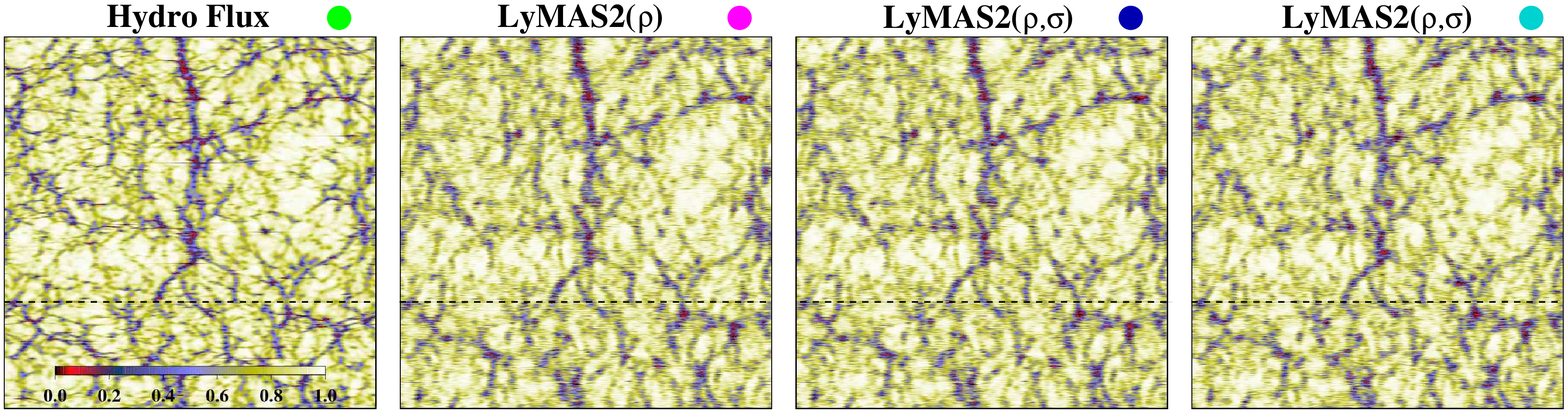}}
\rotatebox{0}{\includegraphics[width=16cm]{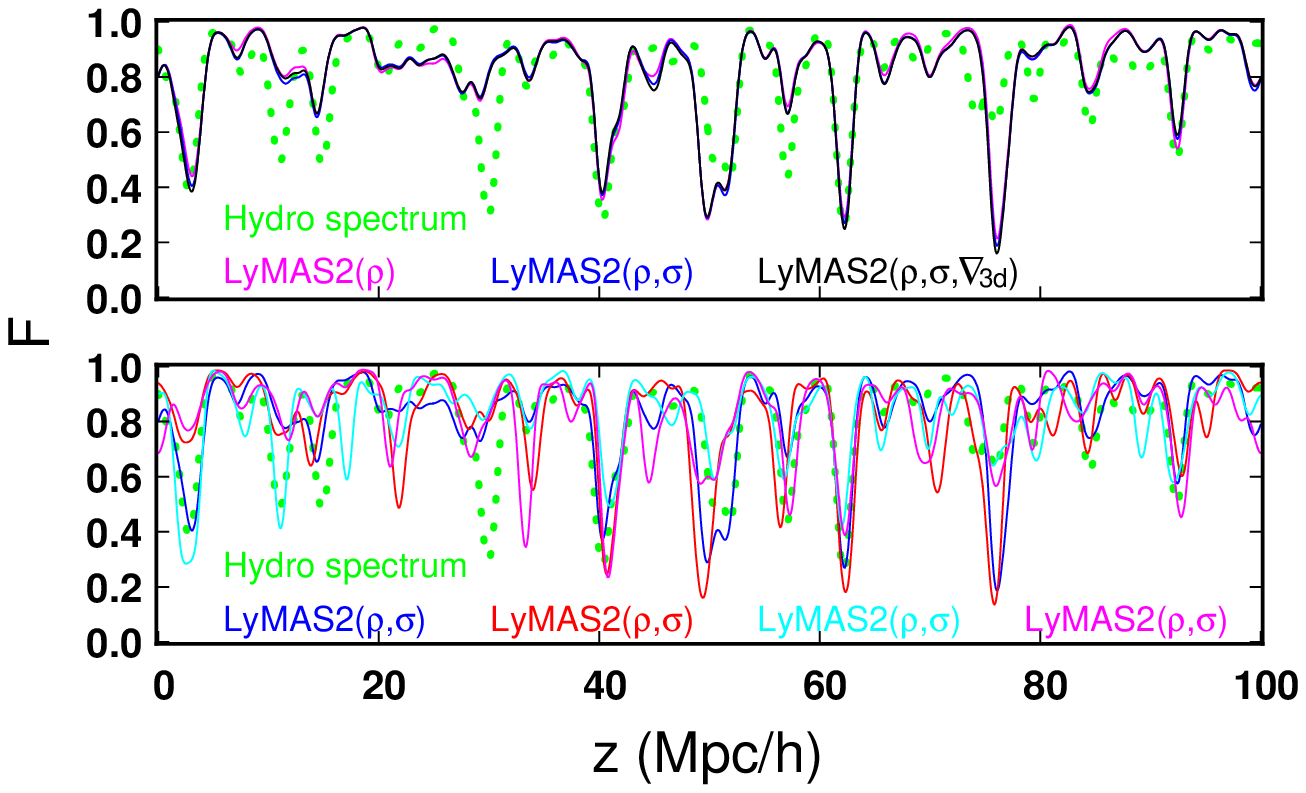}}
\caption{First line: comparison between corresponding slices through the hydro flux (top left
panel), 1-dimensionally smoothed at the BOSS resolution,
and through pseudo-spectra generated with LyMAS2 and different associations
of DM fields (3-dimensionally smoothed at 0.5 Mpc/h).
The direction of redshift distorsion is horizontal. 
Flux and DM fields have been extracted from the \hnoagnn simulation at $z=2.5$.
Visual inspection suggests a very good agreement between the clustering
of the hydro flux and those of  pseudo-spectra.
The two bottom panels show the evolution through a specific line of sight (located
by the dashed line in the different slices)
of the hydro flux (green dotted line)
and pseudo-spectra generated with LyMAS2.
In the top panel, we use the same seed to randomly generate the covariance from
equation~\eqref{eq:sigma} for each LyMAS2 realisation.
We then note that the difference between the different pseudo-spectra is quite weak.
In the bottom panel, we consider LyMAS2($\rho$,$\sigma$) to produce different
realisations of pseudo-spectra using this time different seeds to get the covariance.
In this case, the difference between pseudo-spectra can be much more pronounced.}
\label{ps_los}
\end{center}
 \end{figure*}

\begin{table}
\caption{Summary of the different DM field combinations considered in the LyMAS2 scheme}
\label{tab2}
\begin{tabular}{lllll}
Name & Field 1 & Field 2 & Field 3\\
\hline
LyMAS2($\rho$)& Overdens. &  & \\
LyMAS2($\rho$, $\sigma$)& Overdens. & Vel.disp. & \\
LyMAS2($\rho$, $\Omega$)& Overdens. & Vorticity & \\
LyMAS2($\rho$, $\nabla$$_{1d}$) & Overdens. & 1d div. & \\
LyMAS2($\rho$, $\nabla$$_{3d}$) & Overdens. & 3d div. & \\
LyMAS2($\rho$, $\sigma$, $\Omega$) & Overdens. & Vel.disp. & Vorticity\\
LyMAS2($\rho$, $\sigma$, $\nabla$$_{1d}$) & Overdens. & Vel.disp. & 1d div.\\
LyMAS2($\rho$, $\sigma$, $\nabla$$_{3d}$) & Overdens. & Vel.disp. & 3d div.\\
\end{tabular}
\end{table}
From each specific association of DM fields, and each Fourier mode $k$ along a LOS, 
we have estimated the relevant cross-spectra $P_{ab}$ defined in section~\ref{sec:general}, where $a$ and $b$ refer
either to the transmitted flux, the DM overdensity or a specific DM scalar field derived from the DM velocity field.
For each mode $k$, the covariance matrix $P_{ab}$ is hermitian, and its linear dimension is equal to the total number of fields considered. 
Examples of  cross-power spectra
are shown in Fig.~\ref{fig:cc}. 
 We also derived the relevant 1D power spectrum $P_k$ required to 
 computed the covariance $\Delta f_k$ defined in equation~\eqref{eq:sigma}.
 An example of  $P_k$ is shown in Fig.~\ref{pk_cov}.

\section{Creating pseudo-spectra with LyMAS and LyMAS2}
\label{sec:pseudospectra}

In this section, we apply the LyMAS2 scheme to 
the DM fields extracted from the \hnoagnn simulation
to generate grids of pseudo-spectra at BOSS resolution. The objective is to 
recover the 3d \lyaa clustering statistics of the ``true" hydro spectra. 
For a given skewer, we summarize the main steps to follow to produce
a corresponding pseudo-spectrum { using either LyMAS or LyMAS2}:

\smallskip
\noindent{}
{
\begin{center}
    { The LyMAS scheme:}
\end{center}
\noindent{}
1. Extract the smoothed overdensity field $\rho$ for a specific skewer.

\smallskip
\noindent{}
2. Create a realization $G_\mathrm{per}(x)$ of a 1d Gaussian random field from the 1d power spectrum of the Gaussianized percentile spectra derived from the hydro simulation.

\smallskip
\noindent{}
3. Degaussianize $g_\mathrm{per}(x)=G^{-1}(G_\mathrm{per})$ to get a realization 
of a percentile spectrum.

\smallskip
\noindent{}
4. Create a pseudo spectrum by drawing the flux at each pixel
from the location in $P(F|\rho)$, implied by the value of $g_\mathrm{per}(x)$ (see Eq.6 in P14).

\smallskip
\noindent{}
5. One full iteration.
We first measure the 1d flux power spectrum
$P_\mathrm{ps}(k)$ of the pseudo-spectra created in this way. Then we Fourier 
transform each pseudo-spectrum and multiply each
of its Fourier component by the ratio $[P_\mathrm{F}(k)/P_\mathrm{ps}(k)]^{1/2}$, 
inverse transform to get the same 1d flux power spectrum than
of the true hydro spectra $P_\mathrm{F}(k)$ \citep{weinberg92}.
The second step of the full iteration is to compute the PDF of the pseudo-spectra
after the 1d Pk re-scaling and then monotonically map the flux value to match
the PDF of the true hydro spectra.
This full iteration can be repeated several times. However, 
as we will see, one or two full iterations are enough to get excellent agreement with
the 1d power spectrum up to quite high $k$.  

\begin{center}
    { The LyMAS2 scheme:}
\end{center}

1. Extract and gaussianize the smoothed overdensity field $\rho$ and eventually
one or two additional DM velocity fields (e.g. $\sigma$)
for a specific skewer.

\smallskip
\noindent{}
2. Compute the FFT of each gaussianized field. This gives new (complex) fields,
$\rho_k$, $\sigma_k$, etc.. 

\smallskip
\noindent{}
3. Compute (in Fourier space) the most probable flux 
$\bar{f_k}= T_1\cdot\rho_k + T_2\cdot\sigma_k + ...$,
by applying the relevant filters $T_1$, $T_2$,$...$ (see 
for instance equation~\eqref{eq:fluxmean2d} for the 2-fields case). 

\smallskip
\noindent{}
4. Generate a 1d Gaussian field
of mean 0 and variance defined in equation~\eqref{eq:sigma}
to get the covariance $\Delta f_k$. 

\smallskip
\noindent{}
5. After computing the inverse Fourier transform of $\bar{f_k}+\Delta f_k$ to get $f$,
degaussianize to get the pseudo-spectrum: $F = G^{-1}(f)$. 

\smallskip
\noindent{}
6. One full iteration. Same procedure as 4. in the LyMAS scheme.
}

In Fig.~\ref{ps_los}, we compare the same slice through the
hydro flux  and through different realisations of LyMAS2 using
different combinations of DM fields. The clustering of each  pseudo-spectrum
is in fair agreement with the
clustering of the hydro spectra. Comparing pseudo-spectra and hydro flux
along a specific skewer, also shown in Fig.~\ref{ps_los}, 
confirms that LyMAS2 correctly models low and high absorptions at good locations, 
though amplitudes may differ. 
The second line of Fig.~\ref{ps_los} compares three pseudo-spectra
generated with LyMAS2 using three different field combinations but using the same
seed for the random process to get the variance $\Delta f_k$. It's interesting to see
that these different pseudo-spectra look also the same, which explains
why the slices presented in Fig.~\ref{ps_los} are very
similar.
On the contrary, the third line of Fig.~\ref{ps_los} shows four different
realisations of pseudo-spectrum from LyMAS2 using the DM overdensity and velocity dispersion
fields and different seeds to get the covariance $\Delta f_k$. In this case, 
the amplitude of absorptions can be quite different.

In the next sections, we study in more detail the clustering statistics of
each catalog of pseudo-spectra produced with LyMAS2. We aim at
recovering three observationally relevant statistics of the transmitted 
flux: the probability density function (PDF),
the line-of-sight power spectrum
and the 3d clustering (through the 2-point correlation function). As
we will see, both LyMAS and LyMAS2 reproduce the PDF of the hydro simulations by construction
and nearly reproduce the hydro simulations line-of-sight power spectrum by construction
(step 6 above). The power of LyMAS is to produce accurate large-scale 3d clustering
while also reproducing these line-of-sight statistics.

\begin{figure}
\begin{center}
\rotatebox{0}{\includegraphics[width=\columnwidth]{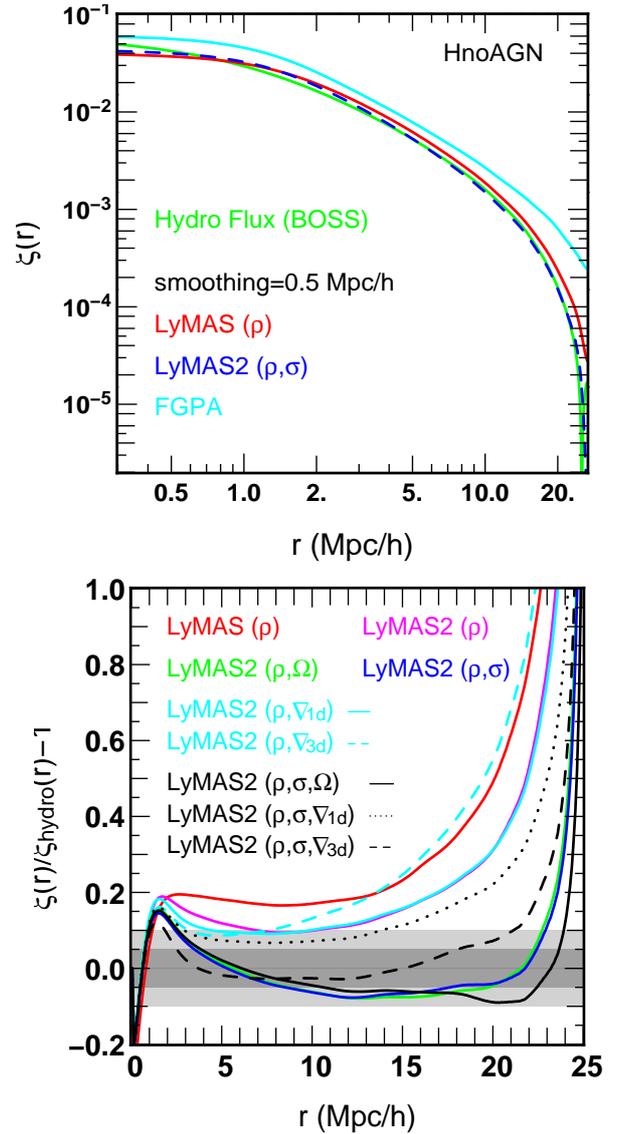}}
\caption{Top panel: the correlation function $\xi$ as a function of the separation $r$.
We present results from the true hydro spectra (green line) and results from
the LyMAS (red line) and the new version LyMAS2 considering the DM overdensity and the velocity dispersion fields. The predictions from FGPA (cyan line) are also shown. DM fields are all smoothed at the scale
0.5 Mpc/h.
Bottom panel: the relative difference as respect to the true hydro spectra (i.e. $\xi(r)/\xi_\mathrm{hydro}(r)-1$) are shown for different combination of
choices in the Wiener filtering. Here all DM fields have been smoothed to $=0.5$ Mph/h. 
Compared to the traditional LyMAS scheme (red lines), the
new version significantly improves the predictions. In particular  the DM overdensity field associated to 
the DM velocity dispersion (blue line) or the vorticity (green line)
leads to relative difference lower than 10\% and close to 5\% in most of
the range we are interested in. On the contrary, using the velocity divergence (cyan lines) does not seem
to improve much the results.}
\label{fig_CF1}
\end{center}
 \end{figure}

\begin{figure*}
\begin{center}
\rotatebox{0}{\includegraphics[width=16cm]{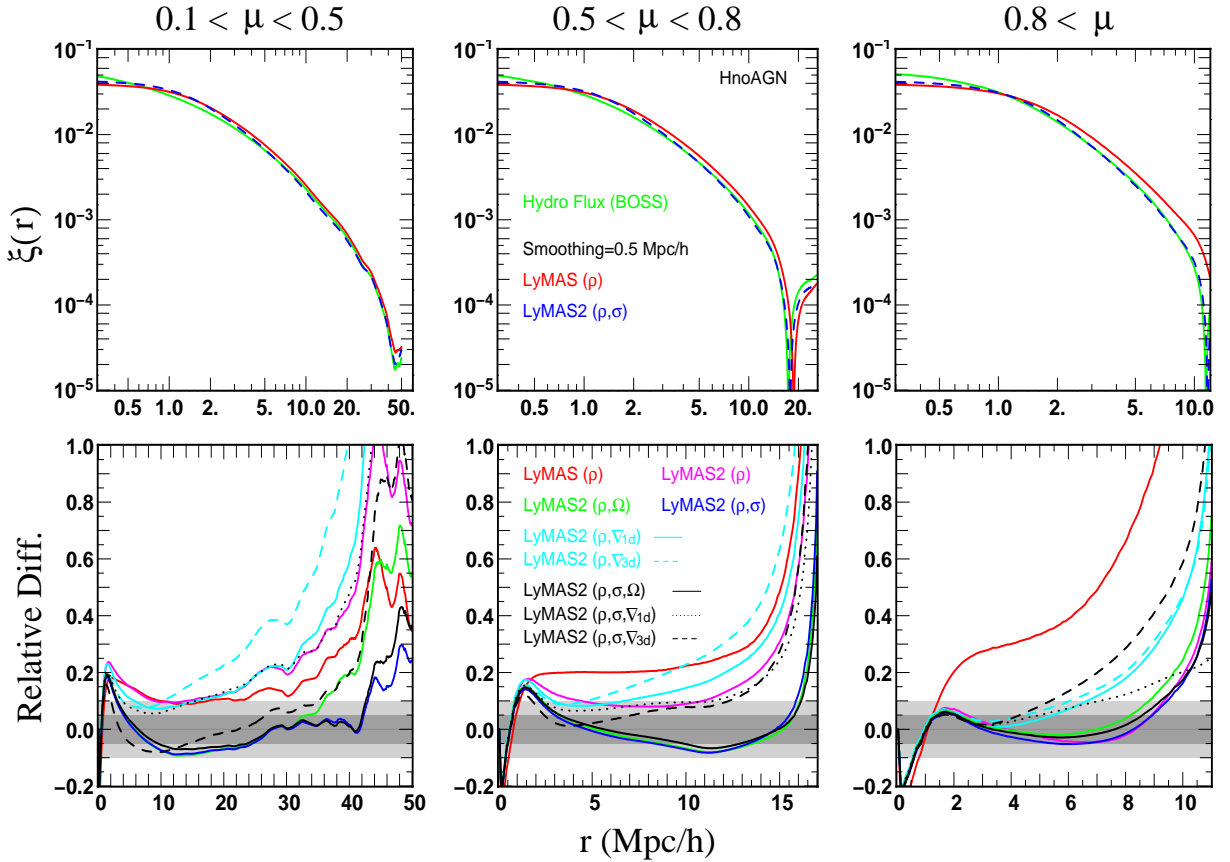}}
\caption{Same as Fig.\ref{fig_CF1} but with a dependency to range of angles $\mu$, as defined in the text.
These plots confirm that the association of the DM overdensity and the velocity dispersion (or vorticity)
 lead to very accurate predictions (see blue or green lines). The corresponding relative differences shown in the bottom panel
 are generally within 10\% and 5\% over a wide range of $r$.
It is also clear that the new LyMAS2 scheme is much more accurate for 
large angles where the 
traditional LyMAS lead to errors that grows quickly with the distance
$r$ (red lines). 
DM fields are again smoothed  at $=0.5$ Mph/h.}
\label{fig_CF2}
\end{center}
 \end{figure*}

\subsection{3d-clustering}
\label{section_3dclustering}

In order to compare the 3d clustering between the hydro and pseudo spectra,
we rely on the 2-point correlation function $\xi(r)$ 
defined by 
\begin{equation}
\xi(r) = \frac{\langle F(x)F(x+r)\rangle}{\langle F\rangle^2}  -1\,,
\end{equation}
as a function of the separation ${\bf r}$. 
To study the effect of redshift distortions, 
we also consider the 2-point correlation function averaged
over bin of angle $\mu$ defined for a pair of pixels ($i$,$j$) by
$(r_i-r_j)_\parallel/{\bf r}$, where ${\bf r} = |r_i-r_j|$ and $(r_i-r_j)_\parallel$
the component along the line of sight.

The top panel of Fig.~\ref{fig_CF1} shows the full 2-point correlation functions
derived from pseudo-spectra using either the first version of LyMAS (red line) or
LyMAS2 using the DM overdensity and velocity dispersion field (blue line). 
Compared to the results of the hydro flux (green line), one can see that LyMAS2 
is significantly improving the predictions which are remarkably close to the hydro spectra results.
In order to estimate the precision of these reconstructions, we plot in 
the bottom panel of Fig.~\ref{fig_CF1} the relative difference,
i.e. $\xi/\xi_\mathrm{hydro}-1$ for different combinations of DM fields.
It appears clearly that LyMAS2 leads in general to much more accurate predictions than LyMAS.
Indeed, when considering the DM overdensity field only, LyMAS2($\rho$) give slightly better results (magenta line)  
but the addition of the velocity dispersion lead to errors that are
generally lower than 10\% and close to 5\% (e.g. blue or black lines).
Also, similar trends are obtained when the vorticity is taken into account (bottom panel, green line), which is not
surprising as these two fields are highly correlated.
On the contrary, the 1d and 3d velocity divergence, when associated to the DM overdensity only (cyan lines), do not seem to 
improve much the predictions as respect to the first version of LyMAS.
Note that in linear theory, the 3d velocity divergence is fully correlated with the density field and therefore adds no additional information. On the other hand, the vorticity and/or velocity dispersion are sourced by the non-linear evolution of the matter fields and therefore add complementary information on small scales.
Finally, we also note that errors are close to 20\% at $r\sim$1-2 Mpc/h probably due to effect of smoothing.

\begin{figure}
\begin{center}
\includegraphics[width=\columnwidth]{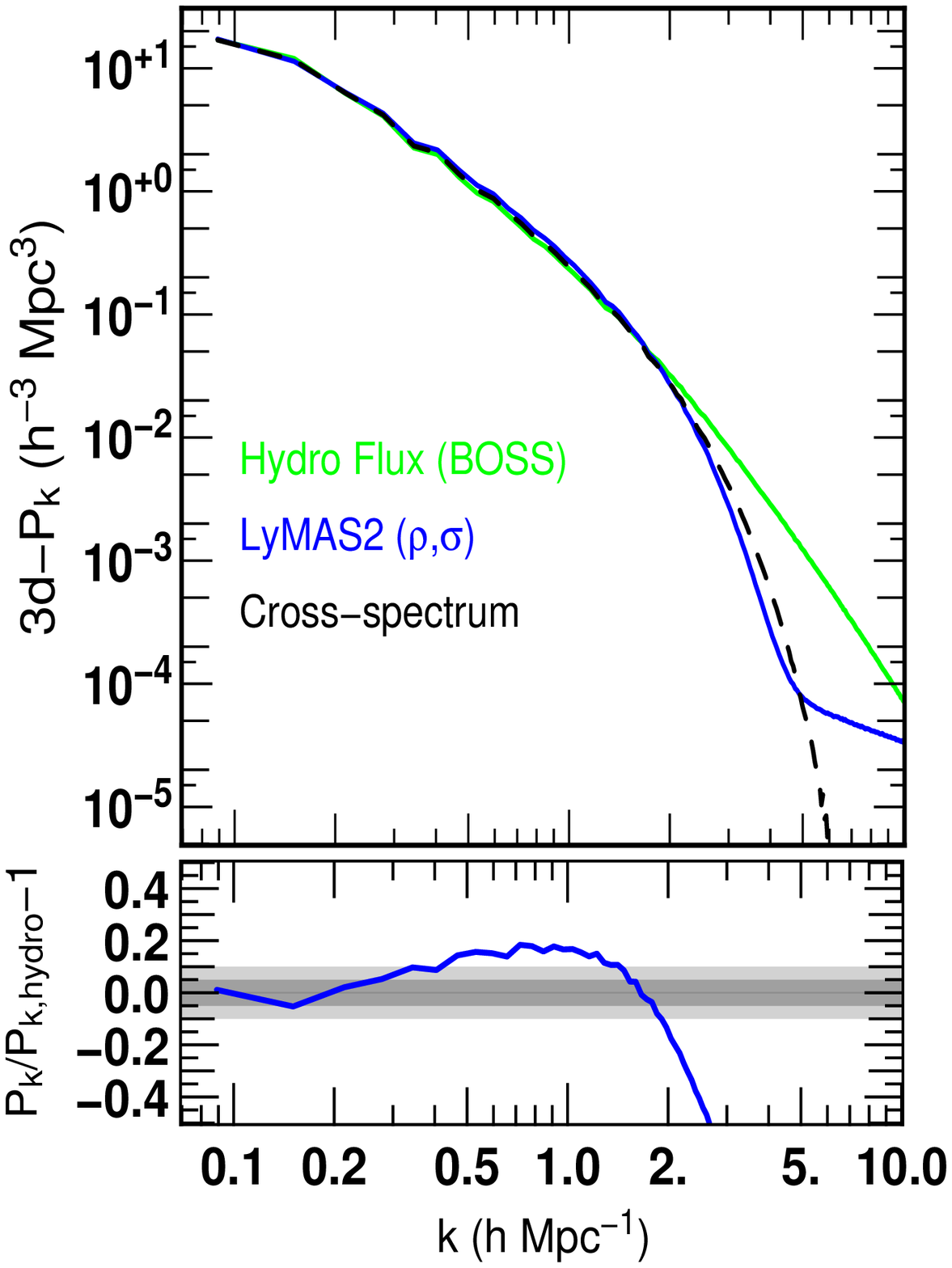}
\vspace{-0.5cm}
\includegraphics[width=\columnwidth]{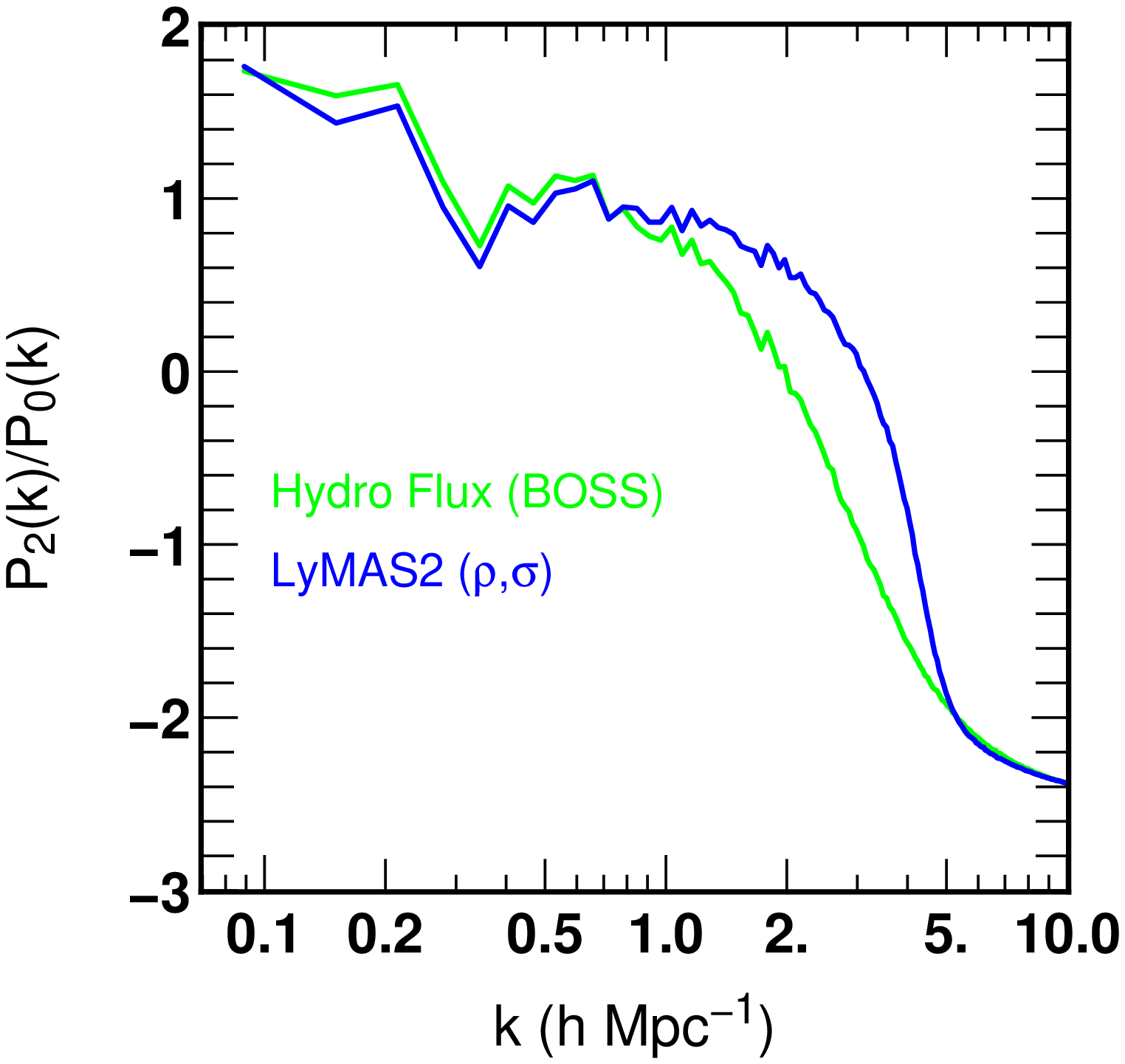}
\caption{Top panel: the 3d power spectrum derived from the (BOSS) hydro 
flux (green line) and from pseudo-spectra generated with
LyMAS2($\rho$,$\sigma$). All DM fields are smoothed at 0.5 Mpc/h.
The black dashed line represents the cross-spectrum.
Bottom panel: The corresponding quadrupole to monopole ratios (same color code).  
}
\label{fig_powerspectrum}
\end{center}
 \end{figure}

We now investigate how the predictions of the two point correlation functions
vary when considering an angle $\mu$. The trends are presented in Fig.~\ref{fig_CF2}
for three ranges of values ($0.1<\mu<0.5$,  $0.5<\mu<0.8$ and $0.8<\mu$) following P14.
The results confirm that LyMAS2 significantly improve the predictions of the \lyaa clustering. 
In particular, some combinations such as ($\rho$,$\sigma$) still lead to errors generally lower than 10\% and most of the time close to 5\%. We also
note that 
LyMAS2 is particularly efficient for reproducing the correlations along transverse separations or high angles (i.e. $\mu>$0.8) 
in which the error is most of the time lower than 5\%.
The top panels of Fig.~\ref{fig_CF2} indicate again a remarkably good agreement between the 2 point correlation functions
of the hydro flux and those derived from pseudo-spectra produced from LyMAS2($\rho$,$\sigma$) even for high angles $\mu>$0.8)
where the previous version of LyMAS is quite inaccurate.
For the two larget $\mu$ bins, the correlation functions eventually 
drop rapidly to zero at large $r$. In this regime, the fractional error in $\xi$ are
inevitably large, even though the absolute errors are small. It is evident that LyMAS2
captures the scale of these zero-crossing more accurately than LyMAS.

As a first conclusion, the LyMAS2 scheme is significantly improving the
predictions of the \lyaa 3d clustering especially
when the DM overdensity field is associated with the velocity
dispersion or the vorticity field.
{ 
For the sake of comparison with results presented in the literature,
we also compare the 3d power spectrum and corresponding quadropole to monopole ratios
in Fig.~\ref{fig_powerspectrum} derived from both the hydro spectra and the 
pseudo-spectra generated with LyMAS2($\rho$,$\sigma$). The (monopole) power spectrum is defined in the usual way as $\langle\tilde{F}(\bmath{k})\tilde{F}(\bmath{k}')\rangle = (2\pi)^3P(k)\delta_{\mathrm{D}}(\bmath{k+k}')$, with $\tilde{F}(\bmath{k}) = \int \mathrm{d}^3x F(\bmath{x})\mathrm{e}^{-i\bmath{k.x}}$. Defined this way, we have the following expression of the variance, $\sigma^2=\int_0^\infty k^3 P(k)\mathrm{dlog}k/(2\pi^2)$.
From Fig.~\ref{fig_powerspectrum},
the 3d power spectrum of the hydro spectra
can be faithfully recovered from the LyMAS2 simulated spectra up to
modes $\sim$2 h\,Mpc$^{-1}$.

For larger modes, however,  the predictions are becoming
less accurate as separations get lower than the considered smoothing scale (0.5 Mpc/h here). Indeed, we observe a lack of power at small scales (2$\leq$k$\leq$10h\,Mpc$^{-1}$) in 
the 3d power spectrum of LyMAS2 simulated spectra, compared to the hydro power spectrum, 
which is mainly explained by the fact that the transverse correlations
are not accounted for in the Wiener filtering scheme, and in particular transverse fluctuations at small scales are not generated in the present scheme. On the other hand, the absence of correlation, between the stochastic realisations at small scales for each line of sight, induces an artificial flattening of the reconstructed power spectrum for modes k$\geq$5h\,Mpc$^{-1}$.
The ratio of the quadrupole to monopole power is an even stricter test as it traces the anisotropic structure of power in the field, and one can see differences in such a ratio already for modes k$\geq$2h\,Mpc$^{-1}$. This test would clearly benefit from accounting for transverse correlations.

It would be interesting to correct this in a forthcoming 
work though this point is not critical. Indeed the transverse separations of spectra from existing surveys are generally much larger than 1 Mpc/h, and on these scales the transverse modes are properly reconstructed. 
Taking into account transverse correlations is straightforward however, and will be worthwile to generalize this method to emission spectra, for which all transverse scales are important. We will therefore include them in future works.

}

\subsection{1d flux power spectrum along LOS}

\begin{figure}
\begin{center}
\rotatebox{0}{\includegraphics[width=\columnwidth]{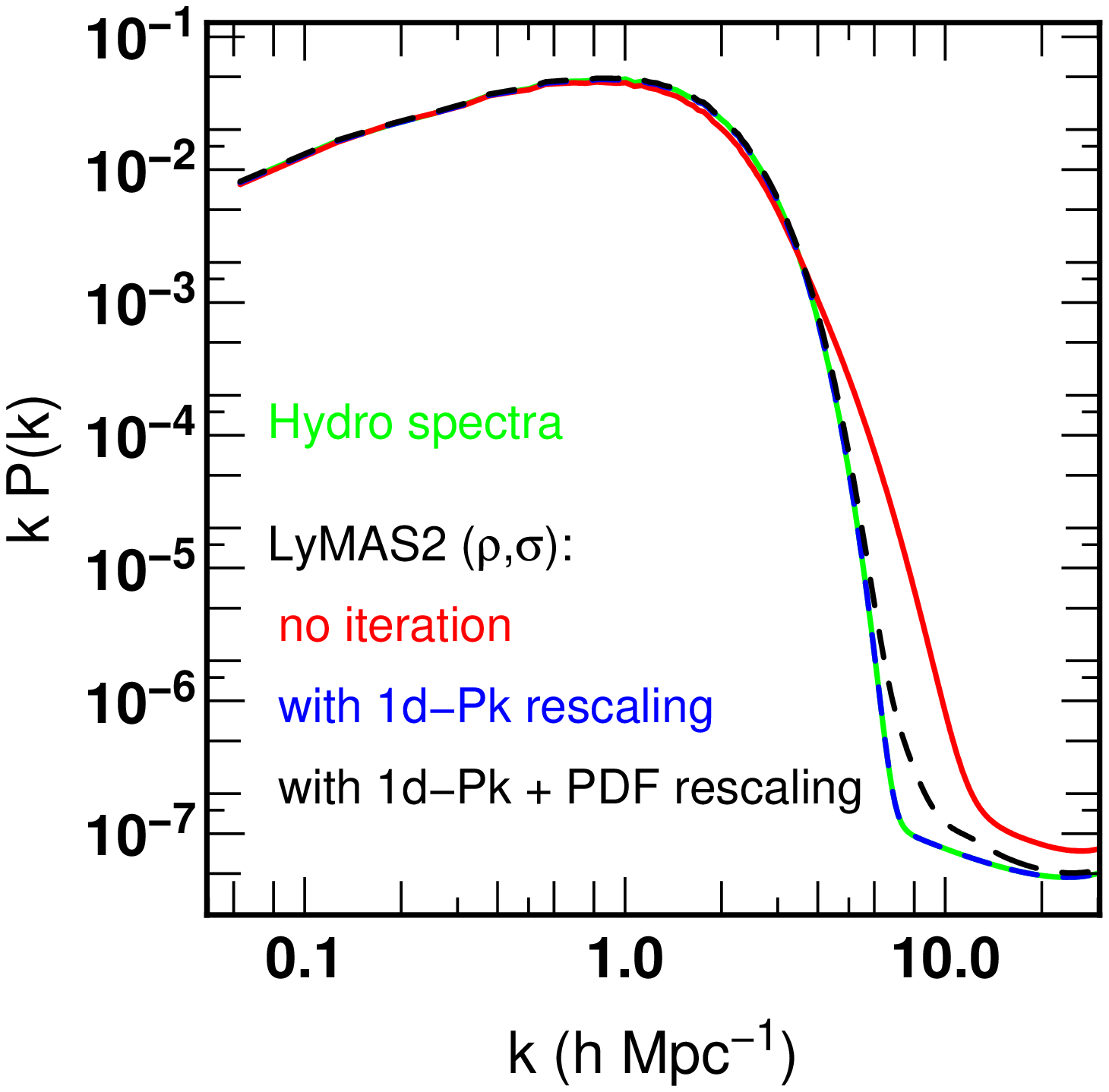}}
\rotatebox{0}{\includegraphics[width=\columnwidth]{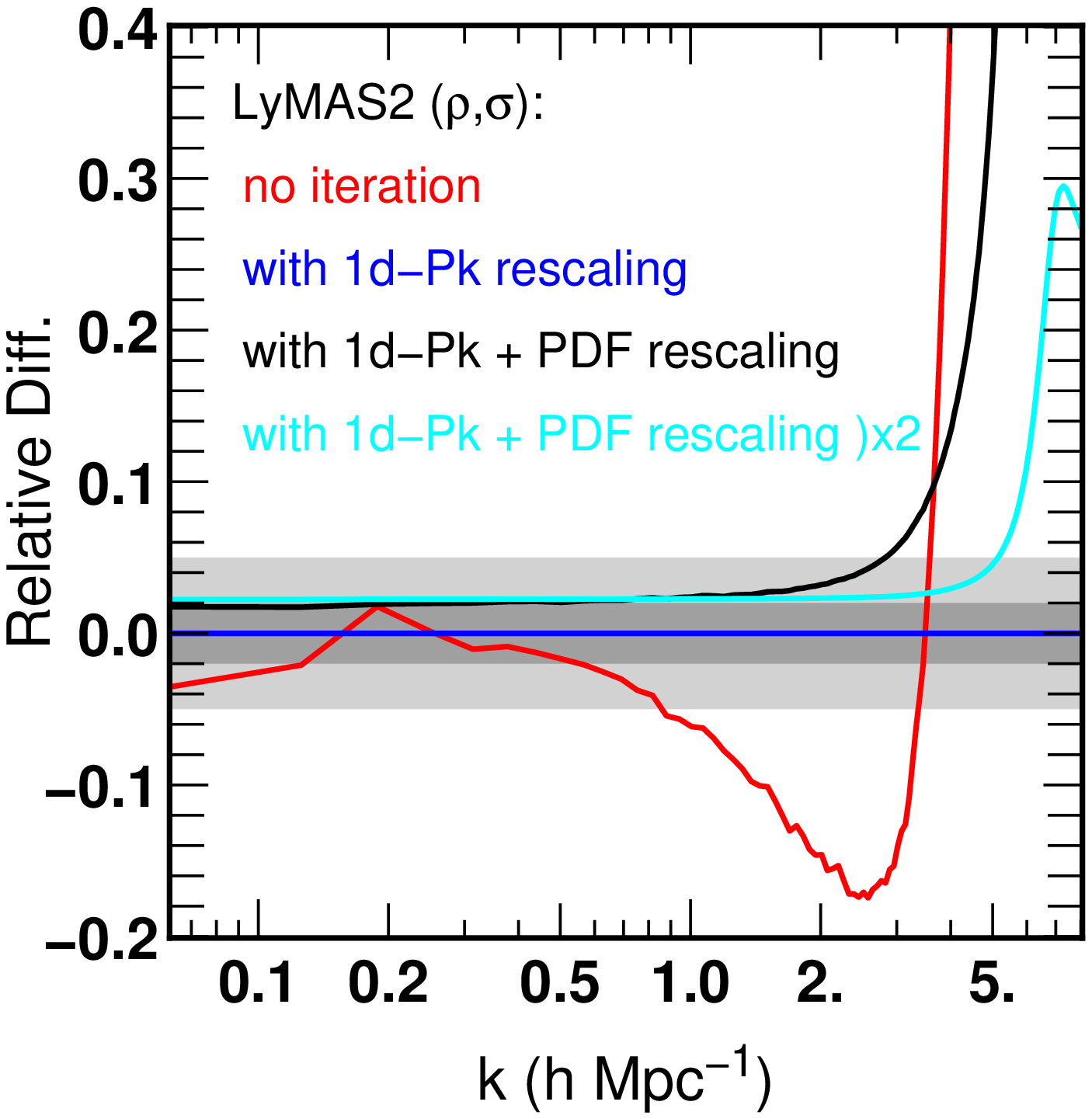}}
\caption{Top panel: the dimensionless 1d power spectrum  of
the true spectra from the Horizon-noAGN simulation (green line) at $z=2.5$ and
from coherent pseudo spectra using \lymasw considering the DM overdensity and
velocity dispersion fields. We show results before (red line) and after
(blue and black lines) 1d power spectrum and PDF transformations (described in
the text). Bottom panel: the relative 
difference as respect to the hydro results (i.e. $P_k/P_{k,\mathrm{hydro}}-1$).
A full iteration (flux 1d-Pk and PDF re-scaling) permits to recover the hydro power spectrum
with an error of $\sim$ 2\% over a wide range of $k$.
The light and dark grey shaded areas indicate regions where
the error is less than 5 and 2\% respectively.}
\label{fig_pkhnoagn}
\end{center}
 \end{figure}

We also aim at producing catalogs of pseudo-spectra that look like
spectra measured by a specific instrument i.e. BOSS in the present study.
Therefore,  the 1d flux power spectrum of each LyMAS mock should be as 
close as possible to the hydro spectra 1d Pk.
In the following, we only present the results derived from LyMAS2($\rho$,$\sigma$)
as same trends are obtained when considering any other combination of
DM fields. 
{ Here, the 1d power spectrum is formally defined as $\langle\hat{F}(k)\hat{F}(k')\rangle = (2\pi)P_{\mathrm{1D}}(k)\delta_\mathrm{D}(k+k')$, where $\hat{F}(k) = \int \mathrm{d}x F(x) \mathrm{e}^{-ikx}$ is the 1d Fourier Transform along the line of sight\footnote{Note that the normalisation is reduced by a factor of 4, as compared to the definition used in P14, which relied on a Fourier series in trigonometric functions.}. When estimating it, we take FFTs along each line of sight and average the result. The expression of the variance is then $\sigma^2 = \int_0^\infty kP_\mathrm{1d}(k)\mathrm{dlog}(k)/\pi$.}
Fig.~\ref{fig_pkhnoagn} shows the dimensionless 1d power spectrum before
power spectrum transformation (red line) and after applying the power
spectrum and PDF transformation described in the text (black line).
We first note that LyMAS2 without iteration reproduces the 1d power spectrum
more accurately than original LyMAS (see Fig.13 of P14). 
Then as expected, the 1d-$P_k$ transformation leads to 
 same power spectrum as the hydro simulation (blue line), by construction.
 The second step of the iteration is to re-scale the flux PDF, and
 this transformation slightly alters the 1d power spectrum.
However, as illustrated in the bottom part of Fig.~\ref{fig_pkhnoagn},
the relative difference is close to 2\% up to high values of $k$ (i.e  $k\sim$2 h\,Mpc$^{-1}$).
If one repeats a full iteration a second time, the same accuracy is reached
for even higher $k$ values ($\sim$4 h\,Mpc$^{-1}$).

\subsection{1-point PDF of the Flux}

\begin{figure}
\begin{center}
\rotatebox{0}{\includegraphics[width=\columnwidth]{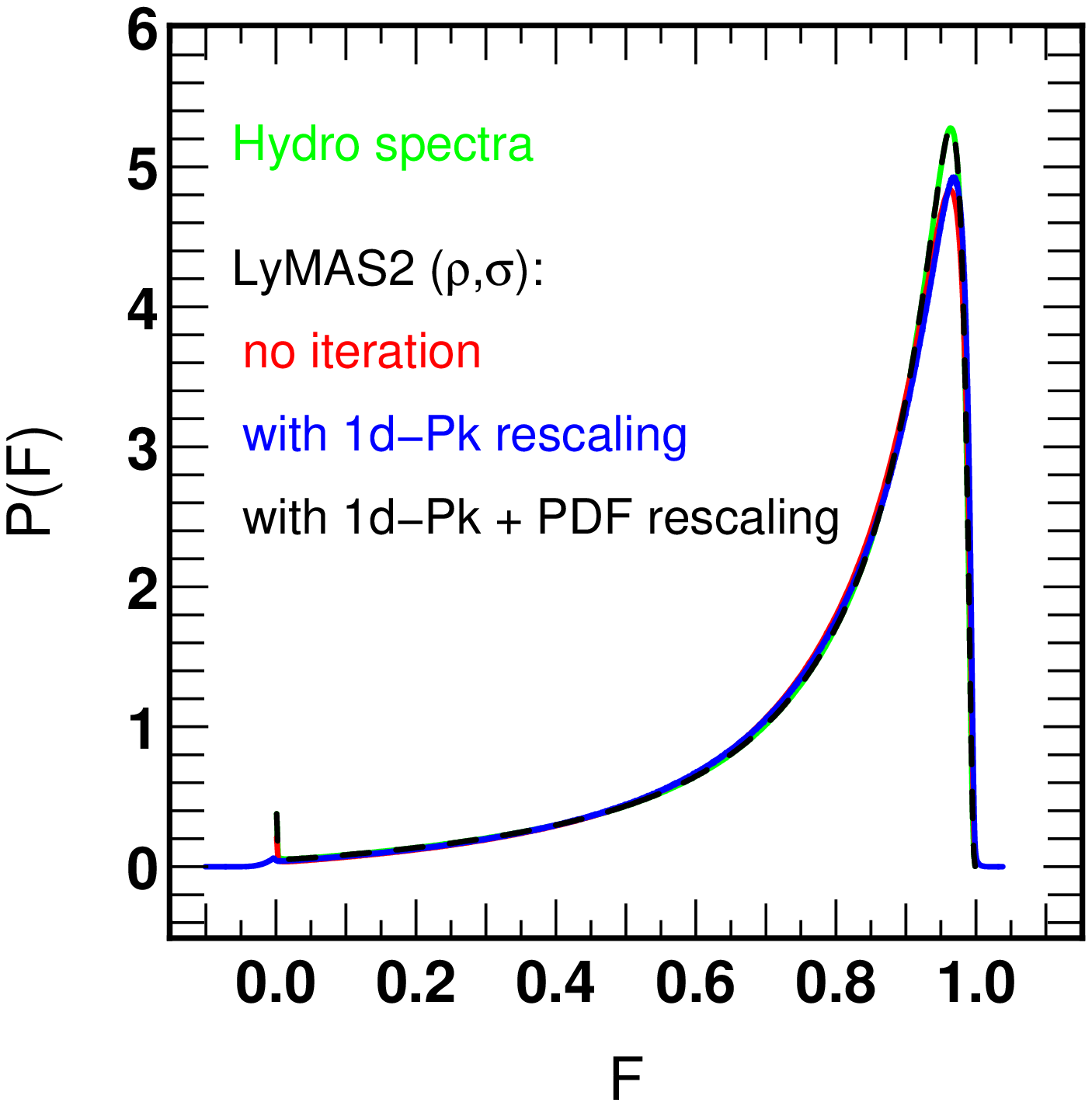}}
\rotatebox{0}{\includegraphics[width=\columnwidth]{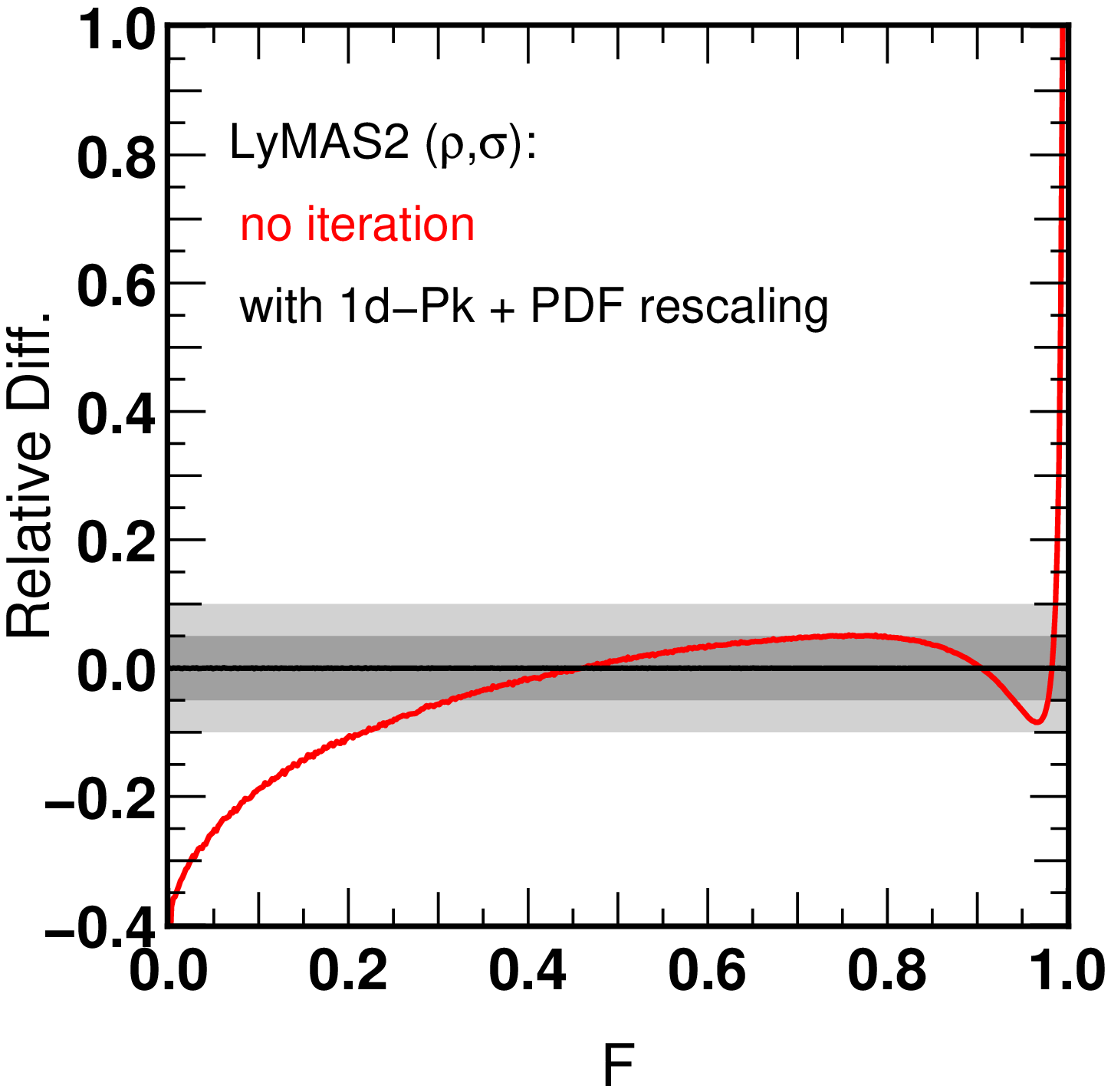}}
\caption{Top panel: the PDF of
the true spectra from the Horizon-noAGN simulation at $z=2.5$ (green line) and
from coherent pseudo spectra using LyMAS2 considering the DM density and
velocity dispersion fields. We show results before (red line) and after
(blue and black lines) 1d power spectrum and PDF transformations (described
in the text). The 1d Pk re-scaling can lead to non-physical Flux values (i.e.
$F<0$ or $F>1$). Bottom panel: the relative 
difference with respect to the hydro results (i.e. $\mathrm{PDF}/\mathrm{PDF}_\mathrm{hydro} -1 $).
The full scheme permits to recover the hydro flux PDF exactly.
The light and dark grey shaded areas indicate regions where
the error is less than 10 and 5\% respectively.}
\label{fig_pdfhnoagn}
\end{center}
 \end{figure}

Since the LyMAS scheme ends after a flux PDF re-scaling (second step of a full iteration),
this ensures that the 1-point PDF of the hydro flux and the pseudo spectra match exactly.
To illustrate, we present 
in Fig.~\ref{fig_pdfhnoagn} the results obtained with LyMAS2($\rho$,$\sigma$) before
(red line) and after (blue line) a full iteration (1d Pk and PDF re-scaling).
Without transformation, the PDF of the pseudo-spectra is already close to 
the PDF of the hydro spectra (green line). But fractional errors on the PDF can be quite large 
for low or high values of $F$.
After the 1d Pk re-scaling, the PDF of the pseudo-spectra has slightly changed 
and has some non-physical values ($F<0$ and $F>1$). But the second step of the
iteration corrects the PDF { (e.g. transforms these unphysical values into physical ones)},
to exactly match that of the hydro spectra.

\subsection{Comparison with the FGPA}
\label{sec:fgpa}
The  FGPA  essentially converts  DM  density  to  optical  depth  using a  physical model motivated  by  photoionization equilibrium, assuming  that  all  gas  contributing  to  the \lyaa  lies  on  a temperature-density  relation
$T\propto(\rho_s/\overline{\rho_s})^{\gamma-1}$. The predicted flux is:
\begin{equation}
F = Ae^{-(\rho_s/\overline{\rho_s})^{2-0.7(\gamma-1)}}\,,
\label{eq:fgpa}
\end{equation}
where $2-0.7(\gamma-1)\approx0.6$ for the values of $\gamma$ expected well after reionization 
\citep{weinberg+98,croft+98,peeples+10,mcquinn09}. 
This relation is reasonable for modeling high-resolution spectra. However,
due to existing non-linear relation between flux and optical depth, it
does not automatically apply at low resolution (though it omits some physical effects
in the high-resolution case).
From the \hnoagnn DM overdensity grid smoothed at 0.5 Mpc/h, we have first generated 1024$\times$1024 pseudo-spectra
using equation~\eqref{eq:fgpa} by estimating $A$ so that $\langle F \rangle=0.795$. Then, we one-dimensionally
smoothed each pseudo spectrum to BOSS resolution. Similarly to the
LyMAS scheme, we end the process by rescaling the flux 1d-Pk and PDF.
The correlation function is shown in the top panel of Fig.~\ref{fig_CF1} and is considerably
overestimated as respect to the hydro flux with a relative error greater than 50\%
(omitted in the bottom panel for the sake of clarity). Such a trend is consistent with the results
of \cite{sorini+16} who found that typical relative errors
in the 3d power spectrum are $\sim$80\% when a DM smoothing scale of 0.4 Mpc/h is considered.
In Appendix~\ref{appendix2}, we will investigate other deterministic mapping than the FGPA. But
will we see that the main conclusion remains unchanged: deterministic sampling
generally tends to significantly overestimate the flux 3d-correlation 
especially when the dark matter density is smoothed to scales greater than $0.3$Mpc/h. 
Note that a similar trend is obtained when studying the correlation  between  the  \lyaa transmitted  flux  and  the mass overdensity (see Fig.~1 of \cite{cai+16}).

\subsection{Horizon-AGN VS Horizon-noAGN}

\begin{figure}
\begin{center}
\rotatebox{0}{\includegraphics[width=\columnwidth]{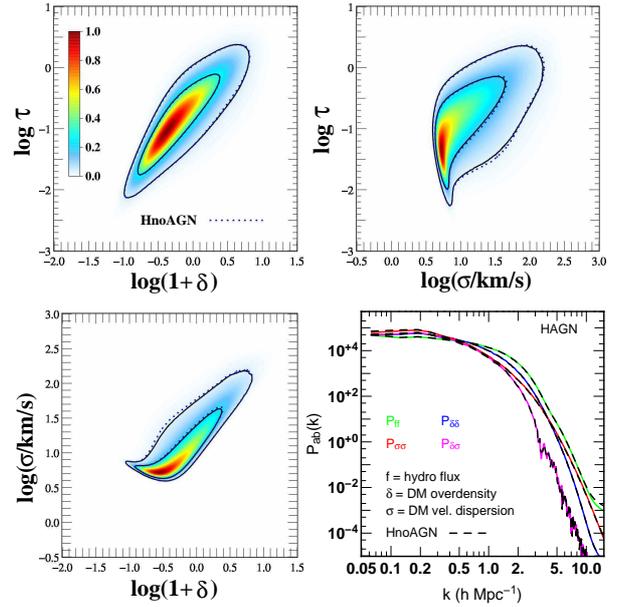}}
\caption{Same as Figs.~\ref{scatter1} and \ref{scatter2} but using \hagnn for the calibration. Here we compare the correlations between the optical depth $\tau =$ -ln$F$ in the hydro spectra
(smoothed at the BOSS resolution) and dark matter quantities smoothed at 0.5 Mpc/h namely 
the overdensity ($1+\delta$) and the velocity dispersion ($\sigma$) at $z=2.5$. We also
show in the lower right panel, some relevant transfer functions (i.e. cross-spectrum)
similarly to Fig.~\ref{fig:cc}. In all panels, the dotted lines correspond to results
from the \hnoagnn simulation. Very similar trends are then obtained
when AGN are included or not.
}
\label{fig_agnvsnoagn1}
\end{center}
 \end{figure}

\begin{figure}
\begin{center}
\rotatebox{0}{\includegraphics[width=\columnwidth]{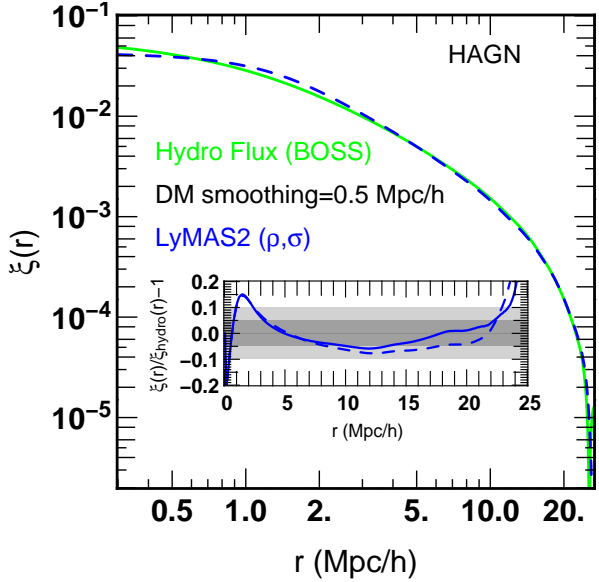}}
\caption{The correlation function $\xi$ as a function of the separation $r$.
We present results from the \hagnn true hydro spectra (green line) and results from
LyMAS2 considering the DM overdensity and the velocity dispersion fields (blue dashed line).
DM fields are all smoothed at the scale 0.5 Mpc/h.
The central small panel indicates the relative difference as respect to the true hydro spectra (i.e. $\xi(r)/\xi_\mathrm{hydro}(r)-1$).
The blue dashed line represents here the result from the \hnoagnn simulation. }
\label{fig_agnvsnoagn2}
\end{center}
 \end{figure}

{
In the present work, we used the \hnoagnn simulation for the calibration of
LyMAS2, mainly to minimise the computational cost as we derived
five additional but lower hydrodynamical simulations to study
both the robustness of the results (see appendix~\ref{appendix1}) as well as the effect of cosmic variance (see section~\ref{sec:cosmic_variance}).  However, since AGN feedback 
may induce subtle modifications in the spatial distribution and in the clustering of the
\lyaa forest, it is important to check if eventual noticeable differences can be
seen in the statistics we present so far.
For this reason, we have repeated to same and whole analysis but considering this time \hagnn
for the calibration. For instance, we plot in Figure~\ref{fig_agnvsnoagn1} some
relevant scatter plots showing the correlations between the optical depth, 
the DM overdensity and the DM velocity dispersion, similarly to Figures~\ref{scatter1} and
\ref{scatter2}. We also show some transfer functions (i.e. cross-spectrum) that we compare
to results from Figure~\ref{fig:cc}. In all cases, the statistics have been derived 
using a DM smoothing of 0.5 Mpc/h.
Compared to results from \hnoagn, we found very similar trends when AGN are included. 
The comparison of the 2-points correlation function of the hydro flux and LyMAS2($\rho$,$\sigma$)
pseudo spectra in Figure~\ref{fig_agnvsnoagn2} confirms this by suggesting predictions
with a very similar accuracy when AGN is included or not.
In conclusion, the inclusion of galactic winds does not seem to affect significantly the clustering statistics of the \lyaa Forest, given our smoothing scales and targeted accuracy,
consistent with results of \cite{bertone+06}.
Recall that we tuned the UV background in the process of producing
the \hnoagnn hydro flux grid, to get the same mean of the Flux. Thus, this conclusion 
is not surprising and is in agreement with previous finding  \citep{lochhaas16}. 
Above all,  this means that the predictions of the 3d clustering from the
LyMAS2 scheme keep the same accuracy,
AGN feedback included or not in the calibration.
}

\subsection{Influence of redshift distorsions?}

\begin{figure}
\begin{center}
\rotatebox{0}{\includegraphics[width=\columnwidth]{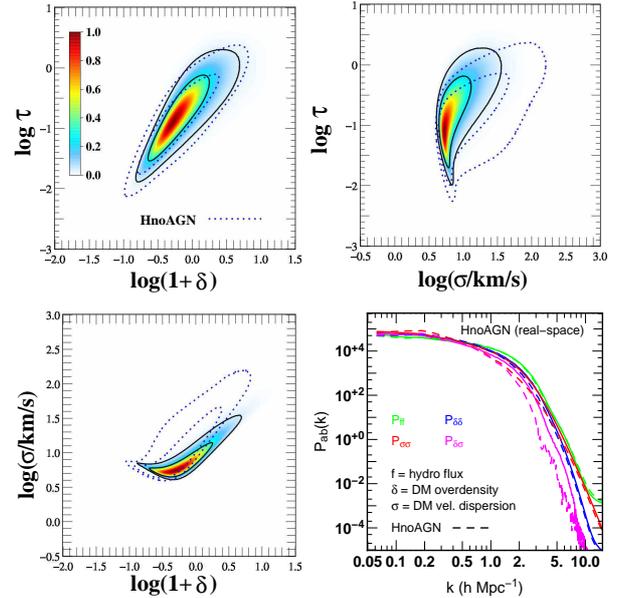}}
\caption{Same as Figs.~\ref{scatter1} and \ref{scatter2} but using \hnoagnn without modelling 
redshift distortions in the LyMAS2 scheme. Here we compare the correlations between the optical depth $\tau =$ -ln$F$ in the hydro spectra
(smoothed at the BOSS resolution) and dark matter quantities smoothed at 0.5 Mpc/h namely 
the overdensity ($1+\delta$) and the velocity dispersion ($\sigma$) at $z=2.5$. We also
show in the lower right panel, some relevant transfer functions (i.e. cross-spectrum)
similarly to Fig.~\ref{fig:cc}. In all panels, the dotted lines correspond to results
from the \hnoagnn simulation with redshift distortions.}
\label{fig_rd1}
\end{center}
 \end{figure}

\begin{figure}
\begin{center}
\rotatebox{0}{\includegraphics[width=\columnwidth]{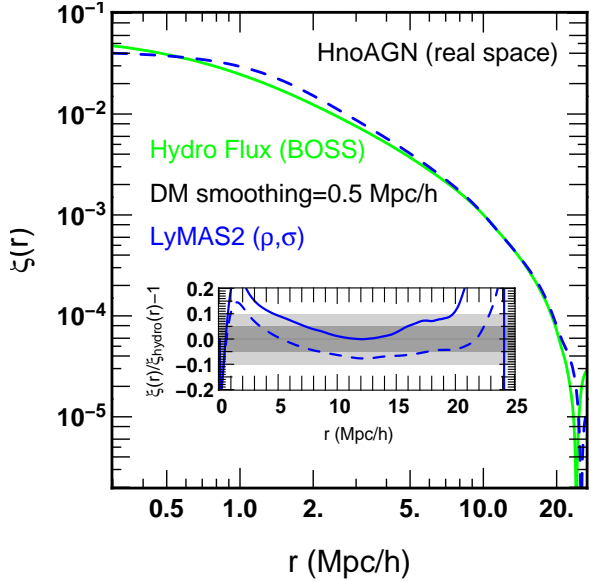}}
\caption{The correlation function $\xi$ as a function of the separation $r$.
We present results from the \hnoagnn (NO redshift distortions)
true hydro spectra (green line) and results from
LyMAS2 considering the DM overdensity and the velocity dispersion fields (blue dashed line).
DM fields are all smoothed at the scale 0.5 Mpc/h and are generated in real-space.
The central small panel indicates the relative difference as respect to the true hydro spectra (i.e. $\xi(r)/\xi_\mathrm{hydro}(r)-1$).
The blue dashed line represents here the result from the \hnoagnn simulation including
redshift distorsions.}
\label{fig_rd2}
\end{center}
 \end{figure}

{
Our results indicate that the inclusion of a DM velocity field in LyMAS2,
especially the velocity dispersion of the vorticity, clearly improves the predictions
of the 3d clustering of the pseudo-spectra. This fact might be understood by
the existing correlations between the DM overdensity and the
velocity fields (see Fig.~\ref{scatter2}) and adding a velocity term in the scheme
may bring additional information. However, since we also model redshift distortions,
this may also introduce or enhance existing correlations between the different considered
fields. 
To estimate the importance of the inclusion of redshift distortion in the process,
using \hnoagn, we have repeated the same analysis in real-space i.e. both hydro flux and DM fields have been generated without redshift distortions. 
Again, we plot in Figure~\ref{fig_rd1} some
relevant scatter plots showing the correlation between the optical depth, 
the DM overdensity and the DM velocity dispersion.
We also show some transfer functions (i.e. cross-spectrum). In all cases, the statistics have been still derived using a DM smoothing of 0.5 Mpc/h.
Compared to results from \hnoagnn including redshift distortion, the new scatter plots and 
transfer functions  show significant differences, especially when a DM velocity field is 
considered. The comparison of the 2-points correlation function of the hydro flux and LyMAS2($\rho$,$\sigma$)
pseudo spectra in Figure~\ref{fig_rd2} suggests however
that errors are still quite low, but a bit higher 
compared to the relative errors obtained when including redshift distortion. 
Then, it appears that the inclusion of redshift distorsion seems to slightly improve
the predictions of the \lyaa clustering statistics.

}

\section{Application to large cosmological dark matter simulations}
\label{sec:largemock}

\subsection{Simulations of 1.0 and 1.5 Gpc/h box side}
\label{sec:largesim}

\begin{figure}
\begin{center}
\rotatebox{0}{\includegraphics[width=\columnwidth]{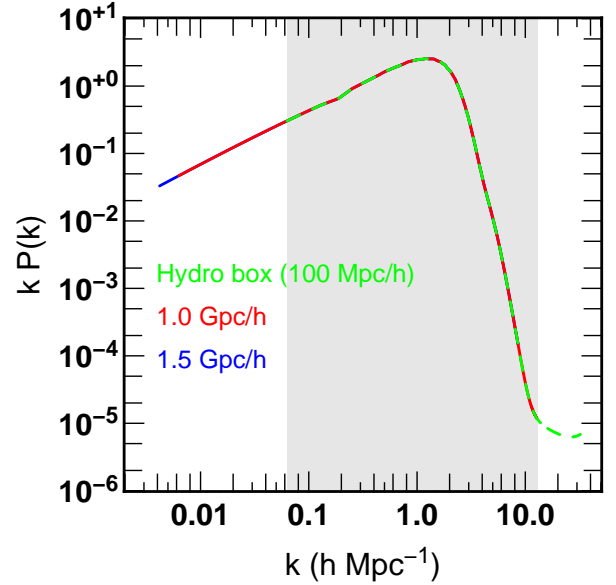}}
\caption{An example of dimensionless 1d Pk required to compute the
covariance $\Delta f_k$ (see Eq.~\eqref{eq:sigma}), derived
from the \hnoagnn simulation at $z=2.5$ and considering 
the overdensity and velocity dispersion in the Wiener filtering (green dashed line).
The resolution of the flux and DM grids are 1024$^3$.
The red and blue lines are interpolations and extrapolation of
the green line to construct the corresponding 1d Pk 
for larger 1.0 and 1.5 Gpc/h boxes (using grids of resolution 4096$^3$). 
The grey shaded area indicates the common $k$ range between the 100 Mpc/h and 1 Gpc/h
boxes.}
\label{pk_cov}
\end{center}
 \end{figure}

In this section we apply our LyMAS2 scheme to large cosmological DM simulations
to produce ensembles of BOSS pseudo-spectra. 
{ We first ran five cosmological N-body simulations using Gadget2 \citep{gadget2},
with a box length of 1.0 Gpc/h with random initial conditions and using the same cosmological parameters as \hnoagn. We additionally run one simulation with a higher volume namely
(1.5 Gpc/h)$^3$. As we discuss in detail in section~\ref{section_limitations}, these latter two values
have been chosen to estimate the performances of LyMAS when using 
 DM smoothing
scales of 0.5 Mpc/h (fiducial) and 1.0 Mpc/h. respectively}
In each simulation, the adopted value of the
Plummer-equivalent force softening is 5\%
of the mean inter-particle distance  (24.4 kpc/h and 36.6 kpc/h for the 1.0 Gpc/h
and 1.5 Gpc/h boxside respectively) and kept constant in comoving units.

From each cosmological simulation, the corresponding DM density and 
velocity dispersion fields are computed and sampled on grids of
4096$^3$ pixels. This allows us to smooth each 1.0 Gpc/h field to 0.5 Mpc/h
and each 1.5 Gpc/h one to 1.0 Mpc/h.
According to section~\ref{sec:simu} and appendix~\ref{appendix1},
the combination of the DM overdensity 
and velocity dispersion fields
leads to accurate and robust \lyaa clustering predictions.
We therefore produce our fiducial large BOSS pseudo-spectra
with LyMAS2($\rho$,$\sigma$). Note that we choose the velocity dispersion
field instead of the vorticity mainly for practical reasons, as the computational and memory costs to compute 
the latter on a large regular grid is much higher.

Once the different dark matter fields are extracted and smoothed 
to the appropriate scales, the last inputs we need are the relevant transfer
functions $T$ defined in equation~\eqref{eq:fluxmean} whose detailed expressions can be found in equation~\eqref{eq:fluxmean1d} and equation~\eqref{eq:fluxmean2d} 
for the 1d or 2d case respectively. We also need the corresponding 1d power spectrum $P_k$ 
to generate the covariance $\Delta f_k$ at the considered box side. Since the calibrations are derived from the \hnoagnn simulation, 
one potential issue arises from the hydrodynamical box being much smaller
than the large DM simulations, so that lower modes are not represented.
For the missing modes ($k\leq$2$\pi$/100), we have extrapolated the
values of $T_1$, $T_2$ and $P_k$, while in the common $k$ range, we
have proceeded with interpolations. As an illustration,
Fig.~\ref{pk_cov} shows the 1d power spectrum required to compute the  covariance  $\Delta f_k$
when considering the DM density and velocity fields extracted  
from the 100 Mpc/h hydrodynamical simulation
as well as the resulting 1d power spectra when considering a 1.0 or 1.5 Gpc/h box side.

\begin{figure*}
\begin{center}
\rotatebox{0}{\includegraphics[width=16cm]{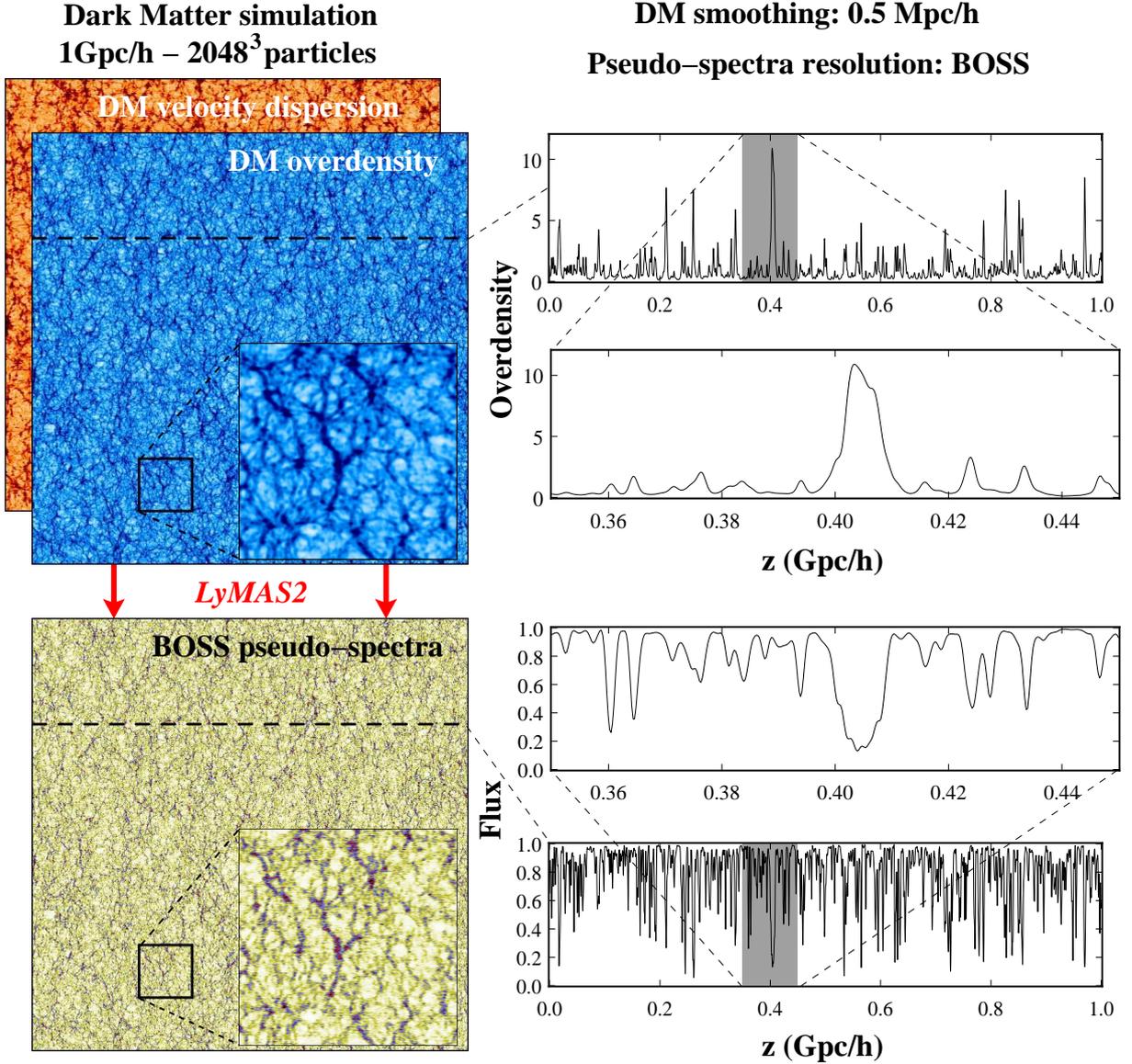}}
\caption{Application of the full LyMAS2 scheme to a large 
dark matter simulation (1 Gpc/h - 2048$^3$ particles) at $z=2.5$. From
the DM overdensity and velocity dispersion fields, both
sampled on 4096$^3$ regular grids and smoothed at 0.5 Mpc/h,
we derive pseudo-sepctra at the BOSS resolution using the full LyMAS2
scheme. The left part of the figure shows corresponding slices which suggest
a fair agreement between the clustering of the DM density and the 
pseudo spectra. The right part shows an individual skewer confirming that 
DM density peaks are associated to high flux absorption.}
\label{fig_mock}
\end{center}
 \end{figure*}

Fig.~\ref{fig_mock} illustrates a reconstruction of pseudo-spectra 
from a given slice of 4096$\times$4096 pixels through  a 1 Gpc/h  
box simulation. It appears clearly that the 2d clustering of the pseudo-spectra
agrees very well with the clustering of the DM overdensity field. 
Another visual inspection of an individual skewer also shows that peaks of density match with high absorption. It is also interesting to see that the specific skewer shown in Fig.~\ref{fig_mock}
has in its center a large absorption that corresponds to a large and high density region. Note that
the study of groups of so called ``Coherently Strong \lyaa Absorption'' (CoSLA) systems imprinted in the absorption spectra of a number of quasars (from e.g. BOSS) is of particular 
interest, as they can potentially detect and trace high redshift proto-clusters 
\citep[see, for instance][]{francis+93,cai+16,lee+18,shi+21}.

\begin{figure}
\begin{center}
\rotatebox{0}{\includegraphics[width=\columnwidth]{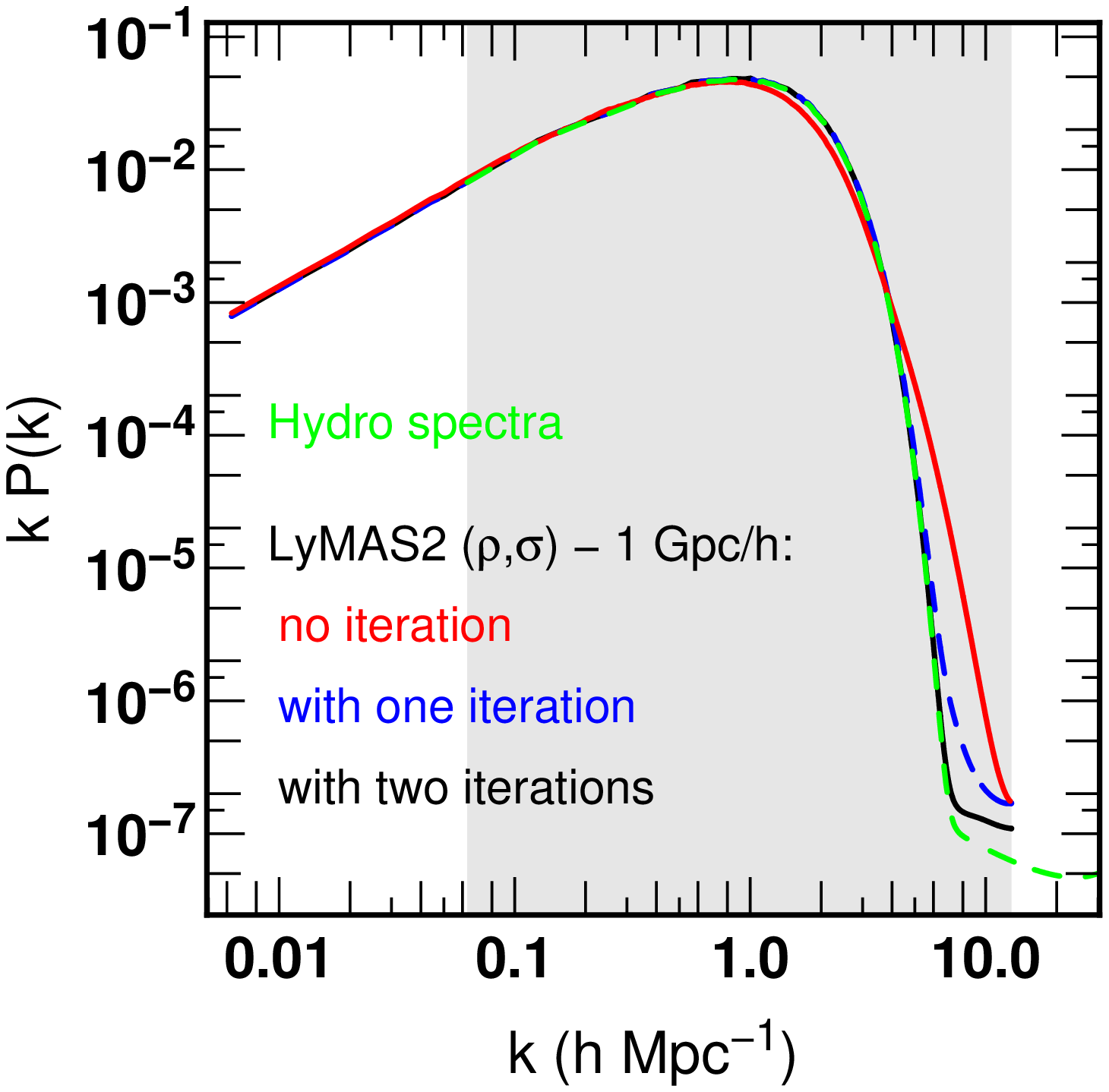}}
\vspace{-0.5cm}
\rotatebox{0}{\includegraphics[width=\columnwidth]{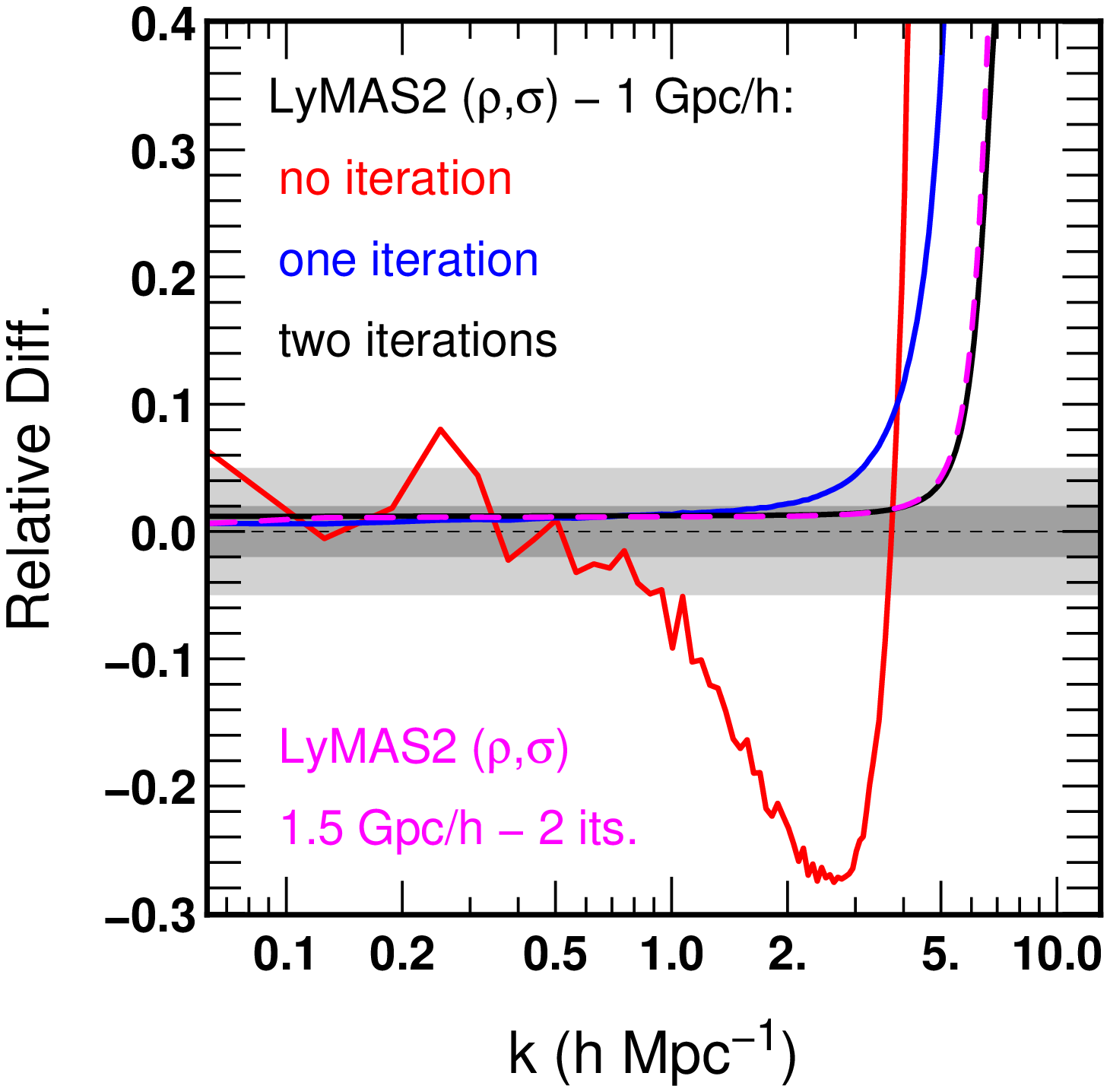}}
\caption{Top panel: the dimensionless, redshift-space, 1d power spectrum of
pseudo
and true spectra from the Horizon-noAGN simulation (green line) and
from coherent pseudo spectra using LyMAS2 considering the DM density and the
velocity dispersion fields from a 1 Gpc/h box side N-body simulation.
Here again, we show results before (red line) and after
(blue line) one full iteration (i.e. flux 1d power spectrum and PDF transformations). We
also show the results when repeating a second iteration (black line).
The grey shape defines the common k-range between hydro spectra and pseudo-spectra.
Bottom panel: the relative 
difference with respect to the hydro results (i.e. $P_{k}/P_{k,\mathbf{hydro}}-1$).
The light grey and dark grey bands define regions where the error is less
than 5 and 2 \% respectively.}
\label{fig_pkmock}
\end{center}
 \end{figure}

Fig.~\ref{fig_pkmock} shows the dimensionless 1d power spectrum of
the pseudo-spectra from a 1 Gpc/h simulation before and after iterations.
First, in the common $k$-range area between the hydro and
the pseudo spectra, we find similar trends to those obtained when applying LyMAS2
to the \hnoagnn simulation (see Fig.~\ref{fig_pkhnoagn}). For instance, after two full iterations, 
the relative difference is close to 2\% even for
high values of $k$ ($\sim$4 h\,Mpc$^{-1}$), and similar results are obtained for the 1.5 Gpc/h pseudo spectra.
For lower values of $k$  ($k\leq$2$\pi/100\sim0.0628$ $h$Mpc$^{-1}$), the power spectrum 
seems to have a natural and consistent extension from the hydro spectra power spectrum.
Note also that the highest values of $k$  for the 1.0 and 1.5 Gpc/h box side simulations
and grid of resolution 4096$^3$ are respectively 
12.87 and 8.58 Mpc$^{-1}$, lower than  (2$\pi/100)\times512\sim32.17$ $h$Mpc$^{-1}$ for 
the hydro box.
But the power at these high values is negligible, and missing them in the
calculations
will not have a noticeable impact on spectra.
Also, since a full iteration ends with a Flux PDF re-scaling, 
this ensures exactly match to the PDF of the hydro flux.

\begin{figure*}
\begin{center}
\rotatebox{0}{\includegraphics[width=\columnwidth]{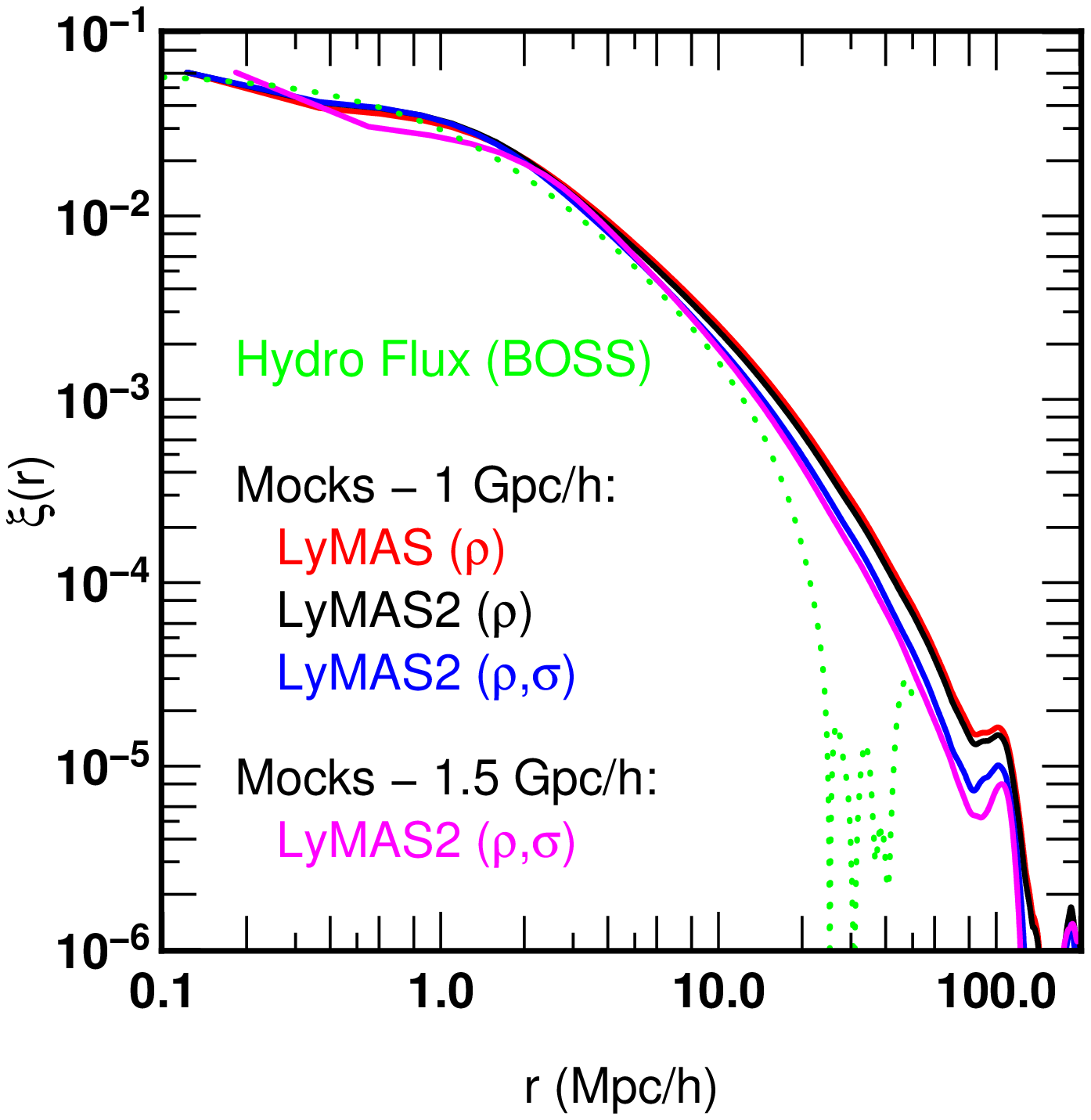}}
\rotatebox{0}{\includegraphics[width=\columnwidth]{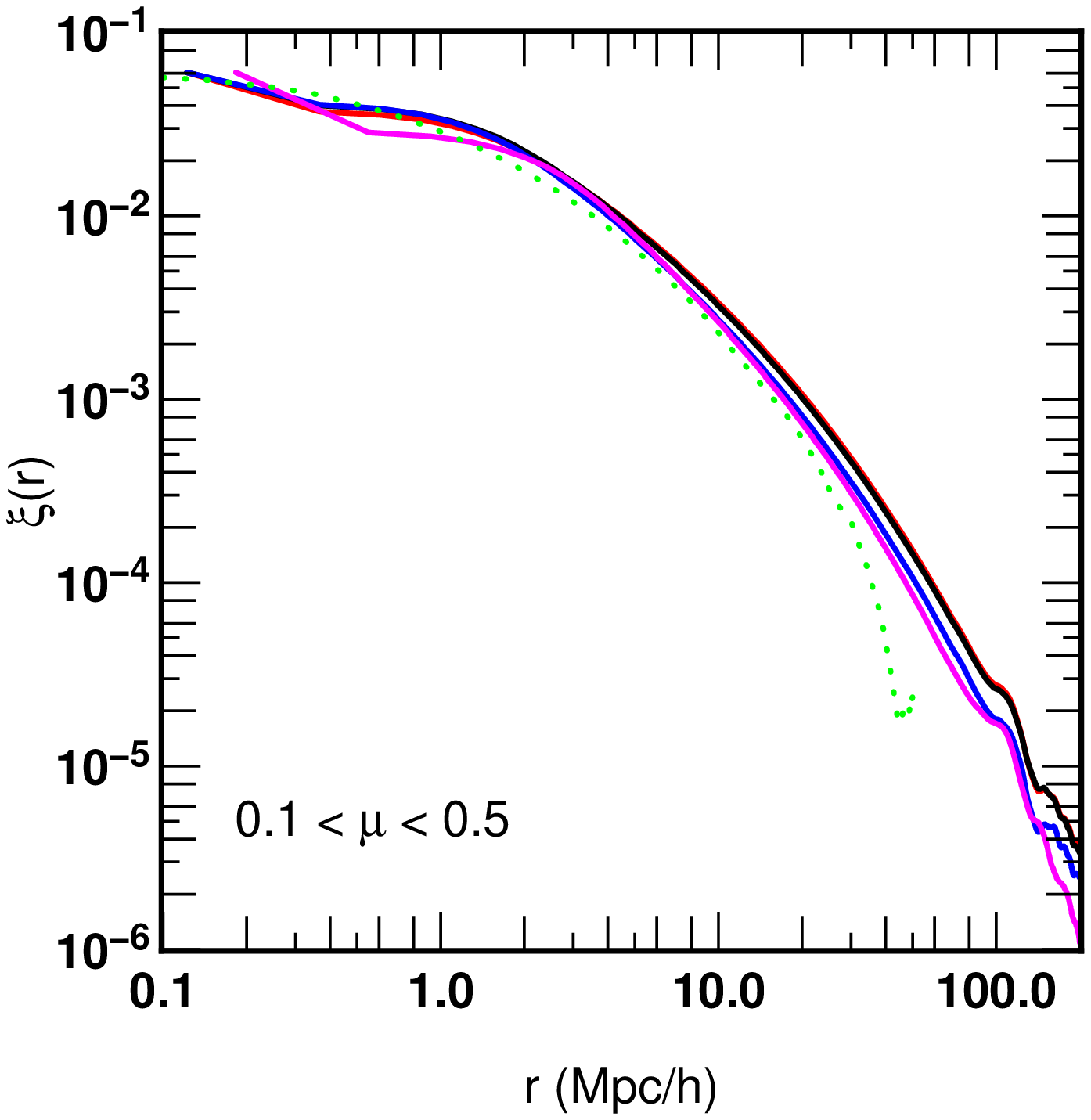}}
\rotatebox{0}{\includegraphics[width=\columnwidth]{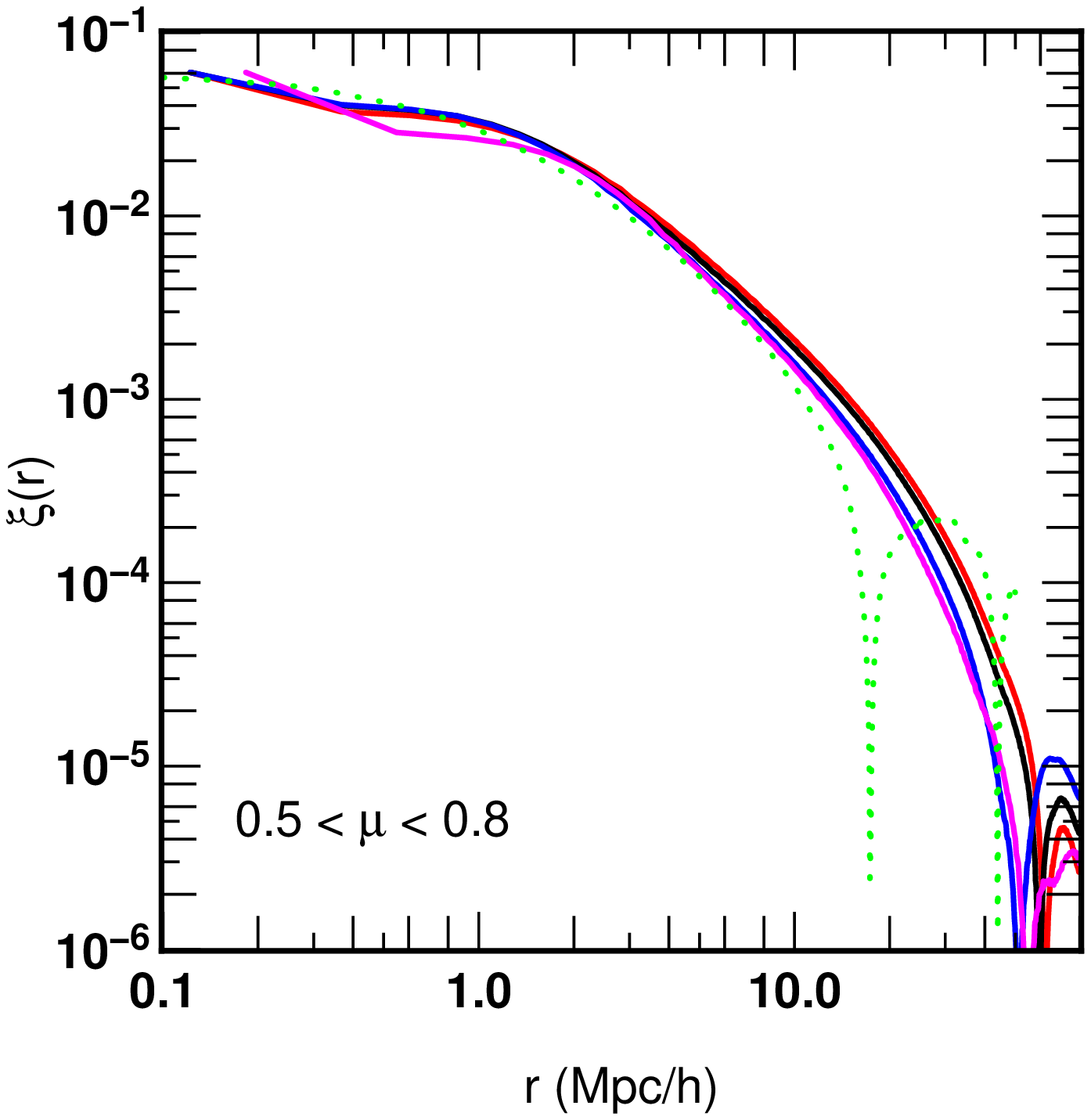}}
\rotatebox{0}{\includegraphics[width=\columnwidth]{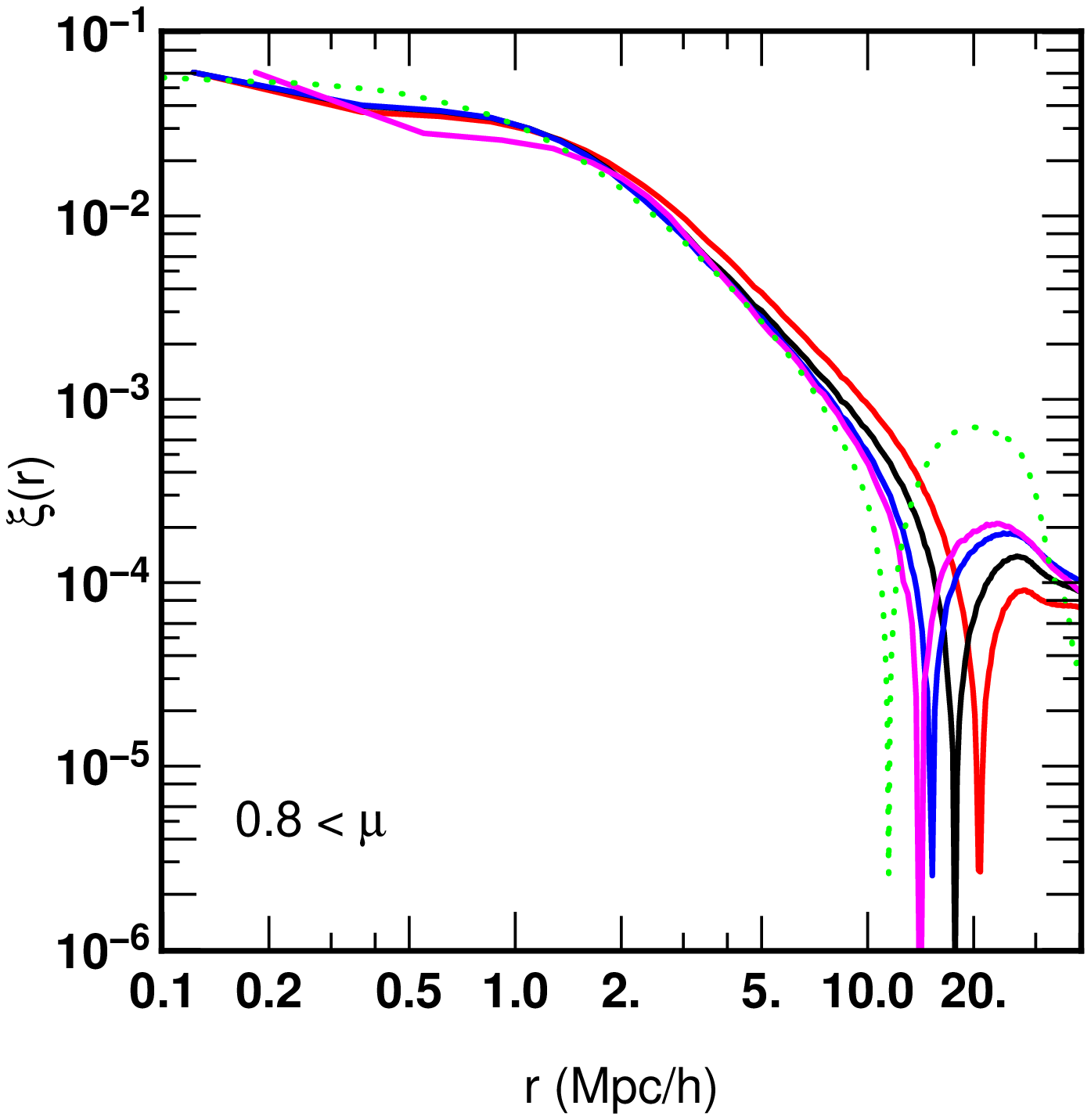}}
\caption{Top left panel: the 2-point correlation functions of the flux
derived from the 100 Mpc/h \hnoagnn simulation (green dotted line)
and from 1 Gpc/h pseudo-spectra using the full LyMAS (red line) and LyMAS2
scheme using the DM overdensity field only (black line) or combined with the DM
velocity field (blue line). We also show the 2-point correlation function
derived from 1.5 Gpc/h pseudo spectra produced with LyMAS2($\rho$,$\sigma$) (magenta line).
In the other panels, we show the
corresponding flux correlation functions averaged over bins of angle $\mu$, as 
labeled. LyMAS tends to overestimate the correlation especially for large separations
and high angles. LyMAS2($\rho$,$\sigma$) tends to significantly reduce such correlations,
which suggests more reliable predictions.}
\label{fig_FC_mock1}
\end{center}
 \end{figure*}

Fig.~\ref{fig_FC_mock1} shows the 2-point correlation
functions derived from  several large-scale pseudo-spectra. 
In particular, we show the predictions from the first version of LyMAS (red lines) and those obtained from
LyMAS2 using the DM density field only (black lines)
and  with additional velocity dispersion field (blue lines).
We also add the predictions derived from the 1.5 Gpc/h simulation (magenta lines), using again LyMAS2($\rho$,$\sigma$).
These plots confirm first that the traditional LyMAS (red lines) tends to
overestimate the correlations and this trend is more pronounced when considering
high angles ($\mu>0.8$), as already noted in section~\ref{sec:simu}.
The result is quite similar with the LyMAS2 scheme when considering the
DM overdensity only. However, the difference from LyMAS is more and more
noticeable as $\mu$ increases. These results are again consistent with those presented
in section~\ref{sec:simu}.
The difference becomes even stronger when adding the DM velocity dispersion field. In this case,
LyMAS2($\rho$,$\sigma$) tends to significantly reduce the correlations
and most probably lead to more reliable predictions.
In the range $2\leq r\leq 10$ Mpc/h, the correlations are very close to those of the hydro simulation.
It is also impressive that the 1.5 Gpc/h mock generated with LyMAS2($\rho$,$\sigma$)
leads to very similar trends (for separations $r\geq 2$ Mpc/h), though the DM fields are now
smoothed to 1.0 Mpc/h. This success is 
consistent with the results presented in the appendix~\ref{appendix1}, where we compare the performance
of LyMAS2 using different DM smoothing scales. 
This robustness is one of the key improvements accomplished with LyMAS2.

\begin{figure}
\begin{center}
\rotatebox{0}{\includegraphics[width=\columnwidth]{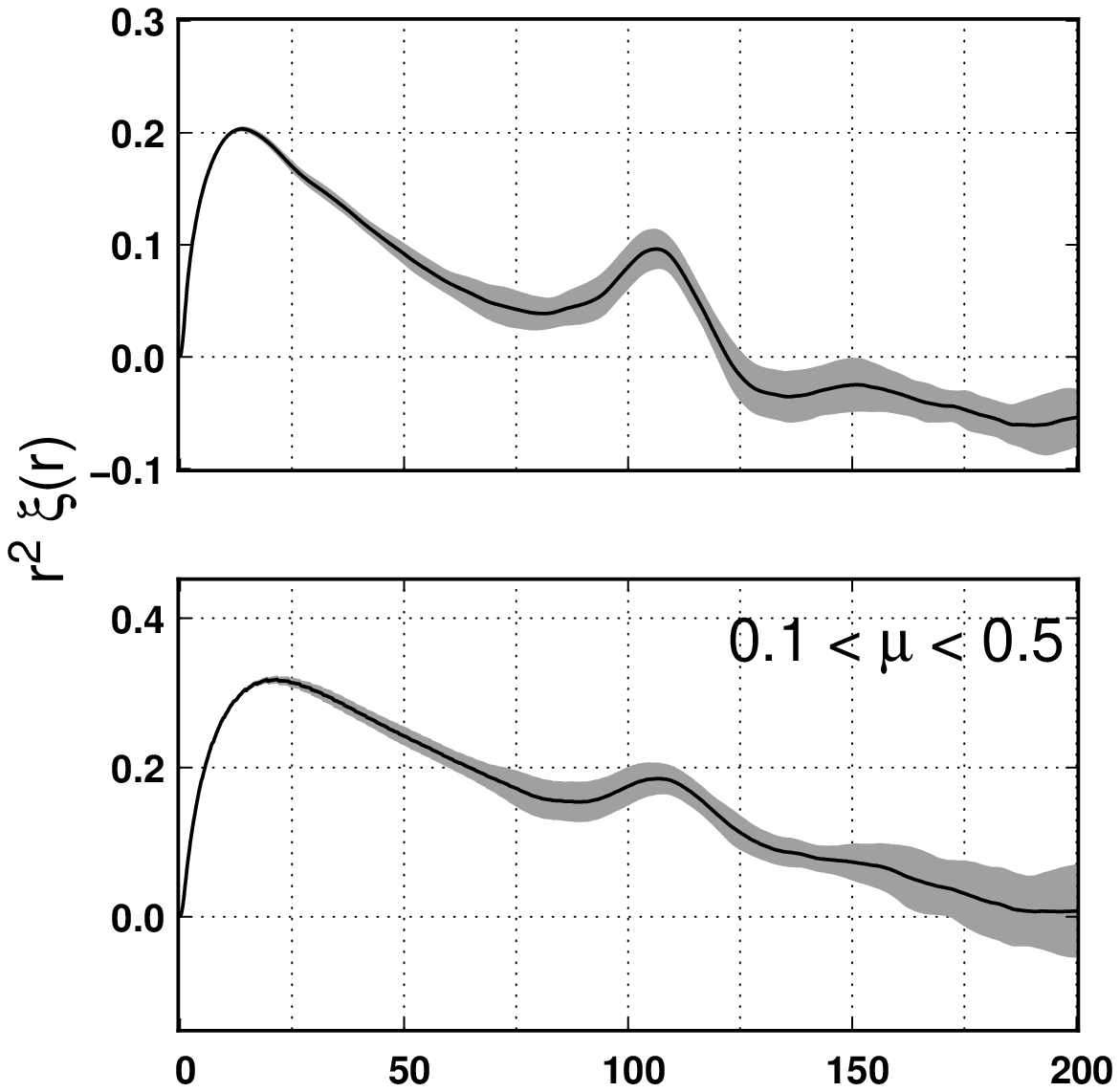}}
\rotatebox{0}{\includegraphics[width=\columnwidth]{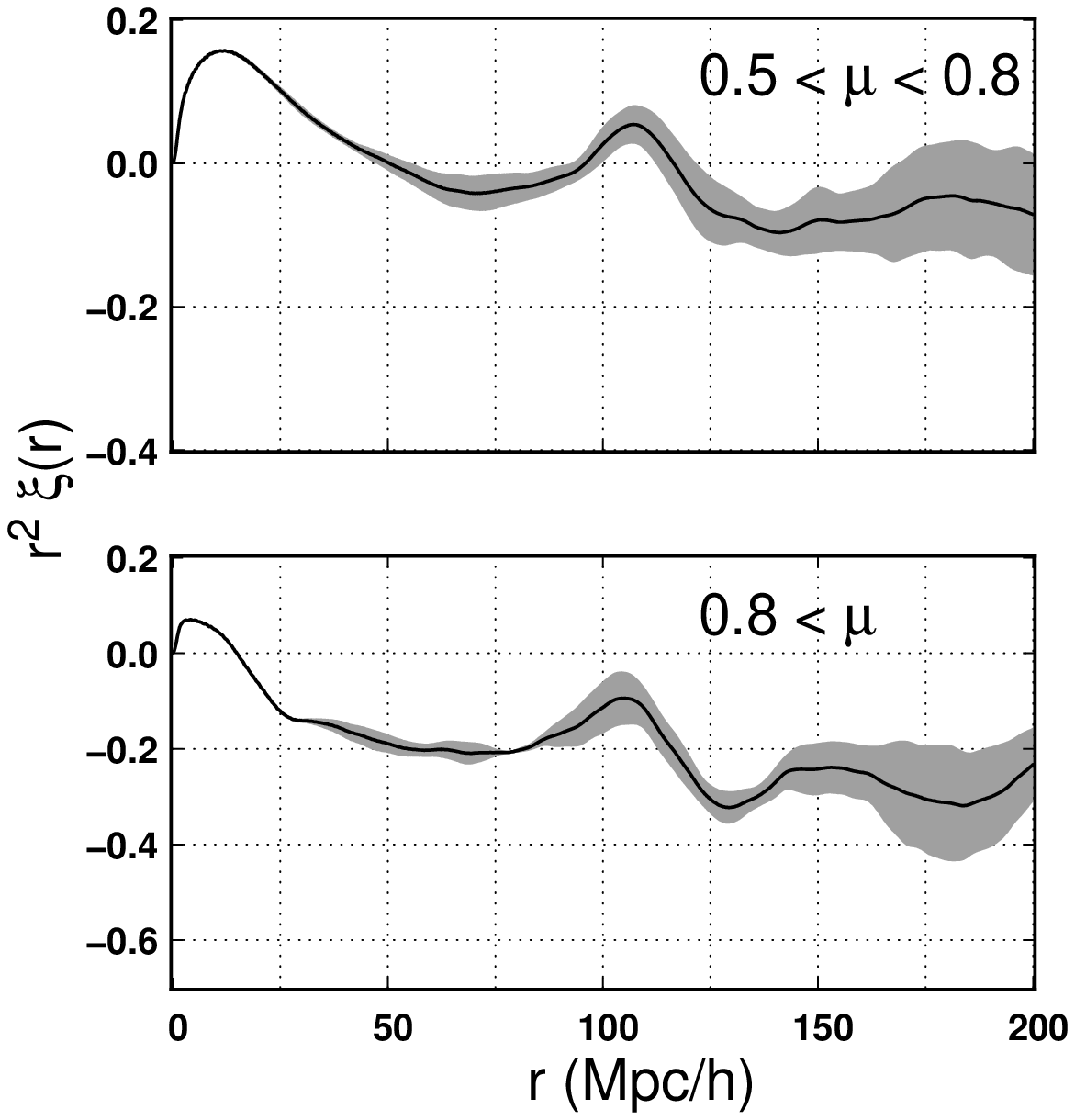}}
\caption{The 2-point correlation function averaged from five different realizations of 1 Gpc/h Ly-$\alpha$ pseudo-spectra obtained by applying LyMAS2($\rho$,$\sigma$) to DM cosmological simulations. Here we use the calibrations obtained from \hnoagn. The shaded areas represent the error on the mean (rms).}
\label{fig_CF_SDSS}
\end{center}
 \end{figure}

Finally, we show in Fig.~\ref{fig_CF_SDSS} the 2-point correlation function 
averaged from five different realizations of 1 Gpc/h Ly-$\alpha$ pseudo-spectra
obtained by applying LyMAS2($\rho$,$\sigma$) to different DM cosmological simulations, at $z=2.5$.
The plots show clear features of BAO at $r\sim$105 Mpc/h and
variations with respect to the angle $\mu$, consistent
with observational trends (see for instance \cite{dumasdesbourboux})
This illustrates the ability of LyMAS to properly describe redshift distorsions
and to model realistic large BOSS \lyaa forest spectra catalogues.

\subsection{Potential limitations of the method}
\subsubsection{Effect of cosmic variance?}
\label{sec:cosmic_variance}

\begin{figure}
\begin{center}
\rotatebox{0}{\includegraphics[width=\columnwidth]{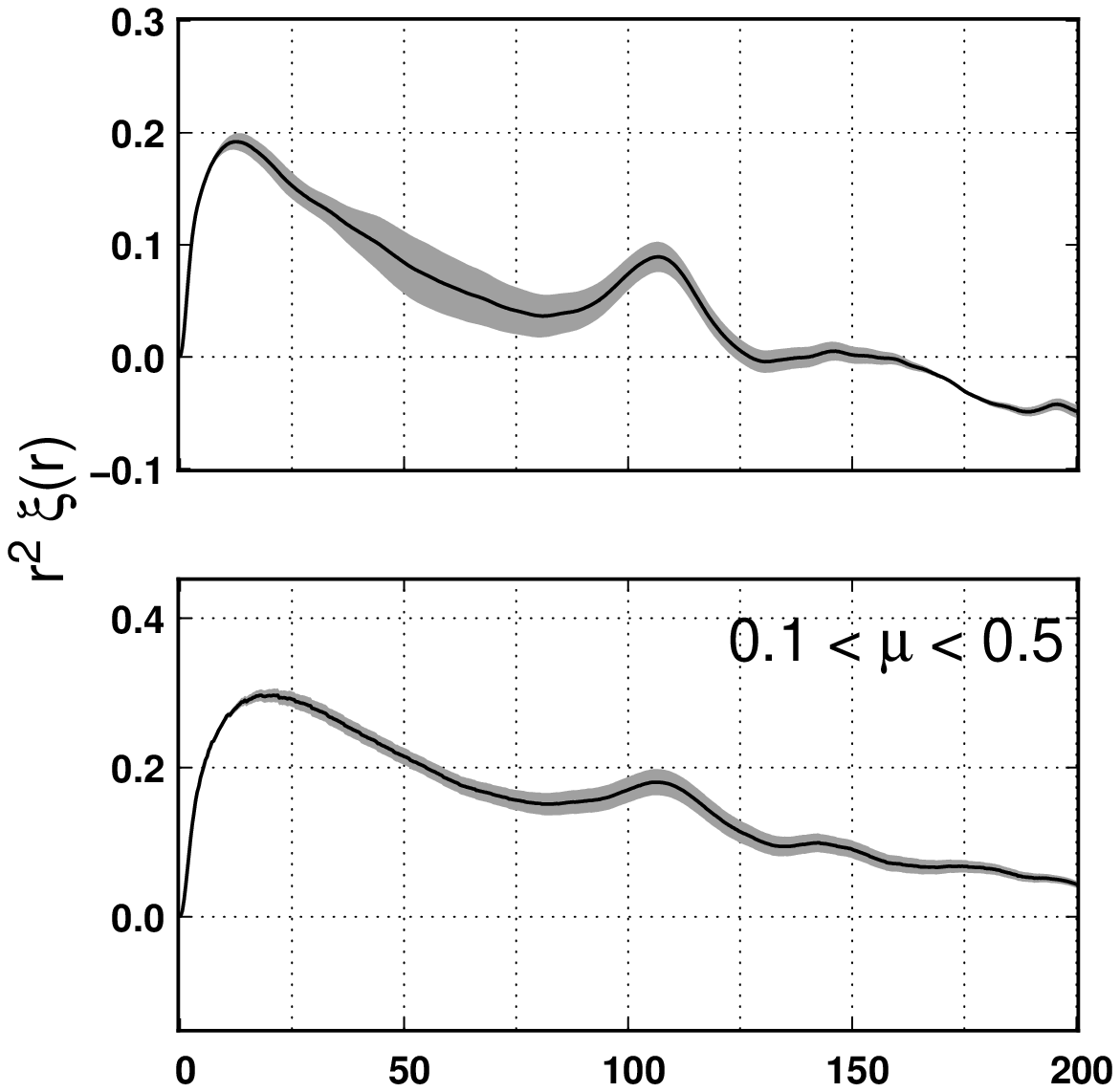}}
\rotatebox{0}{\includegraphics[width=\columnwidth]{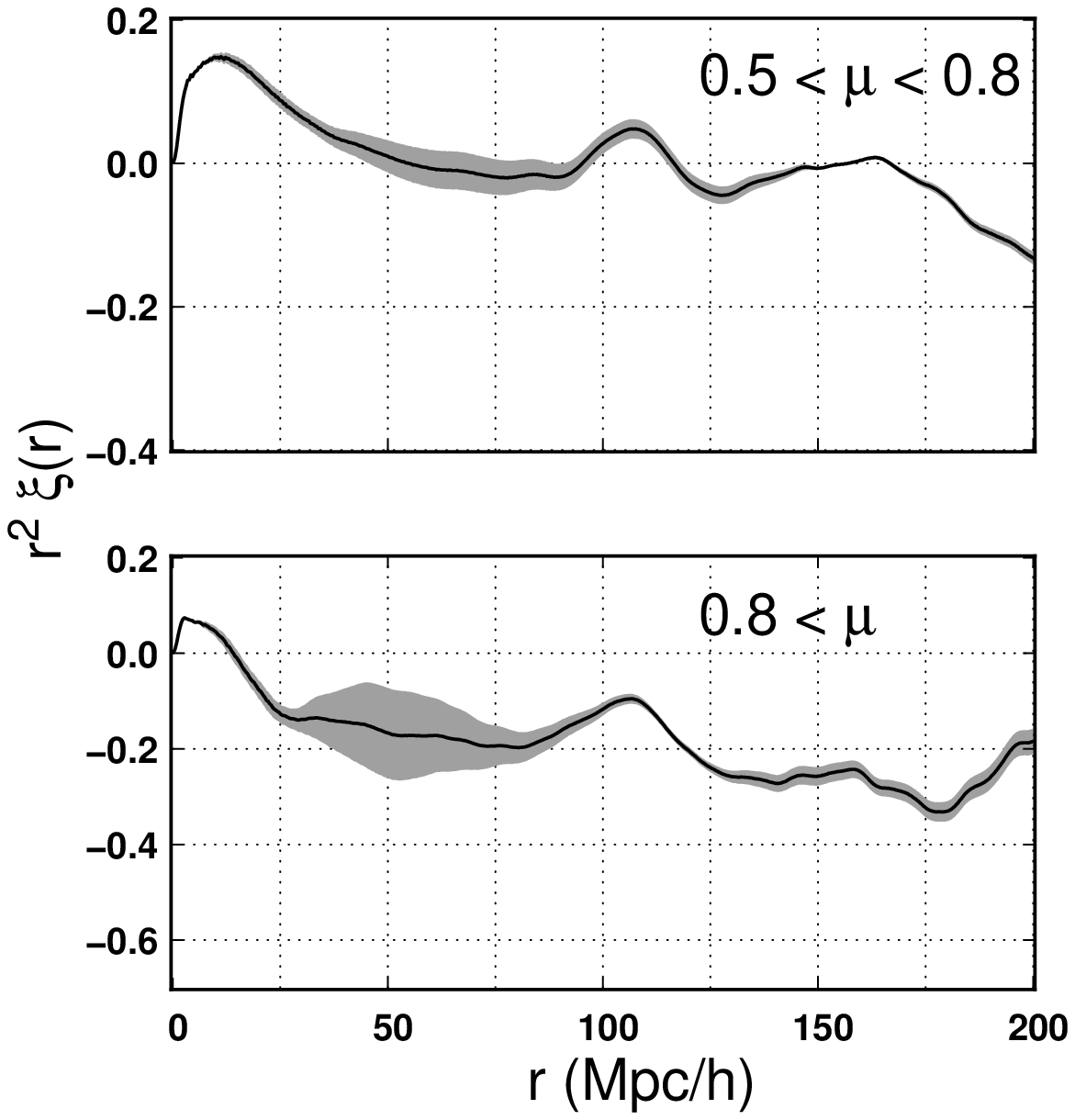}}
\caption{The 2-point correlation function averaged from five different realizations of 1.5 Gpc/h Ly-$\alpha$ pseudo-spectra obtained by applying 
LyMAS2($\rho$,$\sigma$) to DM cosmological simulations and using five different calibrations.
To this regard, we use five lower resolution hydro simulations presented in Appendix~\ref{appendix1}. The shaded areas represent the error on the mean (rms).}
\label{fig_cv}
\end{center}
 \end{figure}

{
One potential limitation in the LyMAS scheme is to use an unique hydro simulation
to generate the calibration. In other words, we assume this hydro simulation to be
fairly representative of the underlying statistics of many simulations which have
$\geq$1000 times larger volumes. This makes the resulting large mocks potentially
affected by the cosmic variance. 
In order to estimate this, we have considered our five lower resolution hydro simulations
presented in Appendix~\ref{appendix1}, originally
produced to estimated the robustness of the LyMAS2 predictions. Here we make good use 
to estimate the effect of cosmic variance by applying each of the five calibration sets
to the (1.5 Gpc/h)$^3$ DM overdensity and velocity grids (of dimension 4096$^3$ each) that we used in section~\ref{sec:largesim}. 
In particular, we have computed the 2-point correlation function
averaged 
from these five realizations and shown  in Fig.~\ref{fig_cv}.
We note that the dispersion tends to be higher for separations between 25 and 100 Mpc/h which makes sense as this
corresponds to the scales probed by the reference hydro simulation of box side 100 Mpc/h. 
We also note that this dispersion tends to be higher for increasing values of the angle $\mu$.
}

\subsubsection{Computational limitations}
\label{section_limitations}

{
As most of the methods presented in the literature to produce large \lyaa mock catalogs, the LyMAS2 scheme can be divided into two main operations. One the one hand,
one needs to generate at a considered 
smoothing scales, a DM overdensity field and eventually associated velocity fields. This task
is generally done using  N-body simulations or log normal 
density fields created from Gaussian initial conditions \citep[e.g.][]{gnedin+96,bi+97}. On the other hand,
one has to ``paint" the \lyaa absorptions from any los using relevant calibrations or recipes.
This latter part is pretty fast in LyMAS since one can treat any los individually and therefore,
the algorithm can be easily and optimally parallelized (with openMP for instance). To give an order of magnitude,
to create a $\geq$1 Gpc/h box side \lyaa mock presented in section~\ref{sec:largesim}.,
using 32 cpu, only $\sim$11 hours are required to generate 4096$\times$4096 BOSS spectra
of resolution 4096 and subsequent 1d-Pk and flux PDF rescaling 
(namely to achieve the six steps 
of the LyMAS2 scheme presented in section~\ref{sec:pseudospectra}).
The main limitation of LyMAS2 is however the ability to generate the DM overdensity or
a velocity field at the appropriated smoothing scales (i.e. 0.5 and 1.0 Mpc/h in our study).
Indeed, the larger the box side of the simulation, the bigger the required dimension of the grid to sample the DM fields.
For instance, a simulation box of side 1.0 Gpc/h or 1.5 Gpc/h can be smoothed at the scale 
of 0.5 Mpc/h and 1.0 Mpc/h respectively if a grid of 4096$^3$ pixels is considered. 
In these cases, the size of individual pixel is respectively
0.244 and 0.366 Mpc/h which is acceptable, though slightly borderline, to produce the smoothing operation.
Moreover, to generate one specific DM field, we used 
a sophisticated scheme, SmoothDens5, presented in the appendix~\ref{appendix3}.  Although SmoothDens5 has been optimized, it needs at least
1 Tb RAM and 50 hours (using 64 cpu) to treat and
produce a single DM field, sampled on a regular  grid of 4096$^3$, 
and from a DM simulation using $2048^3$ particles. 

Technically, it is then rather feasible to generate massive set of mocks
if we limit the studied box side to 1.5 Gpc/h. Beyond this value, 
the computational costs and memory requirement is
becoming an issue.
It would be definitely worth exploring in near future alternative methods
to reduce such costs (e.g. Cell-in-Cloud, ...)  while not altering
the accuracy of the predictions.
Moreover,  although this is a general issue for all the methods based on DM fields described by N-body simulations, N-body simulations can also become too computationally expensive and time-consuming. 
Here also alternative methods do exist to obtain the DM fields using cheap approximate methods (e.g.  LPT, 2LPT, etc...). However, these are typically not able to produce a very accurate velocity field and this may alter the accuracy of the present LyMAS scheme. 
Such investigations are beyond the scope of the present paper and will be considered in 
the next analysis.
}

\section{Conclusions}
\label{sec:conclusions}

We have introduced LyMAS2, an improved version of the LyMAS scheme \citep{peirani14}.
In this new version, we have used the \hnoagnn \citep{peirani17} simulation to characterize 
the relevant cross-correlations between the transmitted flux and the different DM fields.
In particular, we have considered not only the DM overdensity
but also specific DM velocity fields (i.e. velocity dispersion, vorticity, 1d and 3d divergence) and used  Wiener filtering to generate the specific calibrations.
LyMAS2 shares the same philosophy as LyMAS that flux correlations are
mainly driven by the correlations of the underlying DM (over)density, and it uses additional information
from the DM velocity correlations  to refine the
theoretical predictions. 
In a second step, we have applied LyMAS2 to DM fields extracted from the hydrodynamical
or large DM-only simulations to create large ensembles of pseudo-spectra
with redshift distortions,  at $z=2.5$ and at the BOSS resolution. 
Throughout the analysis, we use a DM smoothing of 0.5 Mpc/h to derive the main trends and results.
Our main conclusions can be summarized as follows:

$\bullet$  
LyMAS2 greatly improves the predictions for flux statistics
of the 3d \lyaa forest on small and large scales. More specifically,
we found that the DM overdensity combined with the DM velocity
dispersion (or the vorticity) recovers the
2-point correlation functions of the (reference) hydro flux
within 10\% and (most of the time within 5\%) even when high angles are considered.
This is a major improvement with respect to the original version of LyMAS,
which is rather inaccurate in predicting the \lyaa correlations 
for large separations and high angles.

$\bullet$  
Like LyMAS, LyMAS2 reproduces the 1-point PDF of the flux from the calibrating hydro
simulation exactly, by construction. It also reproduces the 1d (line-of-sight) power
spectrum with en error of about 2\% up high $k$ values. The LyMAS2 pseudo spectra
therefore have realistic observable properties on small scales while also having
accurate large scale 3d clustering when applied to a large volume DM-only simulation.

$\bullet$ 
The trends derived from five different and slightly lower resolution 
hydrodynamical simulations are consistent with those obtained from the fiducial 
\hnoagnn simulation. This suggests that the results presented in this study
are robust. Moreover, this allows us to
estimate error bars on the 2-point correlations functions, 
which are generally low.

$\bullet$ 
We have considered three different DM smoothing scales (0.3, 0.5 and 1.0 Mpc/h)
and found similar trends in the flux clustering predictions.
It is encouraging that a DM smoothing of 1.0 Mpc/h
still leads to very accurate predictions, especially in the 2-point correlation functions 
even at high angles and large separation. Indeed,
the errors are typically lower than 5\%, 
whereas they are generally higher than 30\% with the original version of LyMAS.

$\bullet$ 
LyMAS2 applied to large DM cosmological simulations of box side either 1.0 or 1.5 Gpc/h
indicates that the predicted flux statistics follow the
same trends obtained from the (100 Mpc/h) \hnoagnn DM fields. Indeed, we found
again that the first version of LyMAS tends to overestimate the flux correlations
at large separations and/or at high 
angles. On the contrary, LyMAS2 using for instance 
the DM overdensity and the velocity dispersion clearly reduces the 2-point correlation
functions to lead to more reliable and accurate predictions.
Moreover LyMAS2 adequately models large scale \lyaa absorptions systems which
correspond to massive overdensity regions. 
{ It is also worth mentioning  that these set of mocks were already 
used to asses the ability to recover the connectivity and clustering properties of critical points of the reconstructed
large scale structure from \lyaa tomography in the context
of a realistic quasar survey configuration such
as WEAVE-QSO \citep{kraljic+22}.}


$\bullet$
Deterministic mappings such as the Fluctuating Gunn-Peterson Approximation
tend to considerably overestimate the 3d flux correlations especially at large separation 
or when high angles are considered.

LyMAS2 offers a sophisticated tool to accurately model and predict large scale \lyaa forest
3d statistics. This opens new opportunities to improve diversified 
studies such as \lyaa forest cross-correlation  \citep[e.g.][]{lochhaas16}, 
two-point correlations or three-point correlations analysis \citep[e.g.][]{suk}
or BAO feature predictions.
Moreover, large \lyaa catalogs produced with LyMAS2 can be used
to characterize massive overdensity regions such as proto-clusters through
groups of coherent large absorptions analysis \citep{cai+16,shi+21,lee+18}.
Compared to previous work, 
we recall that the main objective of LyMAS is to create large \lyaa mocks for a specific instrument 
(here BOSS) with 3d flux statistics 
as close as possible to those that would be obtained from a very large volume (but computationally
intractable) hydrodynamical simulation.

The Iteratively Matched Statistics (IMS) developped by \cite{sorini+16}
does not present such predictions and limits their analysis to small simulation boxes ($\leq$100 Mpc/h).
However, when comparing the flux PDF and 1d power spectrum, LyMAS2 and 1D-IMS  (see introduction) lead to similar performances: the 1D-IMS scheme perfectly reproduces these
statistics, while errors of  $\sim$2\% are obtained with LyMAS2 for the flux 1d-Pk.
Regarding the 3D-IMS scheme, errors are much higher, of
order of 15 and 20\% respectively for the flux PDF and 1d power spectrum.
As far as the 3d flux statistics are concerned, at a DM smoothing of 0.4 Mpc/h, 
the 1D-IMS and 3D-IMS present errors of 
20\% and 10-20\% (for a DM smoothing of 0.4 Mpc/h) respectively
regarding the reconstruction of the power spectrum. 
In this study, LyMAS2 mainly considers a DM smoothing of 0.5 Mpc/h, which leads to errors generally 
lower than 5\%  for the 2-point correlation functions. Again, it is worth mentioning that similar (low) errors
are also obtained with LyMAS2 when considering a DM smoothing of 1.0 Mpc/h.
It would be then interesting
to compare the performance of the 1D and 3D-IMS scheme at this specific smoothing scale
in the perspective of creating large ($\geq$ 1.0 Gpc/h) \lyaa mocks. 

Recently, \cite{harrington+21}  have trained a convolutional neural network
from hydrodynamical simulations of side 20 Mpc/h to predict both the density, the temperature and the 
velocity fields. This method is quite flexible and the predictions of the flux PDF and 1d power spectrum 
(i.e. within $\sim$5\% up to k$\sim$10 Mpc/h) are promising and more accurate than the FGPA.
{ Note that in a companion paper \citep{horowitz}, convolutional neural networks
have also been used to synthesize hydrodynamic fields conditioned on dark matter fields from N-body simulations, which might be very useful for the rapid generation of mocks.}
Similarly, \cite{sinigaglia+21} has developed a new physically-motivated supervised machine
learning method (HYDRO-BAM) from a reference hydrodynamical simulation of comoving side
100 Mpc/h. The PDF, 3d power spectrum and bi-spectra can be reconstructed with error of a few percent
up to modes k$=0.9$ Mpc/h.
{ It would be interesting to see how this promising approach performs when considering smoothed spectra and larger boxes. 
}

Improvements can still be done in the LyMAS scheme. For instance,
one main assumption is to consider that the transverse correlations are mainly driven by 
the effect of DM smoothing. In the present study, we stress again that all the approach is based on creating pseudo-spectra individually and independently from each other. 
Because the draws of $\Delta f_k$ are independent on each line-of-sight, spectra at small
but non-zero transverse separations can look quite different on small scales. 
Since the predictions on the clustering of
the flux are already very accurate  with LyMAS2, we haven't considered 
the same approach in volume. This would take into account transverse correlations between lines
of sight that have been neglected in this work: instead of predicting the flux from dark matter
fields independently for each LOS, one would predict the entire cube of flux from the dark matter
field cubes, using the full 3d covariance structure. One would still use an assumption of spatial
homogeneity (stationarity), so that 3d Fourier space coefficients could be computed independently, however
one would need to take care of the statistical anisotropy in the LOS direction, therefore all statistics
in Fourier space would depend on $|k_\perp|$ and $k_\parallel$.
Taking into account transverse correlations would thus further reduce the covariance of the flux conditionally to
the dark matter fields, in other words reduce the noise in the predicted flux field.
Among future prospects, we plan to extend this work to predict 
the flux clustering for other surveys such as 
the {\it Dark Energy Spectroscopic Instrument}  \cite[DESI,][]{desi},
the {\it William Herschel Telescope Enhanced Area Velocity Explorer} \cite[WEAVE-QSO,][]{pieri+16}
or {\it Subaru Prime Focus Spectrograph} \cite[PFS,][]{takada+14}.
They will open new vistas on the high redshift intergalactic medium  probed
by the \lyaa forest.
It would be then interesting to estimate the level of performance of LyMAS2 when the transmitted flux has a higher resolution than BOSS spectra, which might
require reducing the DM smoothing.
Finally, we also intend to use Machine Learning in the process
to see whether we can still improve the predicted flux statistics \citep{chopitan+prep}.

\subsection*{Acknowledgements}
{We warmly thank the referee for an insightful review that considerably improved the quality of the original manuscript}.
This work was carried within the framework of the
Horizon project (\href{http://www.projet-horizon.fr}{http://www.projet-horizon.fr}).
Most of the numerical modeling presented here was done on the Horizon cluster at IAP.
This work was supported by the Programme National Cosmology et Galaxies
(PNCG) of CNRS/INSU with INP and IN2P3, co-funded by CEA and CNES.
DW acknowledges support of U.S. National Foundation grant AST-2009735.
We warmly thank T.\;Sousbie,  B. Wandelt, O. Hahn, M. Buehlmann and S.\;Rouberol
for stimulating discussions. We also thank D.\,Munro for
freely distributing his Yorick programming language (available at
\href{http://yorick.sourceforge.net/}{http://yorick.sourceforge.net/}) which was used during the
course of this work.

\subsection*{Data availability}
The data and numerical codes
underlying this article were produced by the authors.
They will be shared on reasonable request
to the corresponding author.

\bibliographystyle{mnras}
\bibliography{lymas4}

\begin{thebibliography}{}
\makeatletter
\relax
\def\mn@urlcharsother{\let\do\@makeother \do\$\do\&\do\#\do\^\do\_\do\%\do\~}
\def\mn@doi{\begingroup\mn@urlcharsother \@ifnextchar [ {\mn@doi@}
  {\mn@doi@[]}}
\def\mn@doi@[#1]#2{\def\@tempa{#1}\ifx\@tempa\@empty \href
  {http://dx.doi.org/#2} {doi:#2}\else \href {http://dx.doi.org/#2} {#1}\fi
  \endgroup}
\def\mn@eprint#1#2{\mn@eprint@#1:#2::\@nil}
\def\mn@eprint@arXiv#1{\href {http://arxiv.org/abs/#1} {{\tt arXiv:#1}}}
\def\mn@eprint@dblp#1{\href {http://dblp.uni-trier.de/rec/bibtex/#1.xml}
  {dblp:#1}}
\def\mn@eprint@#1:#2:#3:#4\@nil{\def\@tempa {#1}\def\@tempb {#2}\def\@tempc
  {#3}\ifx \@tempc \@empty \let \@tempc \@tempb \let \@tempb \@tempa \fi \ifx
  \@tempb \@empty \def\@tempb {arXiv}\fi \@ifundefined
  {mn@eprint@\@tempb}{\@tempb:\@tempc}{\expandafter \expandafter \csname
  mn@eprint@\@tempb\endcsname \expandafter{\@tempc}}}

\bibitem[\protect\citeauthoryear{{Bautista} et~al.,}{{Bautista}
  et~al.}{2017}]{bautista+17}
{Bautista} J.~E.,  et~al., 2017, \mn@doi [\aap] {10.1051/0004-6361/201730533},
  \href {https://ui.adsabs.harvard.edu/abs/2017A&A...603A..12B} {603, A12}

\bibitem[\protect\citeauthoryear{{Bertone} \& {White}}{{Bertone} \&
  {White}}{2006}]{bertone+06}
{Bertone} S.,  {White} S. D.~M.,  2006, \mn@doi [\mnras]
  {10.1111/j.1365-2966.2005.09936.x}, \href
  {https://ui.adsabs.harvard.edu/abs/2006MNRAS.367..247B} {367, 247}

\bibitem[\protect\citeauthoryear{{Bi} \& {Davidsen}}{{Bi} \&
  {Davidsen}}{1997}]{bi+97}
{Bi} H.,  {Davidsen} A.~F.,  1997, \mn@doi [\apj] {10.1086/303908}, \href
  {https://ui.adsabs.harvard.edu/abs/1997ApJ...479..523B} {479, 523}

\bibitem[\protect\citeauthoryear{{Blanton} et~al.,}{{Blanton}
  et~al.}{2017}]{blanton+17}
{Blanton} M.~R.,  et~al., 2017, \mn@doi [\aj] {10.3847/1538-3881/aa7567}, \href
  {https://ui.adsabs.harvard.edu/abs/2017AJ....154...28B} {154, 28}

\bibitem[\protect\citeauthoryear{{Bolton}, {Puchwein}, {Sijacki}, {Haehnelt},
  {Kim}, {Meiksin}, {Regan}  \& {Viel}}{{Bolton} et~al.}{2017}]{bolton+17}
{Bolton} J.~S.,  {Puchwein} E.,  {Sijacki} D.,  {Haehnelt} M.~G.,  {Kim} T.-S.,
   {Meiksin} A.,  {Regan} J.~A.,   {Viel} M.,  2017, \mn@doi [\mnras]
  {10.1093/mnras/stw2397}, \href
  {https://ui.adsabs.harvard.edu/abs/2017MNRAS.464..897B} {464, 897}

\bibitem[\protect\citeauthoryear{{Buehlmann} \& {Hahn}}{{Buehlmann} \&
  {Hahn}}{2019}]{buehlmann&hahn19}
{Buehlmann} M.,  {Hahn} O.,  2019, \mn@doi [\mnras] {10.1093/mnras/stz1243},
  \href {https://ui.adsabs.harvard.edu/abs/2019MNRAS.487..228B} {487, 228}

\bibitem[\protect\citeauthoryear{{Busca} et~al.,}{{Busca}
  et~al.}{2013}]{busca+13}
{Busca} N.~G.,  et~al., 2013, \mn@doi [\aap] {10.1051/0004-6361/201220724},
  \href {https://ui.adsabs.harvard.edu/abs/2013A&A...552A..96B} {552, A96}

\bibitem[\protect\citeauthoryear{{Cai} et~al.,}{{Cai} et~al.}{2016}]{cai+16}
{Cai} Z.,  et~al., 2016, \mn@doi [\apj] {10.3847/1538-4357/833/2/135}, \href
  {https://ui.adsabs.harvard.edu/abs/2016ApJ...833..135C} {833, 135}

\bibitem[\protect\citeauthoryear{{Caucci}, {Colombi}, {Pichon}, {Rollinde},
  {Petitjean}  \& {Sousbie}}{{Caucci} et~al.}{2008}]{caucci+08}
{Caucci} S.,  {Colombi} S.,  {Pichon} C.,  {Rollinde} E.,  {Petitjean} P.,
  {Sousbie} T.,  2008, \mn@doi [\mnras] {10.1111/j.1365-2966.2008.13016.x},
  \href {https://ui.adsabs.harvard.edu/abs/2008MNRAS.386..211C} {386, 211}

\bibitem[\protect\citeauthoryear{{Chabanier}, {Bournaud}, {Dubois},
  {Palanque-Delabrouille}, {Y{\`e}che}, {Armengaud}, {Peirani}  \&
  {Beckmann}}{{Chabanier} et~al.}{2020}]{chabanier+20}
{Chabanier} S.,  {Bournaud} F.,  {Dubois} Y.,  {Palanque-Delabrouille} N.,
  {Y{\`e}che} C.,  {Armengaud} E.,  {Peirani} S.,   {Beckmann} R.,  2020,
  \mn@doi [\mnras] {10.1093/mnras/staa1242}, \href
  {https://ui.adsabs.harvard.edu/abs/2020MNRAS.495.1825C} {495, 1825}

\bibitem[\protect\citeauthoryear{{Chopitan}, {Lavaux}  \& {Peirani}}{{Chopitan}
  et~al.}{2021}]{chopitan+prep}
{Chopitan} N.,  {Lavaux} G.,   {Peirani} S.,  2021, in prep

\bibitem[\protect\citeauthoryear{{Colombi}, {Chodorowski}  \&
  {Teyssier}}{{Colombi} et~al.}{2007}]{dens_smooth}
{Colombi} S.,  {Chodorowski} M.~J.,   {Teyssier} R.,  2007, \mn@doi [\mnras]
  {10.1111/j.1365-2966.2006.11330.x}, \href
  {https://ui.adsabs.harvard.edu/abs/2007MNRAS.375..348C} {375, 348}

\bibitem[\protect\citeauthoryear{{Croft}, {Weinberg}, {Katz}  \&
  {Hernquist}}{{Croft} et~al.}{1998}]{croft+98}
{Croft} R. A.~C.,  {Weinberg} D.~H.,  {Katz} N.,   {Hernquist} L.,  1998,
  \mn@doi [\apj] {10.1086/305289}, \href
  {https://ui.adsabs.harvard.edu/abs/1998ApJ...495...44C} {495, 44}

\bibitem[\protect\citeauthoryear{{Croft}, {Weinberg}, {Pettini}, {Hernquist}
  \& {Katz}}{{Croft} et~al.}{1999}]{croft+99}
{Croft} R. A.~C.,  {Weinberg} D.~H.,  {Pettini} M.,  {Hernquist} L.,   {Katz}
  N.,  1999, \mn@doi [\apj] {10.1086/307438}, \href
  {https://ui.adsabs.harvard.edu/abs/1999ApJ...520....1C} {520, 1}

\bibitem[\protect\citeauthoryear{{DESI Collaboration} et~al.,}{{DESI
  Collaboration} et~al.}{2016}]{desi}
{DESI Collaboration} et~al., 2016, arXiv e-prints, \href
  {https://ui.adsabs.harvard.edu/abs/2016arXiv161100036D} {p. arXiv:1611.00036}

\bibitem[\protect\citeauthoryear{{Dalton} et~al.,}{{Dalton}
  et~al.}{2016}]{dalton+16}
{Dalton} G.,  et~al., 2016, in {Evans} C.~J.,  {Simard} L.,   {Takami} H.,
  eds,  Society of Photo-Optical Instrumentation Engineers (SPIE) Conference
  Series Vol. 9908, Ground-based and Airborne Instrumentation for Astronomy VI.
  p. 99081G, \mn@doi{10.1117/12.2231078}

\bibitem[\protect\citeauthoryear{{Dalton} et~al.,}{{Dalton}
  et~al.}{2020}]{dalton+20}
{Dalton} G.,  et~al., 2020, in Society of Photo-Optical Instrumentation
  Engineers (SPIE) Conference Series. p. 1144714, \mn@doi{10.1117/12.2561067}

\bibitem[\protect\citeauthoryear{{Dawson} et~al.,}{{Dawson}
  et~al.}{2013}]{dawson+13}
{Dawson} K.~S.,  et~al., 2013, \mn@doi [\aj] {10.1088/0004-6256/145/1/10},
  \href {https://ui.adsabs.harvard.edu/abs/2013AJ....145...10D} {145, 10}

\bibitem[\protect\citeauthoryear{{Dawson} et~al.,}{{Dawson}
  et~al.}{2016}]{dawson+16}
{Dawson} K.~S.,  et~al., 2016, \mn@doi [\aj] {10.3847/0004-6256/151/2/44},
  \href {https://ui.adsabs.harvard.edu/abs/2016AJ....151...44D} {151, 44}

\bibitem[\protect\citeauthoryear{{Delubac} et~al.,}{{Delubac}
  et~al.}{2015}]{delubac+15}
{Delubac} T.,  et~al., 2015, \mn@doi [\aap] {10.1051/0004-6361/201423969},
  \href {https://ui.adsabs.harvard.edu/abs/2015A&A...574A..59D} {574, A59}

\bibitem[\protect\citeauthoryear{{Dubois} et~al.,}{{Dubois}
  et~al.}{2014}]{dubois14}
{Dubois} Y.,  et~al., 2014, \mn@doi [\mnras] {10.1093/mnras/stu1227}, \href
  {https://ui.adsabs.harvard.edu/abs/2014MNRAS.444.1453D} {444, 1453}

\bibitem[\protect\citeauthoryear{{Eisenstein} et~al.,}{{Eisenstein}
  et~al.}{2011}]{eisenstein+11}
{Eisenstein} D.~J.,  et~al., 2011, \mn@doi [\aj] {10.1088/0004-6256/142/3/72},
  \href {https://ui.adsabs.harvard.edu/abs/2011AJ....142...72E} {142, 72}

\bibitem[\protect\citeauthoryear{{Faucher-Gigu{\`e}re}, {Prochaska}, {Lidz},
  {Hernquist}  \& {Zaldarriaga}}{{Faucher-Gigu{\`e}re} et~al.}{2008}]{faucher}
{Faucher-Gigu{\`e}re} C.-A.,  {Prochaska} J.~X.,  {Lidz} A.,  {Hernquist} L.,
  {Zaldarriaga} M.,  2008, \mn@doi [\apj] {10.1086/588648}, \href
  {https://ui.adsabs.harvard.edu/abs/2008ApJ...681..831F} {681, 831}

\bibitem[\protect\citeauthoryear{{Font-Ribera} et~al.,}{{Font-Ribera}
  et~al.}{2012}]{font-ribera+12}
{Font-Ribera} A.,  et~al., 2012, \mn@doi [\jcap]
  {10.1088/1475-7516/2012/11/059}, \href
  {https://ui.adsabs.harvard.edu/abs/2012JCAP...11..059F} {2012, 059}

\bibitem[\protect\citeauthoryear{{Font-Ribera} et~al.,}{{Font-Ribera}
  et~al.}{2013}]{font-ribera+13}
{Font-Ribera} A.,  et~al., 2013, \mn@doi [\jcap]
  {10.1088/1475-7516/2013/05/018}, \href
  {https://ui.adsabs.harvard.edu/abs/2013JCAP...05..018F} {2013, 018}

\bibitem[\protect\citeauthoryear{{Font-Ribera} et~al.,}{{Font-Ribera}
  et~al.}{2014}]{font-ribera+14}
{Font-Ribera} A.,  et~al., 2014, \mn@doi [\jcap]
  {10.1088/1475-7516/2014/05/027}, \href
  {https://ui.adsabs.harvard.edu/abs/2014JCAP...05..027F} {2014, 027}

\bibitem[\protect\citeauthoryear{{Francis} \& {Hewett}}{{Francis} \&
  {Hewett}}{1993}]{francis+93}
{Francis} P.~J.,  {Hewett} P.~C.,  1993, \mn@doi [\aj] {10.1086/116542}, \href
  {https://ui.adsabs.harvard.edu/abs/1993AJ....105.1633F} {105, 1633}

\bibitem[\protect\citeauthoryear{{Gnedin} \& {Hui}}{{Gnedin} \&
  {Hui}}{1996}]{gnedin+96}
{Gnedin} N.~Y.,  {Hui} L.,  1996, \mn@doi [\apjl] {10.1086/310366}, \href
  {https://ui.adsabs.harvard.edu/abs/1996ApJ...472L..73G} {472, L73}

\bibitem[\protect\citeauthoryear{{Harrington}, {Mustafa}, {Dornfest},
  {Horowitz}  \& {Luki{\'c}}}{{Harrington} et~al.}{2021}]{harrington+21}
{Harrington} P.,  {Mustafa} M.,  {Dornfest} M.,  {Horowitz} B.,   {Luki{\'c}}
  Z.,  2021, arXiv e-prints, \href
  {https://ui.adsabs.harvard.edu/abs/2021arXiv210612662H} {p. arXiv:2106.12662}

\bibitem[\protect\citeauthoryear{{Hockney} \& {Eastwood}}{{Hockney} \&
  {Eastwood}}{1988}]{1988csup.book.....H}
{Hockney} R.~W.,  {Eastwood} J.~W.,  1988, {Computer simulation using
  particles}

\bibitem[\protect\citeauthoryear{{Horowitz}, {Dornfest}, {Luki{\'c}}  \&
  {Harrington}}{{Horowitz} et~al.}{2021}]{horowitz}
{Horowitz} B.,  {Dornfest} M.,  {Luki{\'c}} Z.,   {Harrington} P.,  2021, arXiv
  e-prints, \href {https://ui.adsabs.harvard.edu/abs/2021arXiv210612675H} {p.
  arXiv:2106.12675}

\bibitem[\protect\citeauthoryear{{Japelj} et~al.,}{{Japelj}
  et~al.}{2019}]{japelj+19}
{Japelj} J.,  et~al., 2019, \mn@doi [\aap] {10.1051/0004-6361/201936048}, \href
  {https://ui.adsabs.harvard.edu/abs/2019A&A...632A..94J} {632, A94}

\bibitem[\protect\citeauthoryear{{Komatsu} et~al.,}{{Komatsu}
  et~al.}{2011}]{wmap7}
{Komatsu} E.,  et~al., 2011, \mn@doi [\apjs] {10.1088/0067-0049/192/2/18},
  \href {https://ui.adsabs.harvard.edu/abs/2011ApJS..192...18K} {192, 18}

\bibitem[\protect\citeauthoryear{{Kraljic} et~al.,}{{Kraljic}
  et~al.}{2022}]{kraljic+22}
{Kraljic} K.,  et~al., 2022, arXiv e-prints, \href
  {https://ui.adsabs.harvard.edu/abs/2022arXiv220102606K} {p. arXiv:2201.02606}

\bibitem[\protect\citeauthoryear{{Lee} et~al.,}{{Lee} et~al.}{2015}]{lee+15}
{Lee} K.-G.,  et~al., 2015, \mn@doi [\apj] {10.1088/0004-637X/799/2/196}, \href
  {https://ui.adsabs.harvard.edu/abs/2015ApJ...799..196L} {799, 196}

\bibitem[\protect\citeauthoryear{{Lee} et~al.,}{{Lee} et~al.}{2018}]{lee+18}
{Lee} K.-G.,  et~al., 2018, \mn@doi [\apjs] {10.3847/1538-4365/aace58}, \href
  {https://ui.adsabs.harvard.edu/abs/2018ApJS..237...31L} {237, 31}

\bibitem[\protect\citeauthoryear{{Lochhaas} et~al.,}{{Lochhaas}
  et~al.}{2016}]{lochhaas16}
{Lochhaas} C.,  et~al., 2016, \mn@doi [\mnras] {10.1093/mnras/stw1646}, \href
  {https://ui.adsabs.harvard.edu/abs/2016MNRAS.461.4353L} {461, 4353}

\bibitem[\protect\citeauthoryear{{Lynds}}{{Lynds}}{1971}]{lynds71}
{Lynds} R.,  1971, \mn@doi [\apjl] {10.1086/180695}, \href
  {https://ui.adsabs.harvard.edu/abs/1971ApJ...164L..73L} {164, L73}

\bibitem[\protect\citeauthoryear{{McQuinn}}{{McQuinn}}{2009}]{mcquinn09}
{McQuinn} M.,  2009, \mn@doi [\apjl] {10.1088/0004-637X/704/2/L89}, \href
  {https://ui.adsabs.harvard.edu/abs/2009ApJ...704L..89M} {704, L89}

\bibitem[\protect\citeauthoryear{{Monaghan} \& {Lattanzio}}{{Monaghan} \&
  {Lattanzio}}{1985}]{1985A&A...149..135M}
{Monaghan} J.~J.,  {Lattanzio} J.~C.,  1985, \aap, \href
  {https://ui.adsabs.harvard.edu/abs/1985A&A...149..135M} {149, 135}

\bibitem[\protect\citeauthoryear{{Ozbek}, {Croft}  \& {Khandai}}{{Ozbek}
  et~al.}{2016}]{ozbek+16}
{Ozbek} M.,  {Croft} R. A.~C.,   {Khandai} N.,  2016, \mn@doi [\mnras]
  {10.1093/mnras/stv2894}, \href
  {https://ui.adsabs.harvard.edu/abs/2016MNRAS.456.3610O} {456, 3610}

\bibitem[\protect\citeauthoryear{{Palanque-Delabrouille}
  et~al.,}{{Palanque-Delabrouille} et~al.}{2013}]{palanque+13}
{Palanque-Delabrouille} N.,  et~al., 2013, \mn@doi [\aap]
  {10.1051/0004-6361/201322130}, \href
  {https://ui.adsabs.harvard.edu/abs/2013A&A...559A..85P} {559, A85}

\bibitem[\protect\citeauthoryear{{Peeples}, {Weinberg}, {Dav{\'e}}, {Fardal}
  \& {Katz}}{{Peeples} et~al.}{2010}]{peeples+10}
{Peeples} M.~S.,  {Weinberg} D.~H.,  {Dav{\'e}} R.,  {Fardal} M.~A.,   {Katz}
  N.,  2010, \mn@doi [\mnras] {10.1111/j.1365-2966.2010.16383.x}, \href
  {https://ui.adsabs.harvard.edu/abs/2010MNRAS.404.1281P} {404, 1281}

\bibitem[\protect\citeauthoryear{{Peirani}, {Weinberg}, {Colombi}, {Blaizot},
  {Dubois}  \& {Pichon}}{{Peirani} et~al.}{2014}]{peirani14}
{Peirani} S.,  {Weinberg} D.~H.,  {Colombi} S.,  {Blaizot} J.,  {Dubois} Y.,
  {Pichon} C.,  2014, \mn@doi [\apj] {10.1088/0004-637X/784/1/11}, \href
  {https://ui.adsabs.harvard.edu/abs/2014ApJ...784...11P} {784, 11}

\bibitem[\protect\citeauthoryear{{Peirani} et~al.,}{{Peirani}
  et~al.}{2017}]{peirani17}
{Peirani} S.,  et~al., 2017, \mn@doi [\mnras] {10.1093/mnras/stx2099}, \href
  {https://ui.adsabs.harvard.edu/abs/2017MNRAS.472.2153P} {472, 2153}

\bibitem[\protect\citeauthoryear{{Pichon}, {Vergely}, {Rollinde}, {Colombi}  \&
  {Petitjean}}{{Pichon} et~al.}{2001}]{pichon+01}
{Pichon} C.,  {Vergely} J.~L.,  {Rollinde} E.,  {Colombi} S.,   {Petitjean} P.,
   2001, \mn@doi [\mnras] {10.1046/j.1365-8711.2001.04595.x}, \href
  {https://ui.adsabs.harvard.edu/abs/2001MNRAS.326..597P} {326, 597}

\bibitem[\protect\citeauthoryear{{Pieri} et~al.,}{{Pieri}
  et~al.}{2016}]{pieri+16}
{Pieri} M.~M.,  et~al., 2016, in {Reyl{\'e}} C.,  {Richard} J.,  {Cambr{\'e}sy}
  L.,  {Deleuil} M.,  {P{\'e}contal} E.,  {Tresse} L.,   {Vauglin} I.,  eds,
  SF2A-2016: Proceedings of the Annual meeting of the French Society of
  Astronomy and Astrophysics. pp 259--266 (\mn@eprint {arXiv} {1611.09388})

\bibitem[\protect\citeauthoryear{{Ravoux} et~al.,}{{Ravoux}
  et~al.}{2020}]{ravoux+20}
{Ravoux} C.,  et~al., 2020, \mn@doi [\jcap] {10.1088/1475-7516/2020/07/010},
  \href {https://ui.adsabs.harvard.edu/abs/2020JCAP...07..010R} {2020, 010}

\bibitem[\protect\citeauthoryear{{Sargent}, {Young}, {Boksenberg}  \&
  {Tytler}}{{Sargent} et~al.}{1980}]{sargent+80}
{Sargent} W.~L.~W.,  {Young} P.~J.,  {Boksenberg} A.,   {Tytler} D.,  1980,
  \mn@doi [\apjs] {10.1086/190644}, \href
  {https://ui.adsabs.harvard.edu/abs/1980ApJS...42...41S} {42, 41}

\bibitem[\protect\citeauthoryear{{Schaye} et~al.,}{{Schaye}
  et~al.}{2015}]{eagle}
{Schaye} J.,  et~al., 2015, \mn@doi [\mnras] {10.1093/mnras/stu2058}, \href
  {https://ui.adsabs.harvard.edu/abs/2015MNRAS.446..521S} {446, 521}

\bibitem[\protect\citeauthoryear{{Shi}, {Cai}, {Fan}, {Zheng}, {Huang}  \&
  {Xu}}{{Shi} et~al.}{2021}]{shi+21}
{Shi} D.,  {Cai} Z.,  {Fan} X.,  {Zheng} X.,  {Huang} Y.-H.,   {Xu} J.,  2021,
  arXiv e-prints, \href {https://ui.adsabs.harvard.edu/abs/2021arXiv210502248S}
  {p. arXiv:2105.02248}

\bibitem[\protect\citeauthoryear{{Sinigaglia}, {Kitaura},
  {Balaguera-Antol{\'\i}nez}, {Shimizu}, {Nagamine}, {S{\'a}nchez-Benavente}
  \& {Ata}}{{Sinigaglia} et~al.}{2021}]{sinigaglia+21}
{Sinigaglia} F.,  {Kitaura} F.-S.,  {Balaguera-Antol{\'\i}nez} A.,  {Shimizu}
  I.,  {Nagamine} K.,  {S{\'a}nchez-Benavente} M.,   {Ata} M.,  2021, arXiv
  e-prints, \href {https://ui.adsabs.harvard.edu/abs/2021arXiv210707917S} {p.
  arXiv:2107.07917}

\bibitem[\protect\citeauthoryear{{Slosar} et~al.,}{{Slosar}
  et~al.}{2011}]{slosar+11}
{Slosar} A.,  et~al., 2011, \mn@doi [\jcap] {10.1088/1475-7516/2011/09/001},
  \href {https://ui.adsabs.harvard.edu/abs/2011JCAP...09..001S} {2011, 001}

\bibitem[\protect\citeauthoryear{{Slosar} et~al.,}{{Slosar}
  et~al.}{2013}]{slosar+13}
{Slosar} A.,  et~al., 2013, \mn@doi [\jcap] {10.1088/1475-7516/2013/04/026},
  \href {https://ui.adsabs.harvard.edu/abs/2013JCAP...04..026S} {2013, 026}

\bibitem[\protect\citeauthoryear{{Sorini}, {O{\~n}orbe}, {Luki{\'c}}  \&
  {Hennawi}}{{Sorini} et~al.}{2016}]{sorini+16}
{Sorini} D.,  {O{\~n}orbe} J.,  {Luki{\'c}} Z.,   {Hennawi} J.~F.,  2016,
  \mn@doi [\apj] {10.3847/0004-637X/827/2/97}, \href
  {https://ui.adsabs.harvard.edu/abs/2016ApJ...827...97S} {827, 97}

\bibitem[\protect\citeauthoryear{{Springel}}{{Springel}}{2005}]{gadget2}
{Springel} V.,  2005, \mn@doi [\mnras] {10.1111/j.1365-2966.2005.09655.x},
  \href {https://ui.adsabs.harvard.edu/abs/2005MNRAS.364.1105S} {364, 1105}

\bibitem[\protect\citeauthoryear{{Takada} et~al.,}{{Takada}
  et~al.}{2014}]{takada+14}
{Takada} M.,  et~al., 2014, \mn@doi [\pasj] {10.1093/pasj/pst019}, \href
  {https://ui.adsabs.harvard.edu/abs/2014PASJ...66R...1T} {66, R1}

\bibitem[\protect\citeauthoryear{{Teyssier}}{{Teyssier}}{2002}]{ramses}
{Teyssier} R.,  2002, \mn@doi [\aap] {10.1051/0004-6361:20011817}, \href
  {https://ui.adsabs.harvard.edu/abs/2002A&A...385..337T} {385, 337}

\bibitem[\protect\citeauthoryear{{Tie}, {Weinberg}, {Martini}, {Zhu},
  {Peirani}, {Suarez}  \& {Colombi}}{{Tie} et~al.}{2019}]{suk}
{Tie} S.~S.,  {Weinberg} D.~H.,  {Martini} P.,  {Zhu} W.,  {Peirani} S.,
  {Suarez} T.,   {Colombi} S.,  2019, \mn@doi [\mnras] {10.1093/mnras/stz1632},
  \href {https://ui.adsabs.harvard.edu/abs/2019MNRAS.487.5346T} {487, 5346}

\bibitem[\protect\citeauthoryear{{Viel}, {Schaye}  \& {Booth}}{{Viel}
  et~al.}{2013}]{viel+13}
{Viel} M.,  {Schaye} J.,   {Booth} C.~M.,  2013, \mn@doi [\mnras]
  {10.1093/mnras/sts465}, \href
  {https://ui.adsabs.harvard.edu/abs/2013MNRAS.429.1734V} {429, 1734}

\bibitem[\protect\citeauthoryear{{Vogelsberger} et~al.,}{{Vogelsberger}
  et~al.}{2014}]{illustris}
{Vogelsberger} M.,  et~al., 2014, \mn@doi [\mnras] {10.1093/mnras/stu1536},
  \href {https://ui.adsabs.harvard.edu/abs/2014MNRAS.444.1518V} {444, 1518}

\bibitem[\protect\citeauthoryear{{Weinberg} \& {Cole}}{{Weinberg} \&
  {Cole}}{1992}]{weinberg92}
{Weinberg} D.~H.,  {Cole} S.,  1992, \mn@doi [\mnras]
  {10.1093/mnras/259.4.652}, \href
  {https://ui.adsabs.harvard.edu/abs/1992MNRAS.259..652W} {259, 652}

\bibitem[\protect\citeauthoryear{{Weinberg}, {Hernsquit}, {Katz}, {Croft}  \&
  {Miralda-Escud{\'e}}}{{Weinberg} et~al.}{1997}]{weinberg+97}
{Weinberg} D.~H.,  {Hernsquit} L.,  {Katz} N.,  {Croft} R.,
  {Miralda-Escud{\'e}} J.,  1997, in {Petitjean} P.,  {Charlot} S.,  eds,
  Structure and Evolution of the Intergalactic Medium from QSO Absorption Line
  System. p.~133 (\mn@eprint {arXiv} {astro-ph/9709303})

\bibitem[\protect\citeauthoryear{{Weinberg}, {Katz}  \& {Hernquist}}{{Weinberg}
  et~al.}{1998}]{weinberg+98}
{Weinberg} D.~H.,  {Katz} N.,   {Hernquist} L.,  1998, in {Woodward} C.~E.,
  {Shull} J.~M.,   {Thronson} Harley~A. J.,  eds,  Astronomical Society of the
  Pacific Conference Series Vol. 148, Origins. p.~21 (\mn@eprint {arXiv}
  {astro-ph/9708213})

\bibitem[\protect\citeauthoryear{{du Mas des Bourboux} et~al.,}{{du Mas des
  Bourboux} et~al.}{2020}]{dumasdesbourboux}
{du Mas des Bourboux} H.,  et~al., 2020, \mn@doi [\apj]
  {10.3847/1538-4357/abb085}, \href
  {https://ui.adsabs.harvard.edu/abs/2020ApJ...901..153D} {901, 153}

\makeatother
\end{thebibliography}

\appendix

\section{General trends}
\label{appendix1}
In section~\ref{section_3dclustering}, we have presented
the predictions regarding the 2-point correlation functions
 for eight different combinations of DM fields, with calibrations
 derived from \hnoagn. To check the robustness of the results,
 the analysis of other similar hydrodynamical simulations
 is definitely required. To limit the computational time, we
 ran five additional hydrodynamical simulations with the same 
 box side and same physics than \hnoagnn but with two times 
 lower resolution (i.e. 512$^3$ DM particles instead
 of 1024$^3$ and a minimal cell size of $\Delta$x=2 kpc instead of 1 kpc).
 The first simulation uses degraded \hnoagnn initial conditions while
 the other ones have different initial phases.
For each of the five new simulations, we generated the corresponding grids of transmitted 
flux, DM overdensity and velocity fields and calibrations folowing
the same methodology presented in section~\ref{sec:simu}. 
We consider here flux and DM fields sampled on grids of 512$\times$512$\times$1024 namely
512$\times$512 spectra of resolution 1024.

\begin{figure*}
\begin{center}
\rotatebox{0}{\includegraphics[width=\columnwidth]{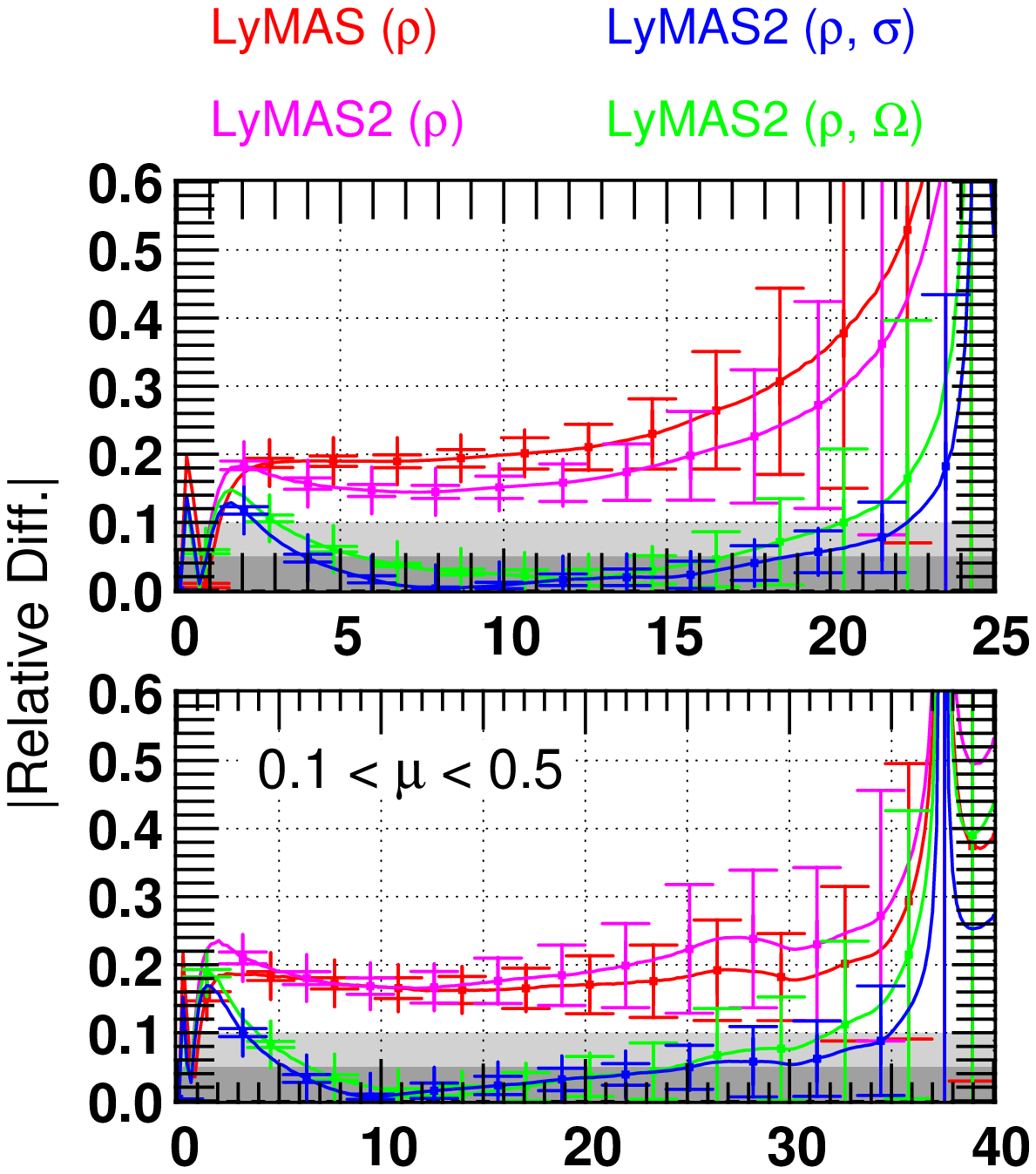}}
\rotatebox{0}{\includegraphics[width=\columnwidth]{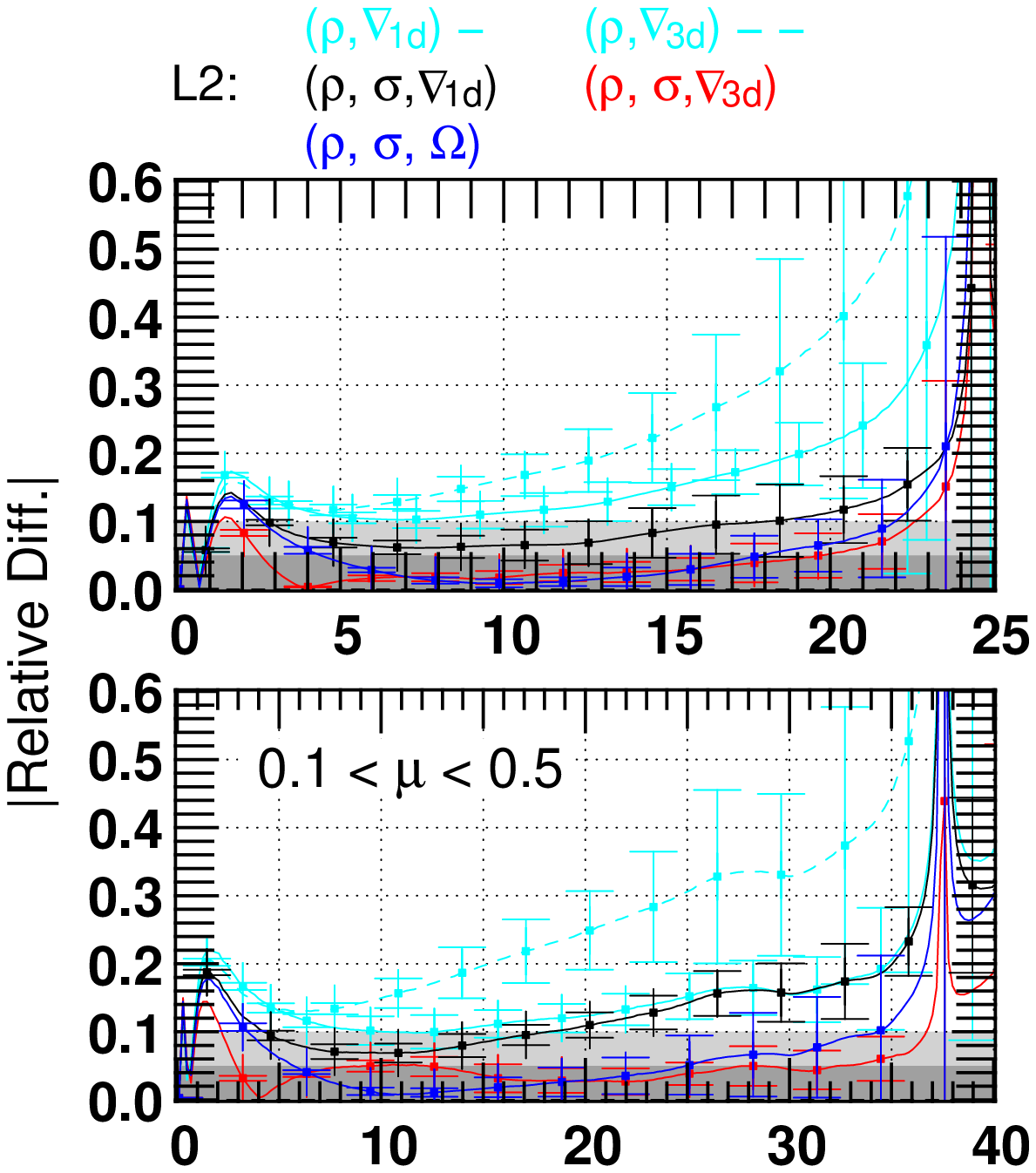}}
\rotatebox{0}{\includegraphics[width=\columnwidth]{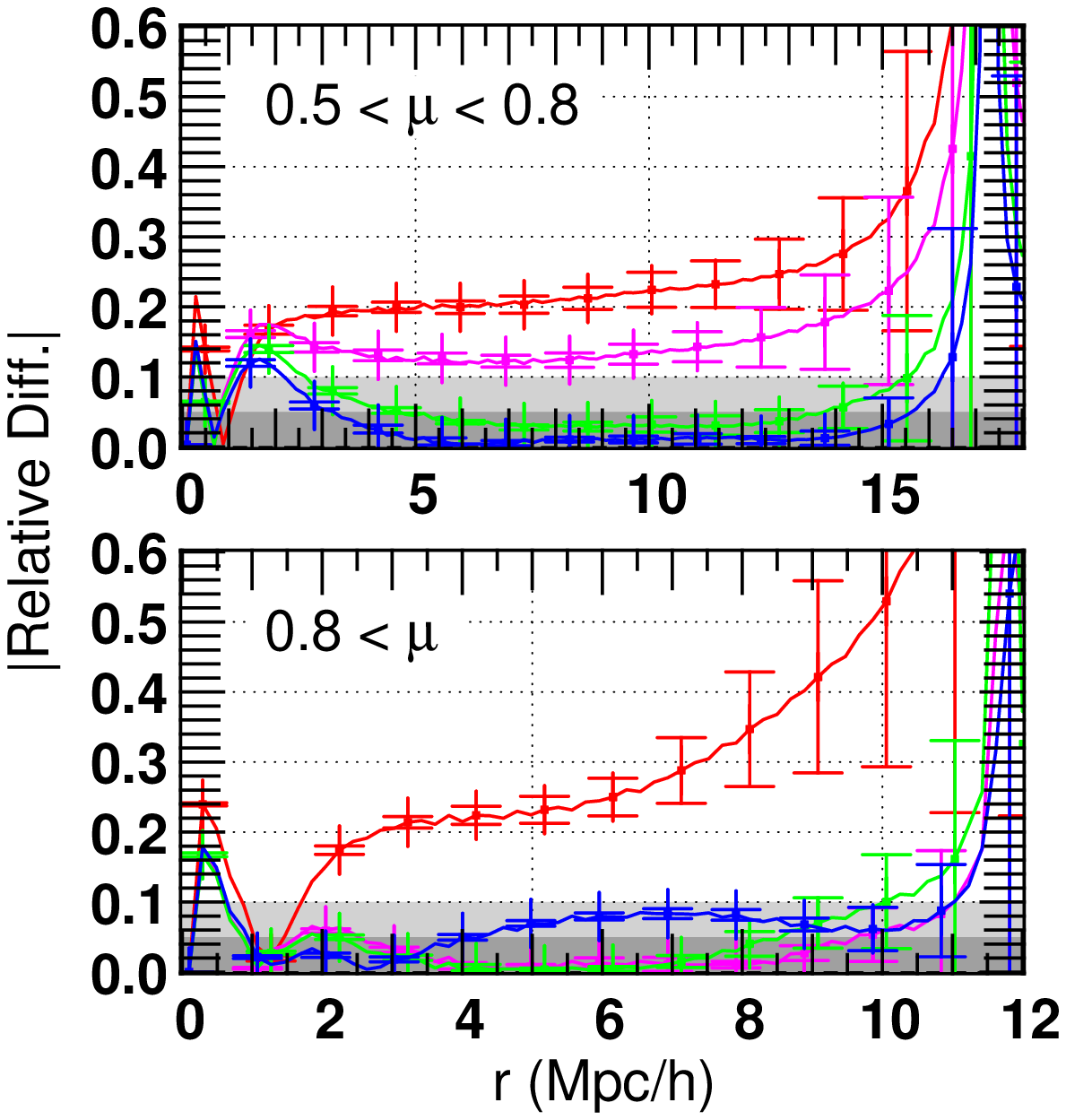}}
\rotatebox{0}{\includegraphics[width=\columnwidth]{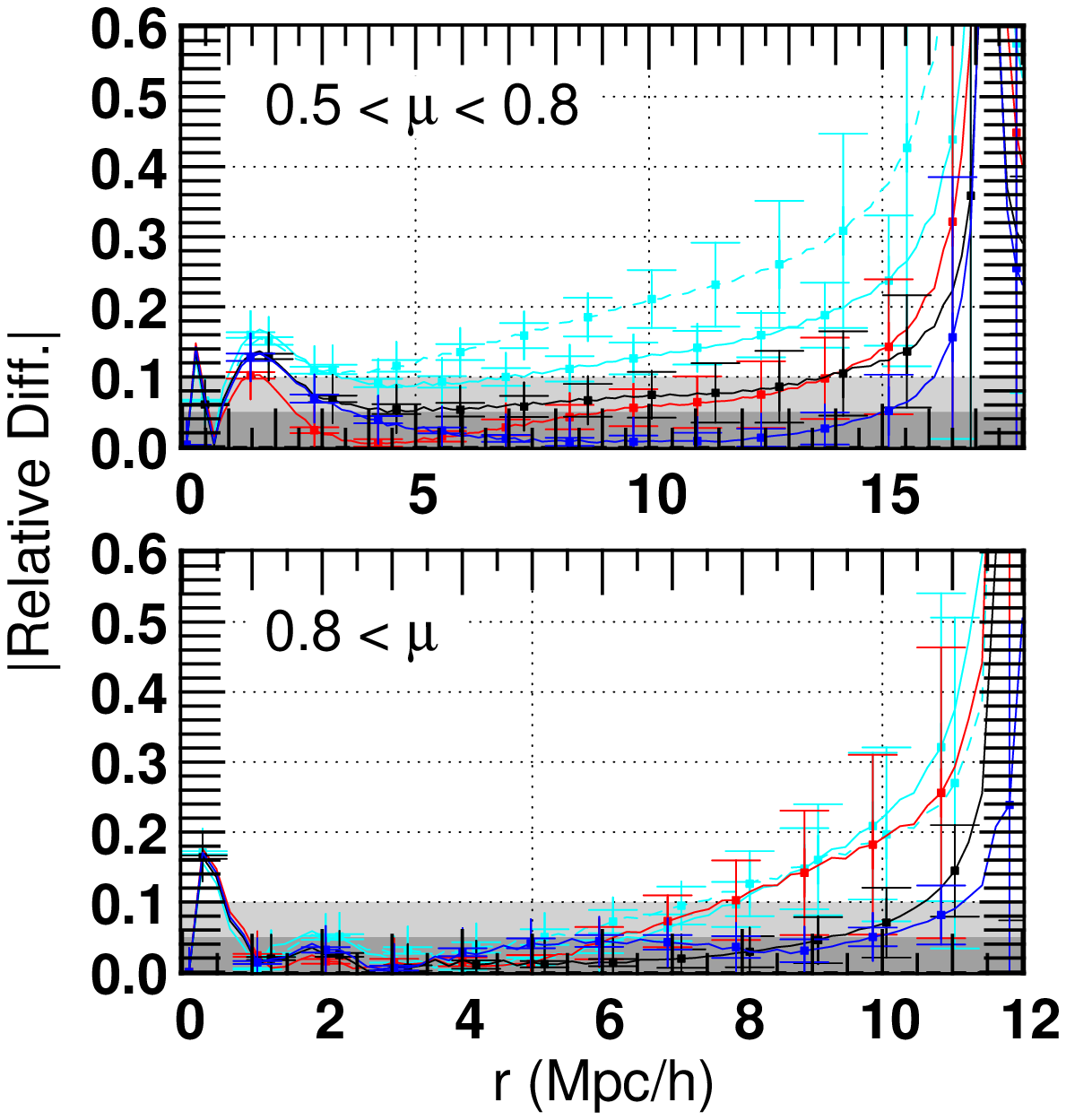}}
\caption{The evolution  of the mean of the absolute relative difference 
in the two point correlation functions (1/5)$|\sum_{i=1}^5$($\xi_i/\xi_{hydro,i}$-1)$|$
derived from 5 different hydrodynamical simulations at $z=2.5$.
All dark matter fields are smoothed to 0.5 Mpc/h.
The error bars correspond to the dispersion.
Comparison with results from Figures \ref{fig_CF1} and \ref{fig_CF2}
suggest a very good agreement and therefore robust trends.}
\label{fig_appendix_1}
\end{center}
 \end{figure*}

In the first step, we consider all DM fields smoothed at 0.5 Mpc/h.
After checking first that the ``high" and ``low" resolution \hnoagnn simulations
give consistent trends, we took an interest in the variations of
then mean of the absolute relative difference (1/5)$|\sum_{i=1}^5$($\xi_i/\xi_{hydro,i}$-1)$|$,
where we compare the 2-point correlation function of the hydro spectra $\xi_{hydro,i}$ from
a given simulation ``$i$"  to those derived from pseudo-spectra 
generated with LyMAS2  $\xi_{hydro,i}$.
In Fig.~\ref{fig_appendix_1}, we summarize the results obtained with
the original LyMAS and LyMAS2 considering the same DM field combinations
than in Figures~\ref{fig_CF1} and \ref{fig_CF2}.
The main conclusion is that we do find very similar trends than those obtained
with \hnoagnn, which strongly suggest
 that our results are robust. In particular, the use of the velocity dispersion
($\sigma$) or the vorticity ($\Omega$) lead to relative errors that 
are remarkably low i.e. in general lower
than 5\% even for the different ranges of angle.
The plots also confirm  that the 1d and 3d velocity divergence fields
do no permit to reach the same level of accuracy.

\begin{figure*}
\begin{center}
\rotatebox{0}{\includegraphics[width=\columnwidth]{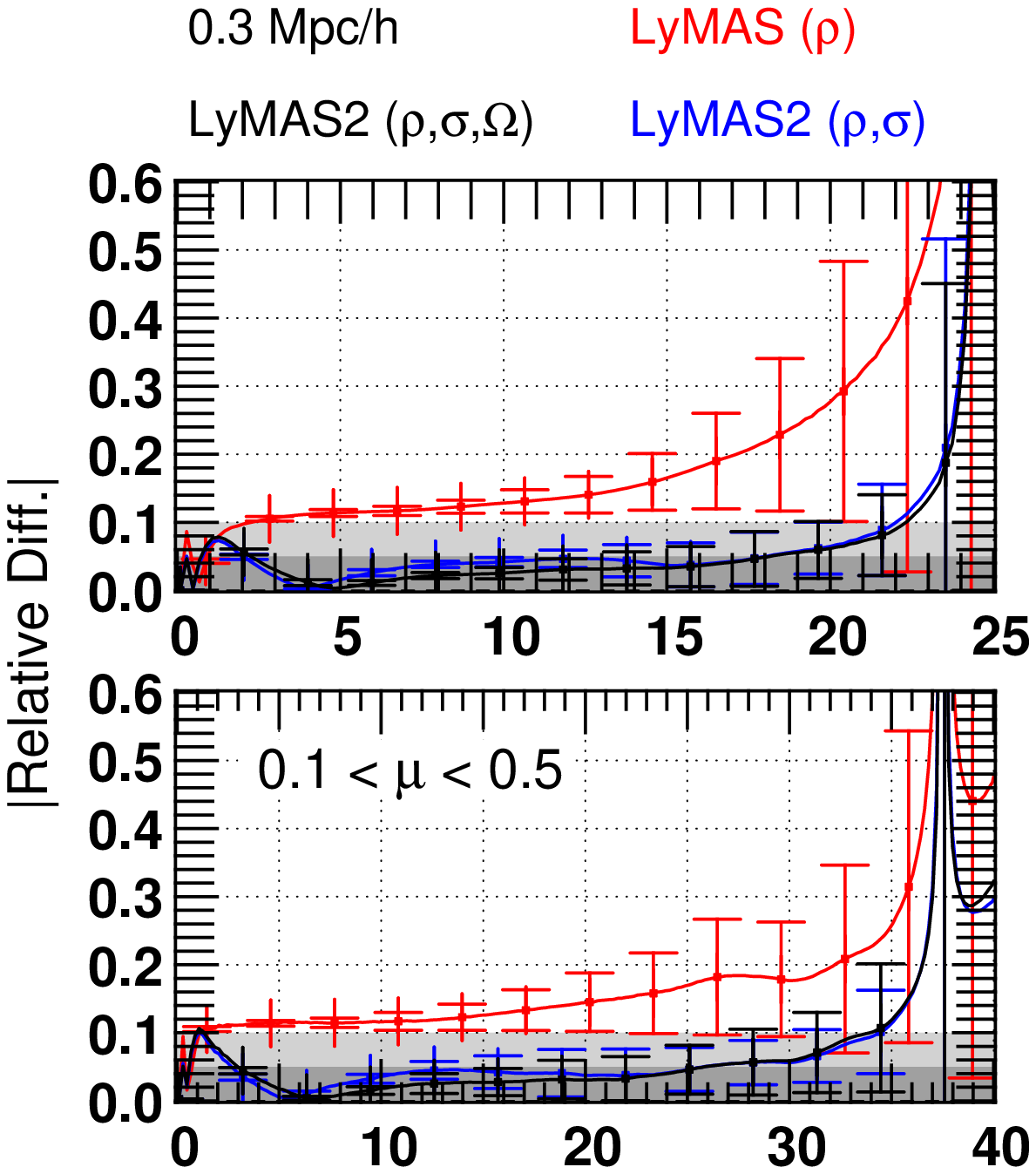}}
\rotatebox{0}{\includegraphics[width=\columnwidth]{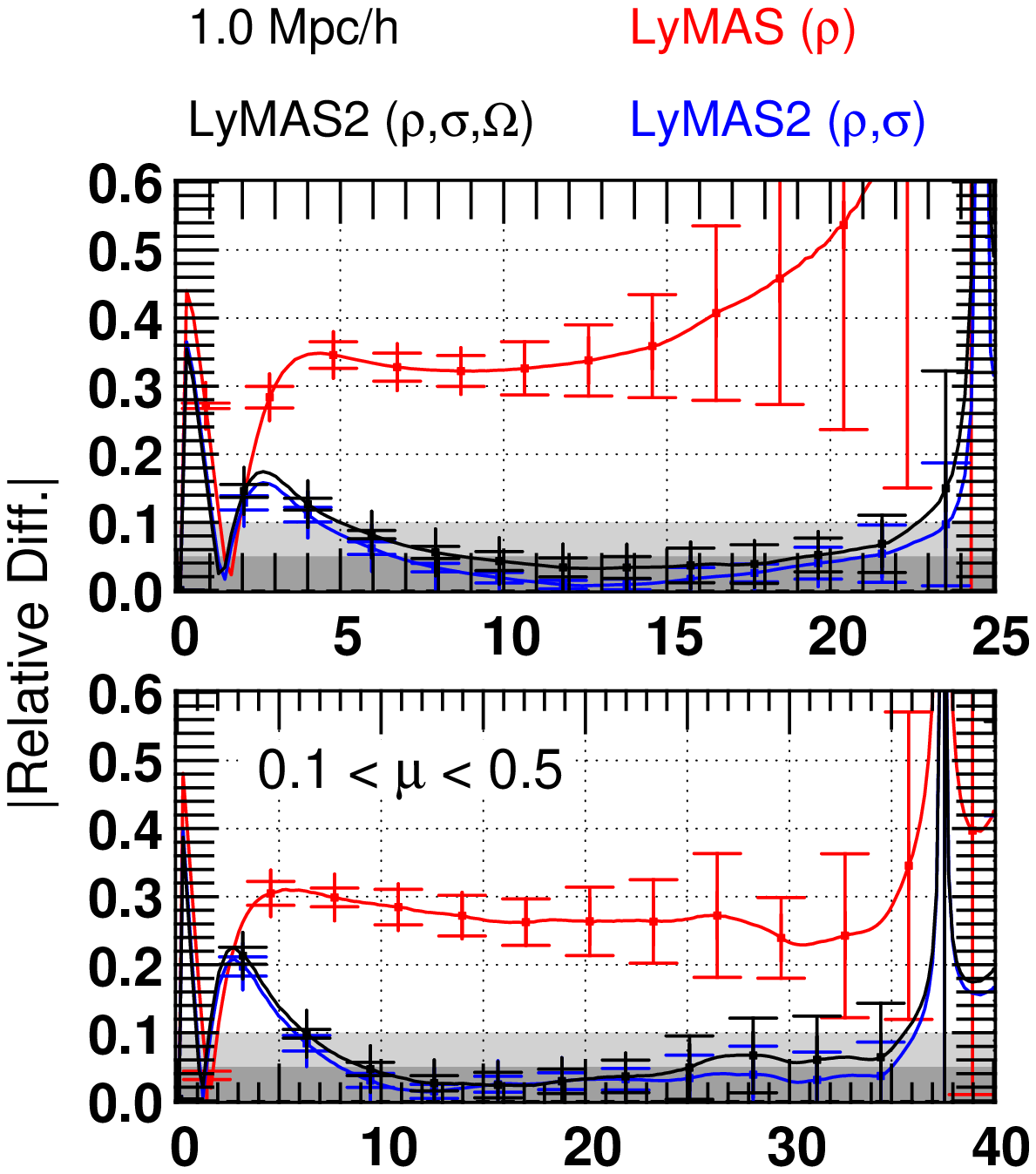}}
\rotatebox{0}{\includegraphics[width=\columnwidth]{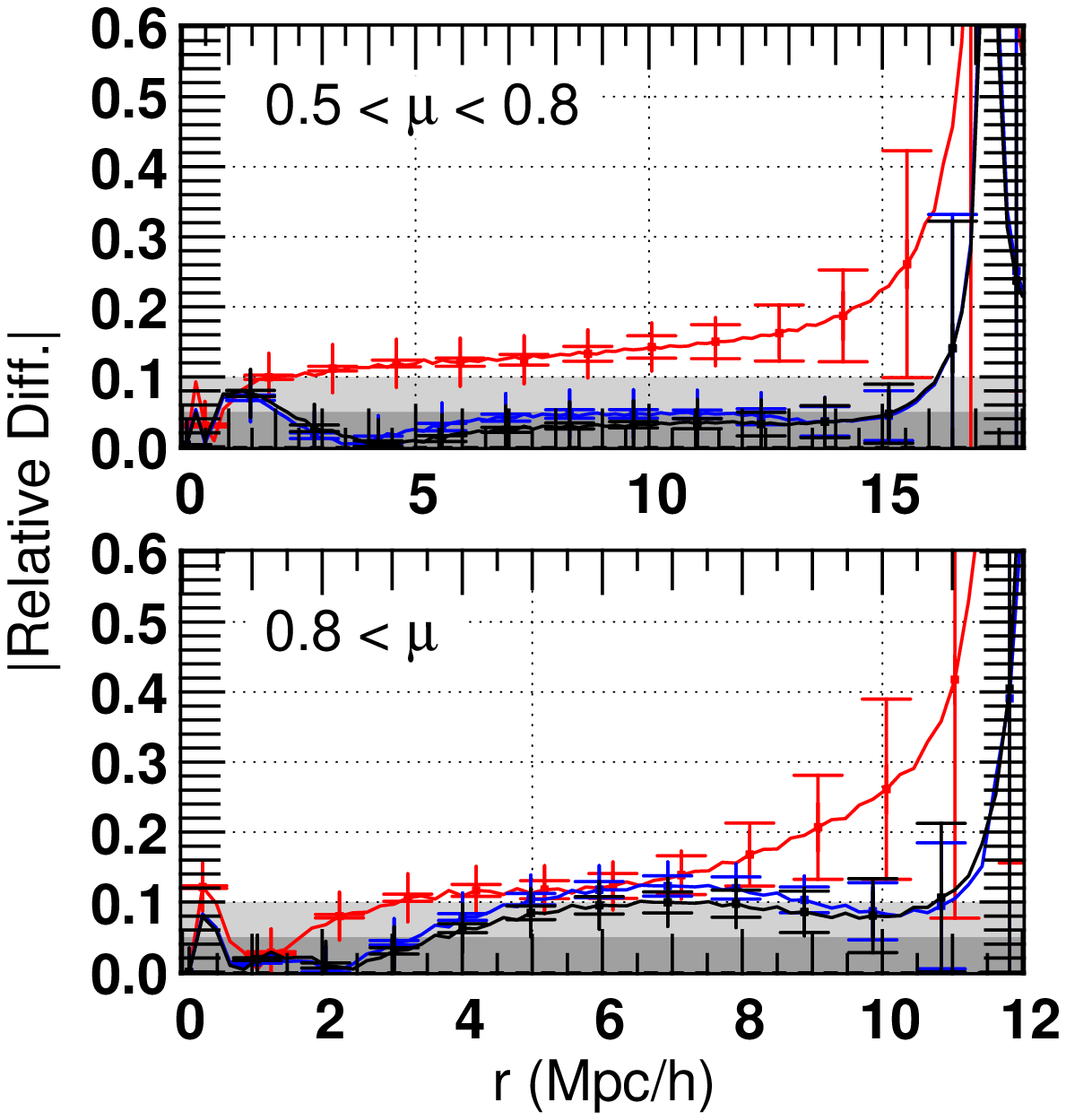}}
\rotatebox{0}{\includegraphics[width=\columnwidth]{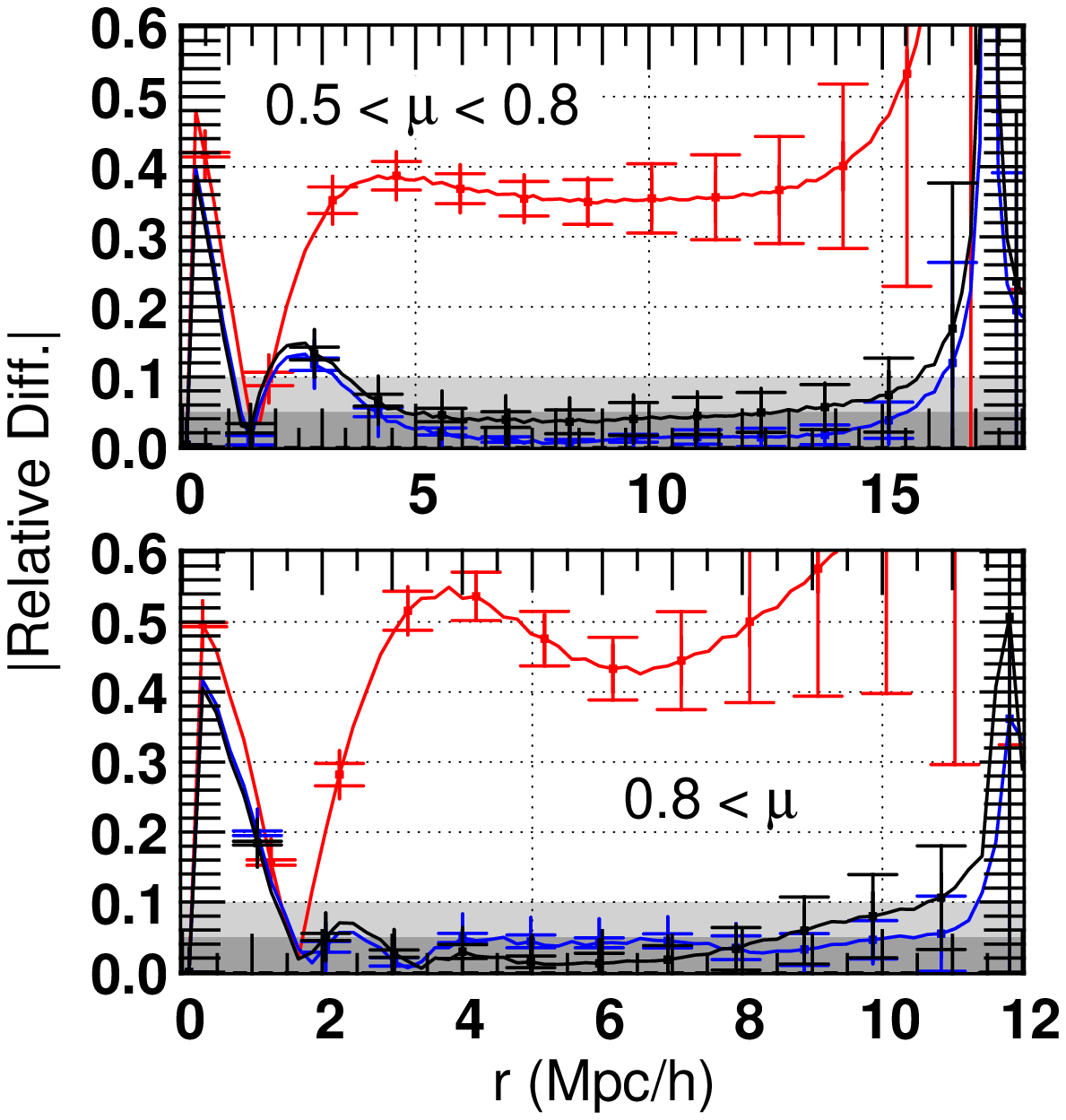}}
\caption{Same as Figure \ref{appendix1} but for DM smoothing of 0.3 Mpc/h (left column)
or 1.0 Mpc/h (righ column). For the sake of clarity, we only show results for
the original LyMAS (red curves), LyMAS2($\rho$,$\sigma$) (blue curves) and LyMAS2($\rho$,$\sigma$,$\Omega$) (black curves). Note that even with a DM smoothing of 1.0 Mpc/h,
LyMAS2 still permits to reach a high level accuracy, even for high angles ($mu>0.8)$.}
\label{fig_appendix_2}
\end{center}
 \end{figure*}

In the next step, we present the trends obtained
when the DM fields are smoothed to 0.3 or 1.0 Mpc/h.
We only present in Fig.~\ref{fig_appendix_2}
the results for LyMAS, LyMAS2($\rho$,$\sigma$) and LyMAS2($\rho$,$\sigma$,$\Omega$) to
have a clear overview of the general trends.
In P14, we found that a DM smoothing of 0.3 Mpc/h was an optimal value to reach the 
highest accuracy in the predictions. This is confirmed here since
we get errors of $\geq$10\% compared to $\geq$20\% and $\geq$30\% with values
0.5 and 1.0 Mpc/h respectively. 
As expected,  LyMAS2 permits to reduces such errors that are in general much
lower than 10\% and most of the time lower than 5\%.
It is also very promising that LyMAS2 applied to DM fields smoothed at
1 Mpc/h  gives such accurate predictions
even for high values of $\mu$. This is definitely not the case with the original LyMAS
leading to very high errors.
Note also that due the smoothing scale, the predictions are less accurate for distance lower 
than 2 Mpc/h but acceptable for large scale analysis.

\section{Deterministic mapping}
\label{appendix2}

One commonly way to produce large mocks of \lyaa forest,
from Gaussian fields or DM distributions extracted
from cosmological simulations, is to use
a physically motivated deterministic relation which links the \lyaa optical depth (or transmitted flux) to the dark matter overdensity. 
This is the case with the so-called Fluctuating Gunn-Peterson Approximation 
which has been extensively used in the literature. 
However, the FGPA is supposed to be more suitable for modelling high-resolution spectra
and can be strongly limited when the DM density field is smoothed to a scale greater than 0.1 Mpc/h (see for instance the analysis of \cite{sorini+16}) and confirmed by our results in section~\ref{sec:fgpa}.
For this reason, we have derived in P14 an ``optimal''  deterministic relations by matching the corresponding cumulative distributions of the smoothed
transmitted Flux $F_s$ and dark matter overdensity $\rho_s$ as:
\begin{equation*}
\int_0^{F_s} P(F'_s)dF'_s = \int_{\rho_s}^\infty P(\rho'_s)d\rho'_s  \,,  
\end{equation*}
where $P(F_s)$ and $P(\rho_s)$ are the 1-point PDFs of the flux and DM overdensity 
measure from the simulation.
One advantage of choosing such deterministic relation is to recover by construction
the PDF of the hydro flux. However, the 2-point correlation function of pseudo-spectra
generated with this approach is still highly overestimated (see for instance Figures 10 and 19 in P14). 

\begin{figure}
\begin{center}
\rotatebox{0}{\includegraphics[width=\columnwidth]{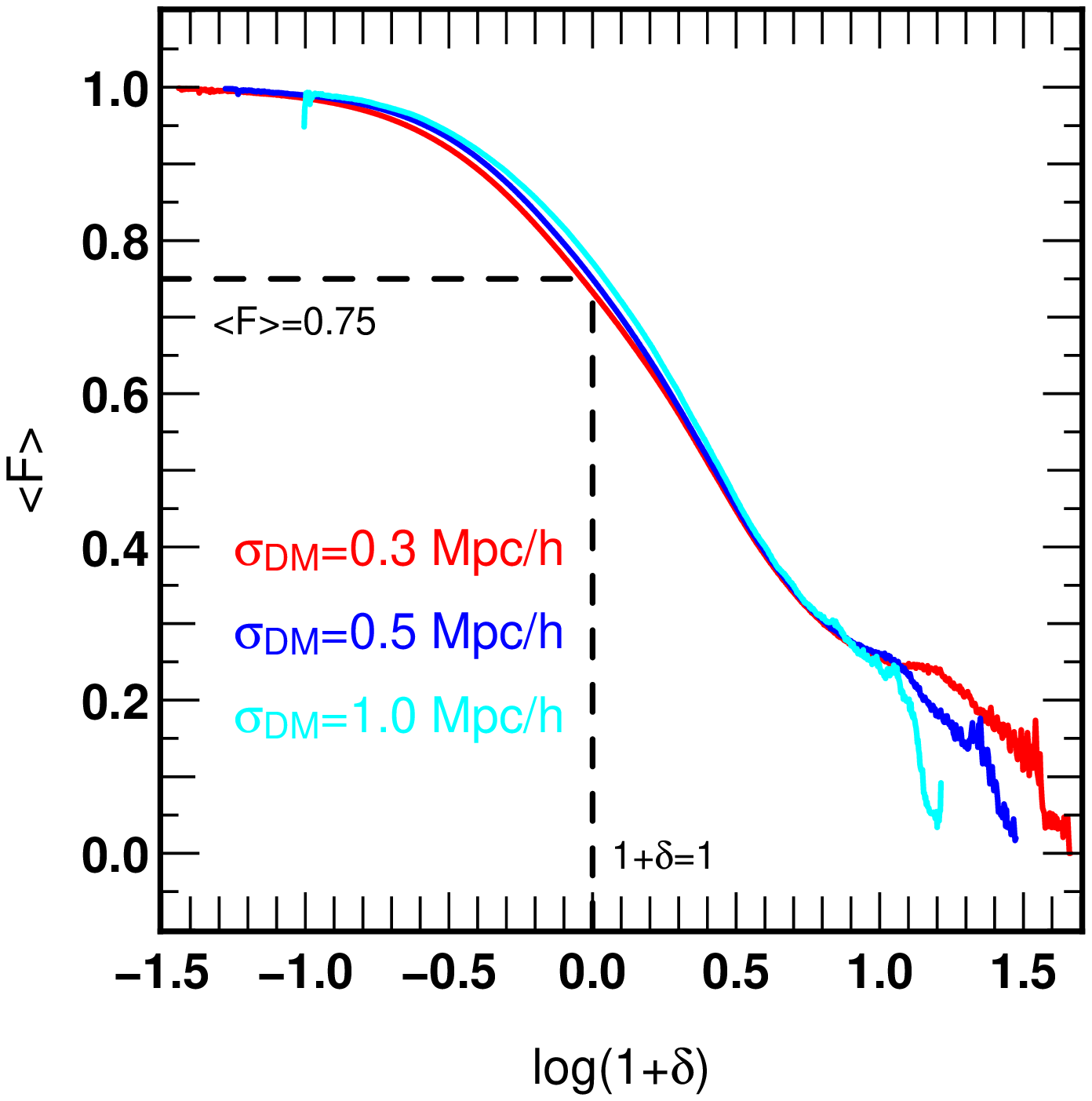}}
\rotatebox{0}{\includegraphics[width=\columnwidth]{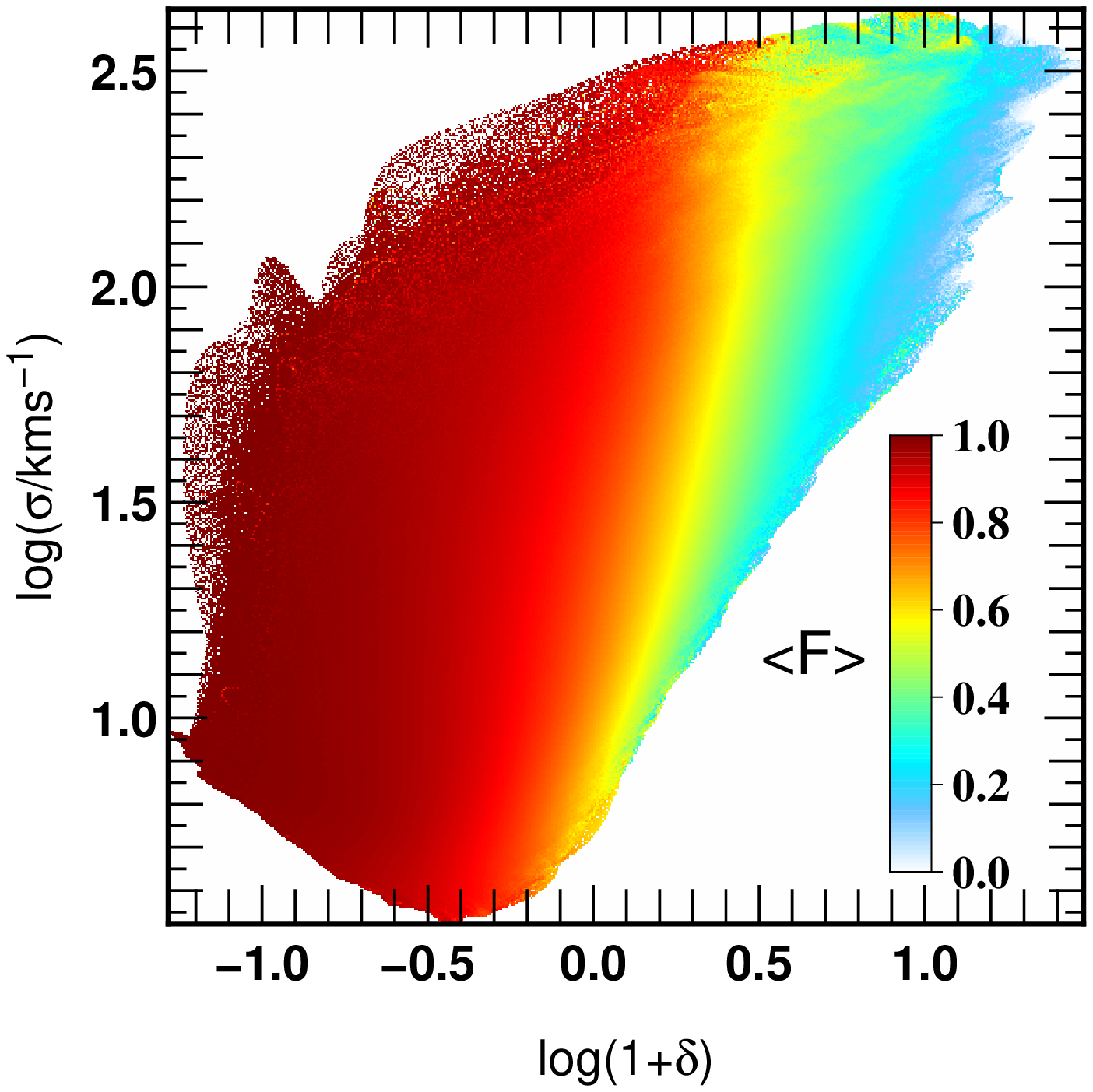}}
\caption{Examples of 1d and 2d deterministic relations derived from
the \hnoagnn simulation at $z=2.5$. Top panel: The mean flux $<F>$ only depends
on the DM overdensity $1+\delta$. Three different DM smoothing have been considered.
For example,  the mean flux has a value of 0.75 for an overdensity of 1
(DM smoothing$=$0.5 Mpc/h). Bottom panel: scatter plot showing the mean flux $<F>$
as respect to the DM overdensity and velocity dispersion field ($\sigma$).
In this case, $<F>$ can have a wide range of values for $1+\delta$=1 (see also
Figure \ref{fig_appendix_4}).
}
\label{fig_appendix_3}
\end{center}
 \end{figure}

\begin{figure}
\begin{center}
\rotatebox{0}{\includegraphics[width=\columnwidth]{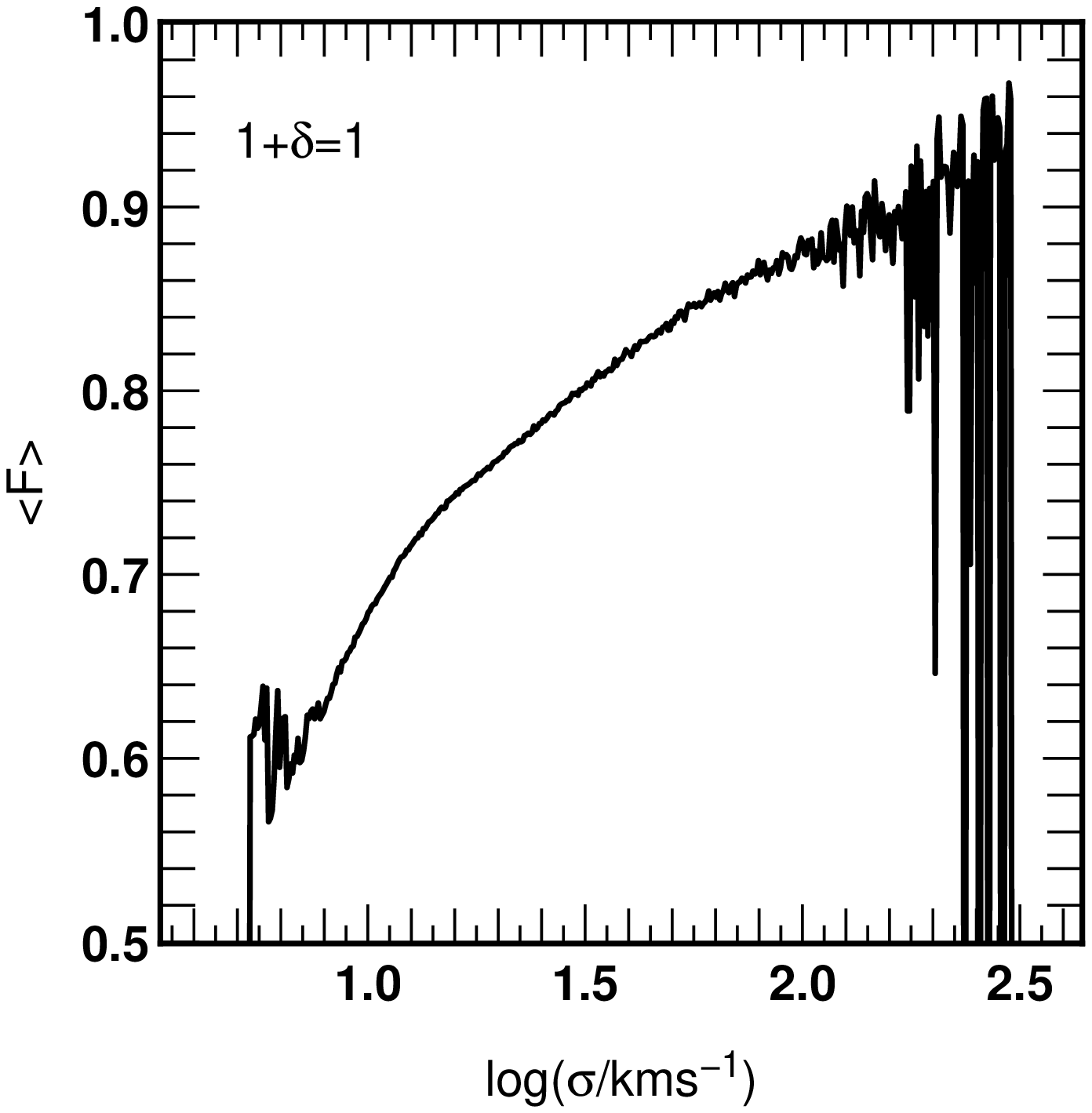}}
\caption{The evolution of the mean flux as respect to the velocity dispersion
($\sigma$) and for a given overdensity (i.e. $1+\delta$=1). The DM fields 
are smoothed at 0.5 Mpc/h.
This variation
is directly derived from the 2d deterministic sampling presented in 
Figure \ref{fig_appendix_3}. Compared to the 1d deterministic relation,
a wide range of values of $<F>$ is obtained and should refine the predictions.
}
\label{fig_appendix_4}
\end{center}
 \end{figure}

In this section, we consider another choices of deterministic relations.
In particular, \cite{suk} have used the conditional probability $P(F|1+\delta)$
of the transmitted flux on the DM overdensity to get the conditional mean flux:
\begin{equation}
\overline{F}(1+\delta) = \int F.P(F|1+\delta)dF    \,.
\end{equation}
It can be analytically demonstrated that the 2-point correlation function 
of pseudo-spectra obtained from such a determinisic mapping
is the same that the one  obtained with the first version of LyMAS  \citep{suk}. 
Since this deterministic relation can be easily extended to several DM fields,
our aim is to investigate whether the inclusion of different DM velocity fields in 
such deterministic mappings may improve the trends or not.
As an illustration, Fig.~\ref{fig_appendix_3} shows examples of 1d-deterministic relations
constructed for different smoothing of the DM overdensity field and
and one example of 2d-deterministic relation $\overline{F}(1+\delta,\sigma)$ using 
both the DM overdensity and the velocity dispersion fields (smoothed at 0.5 Mpc/h).
All  relations are derived from the \hnoagnn simulation at $z=2.5$ (in redshift space).
In principle, the 2d-deterministic relation is supposed to refine the results as repect to
the 1d deterministic one. Indeed,
let's take for instance a DM overdensity of $1+\delta=1$. This leads to an unique mean flux 
of $\langle F\rangle$=0.75 from the 1d deterministic relation (using a DM smoothing of 0.5 Mpc/h).
The 2d-determinisic relation provides, however, a wide range of possible values of $\langle F\rangle$
depending this time on the velocity dispersion (see figure~\ref{fig_appendix_4}).

\begin{figure}
\begin{center}
\rotatebox{0}{\includegraphics[width=\columnwidth]{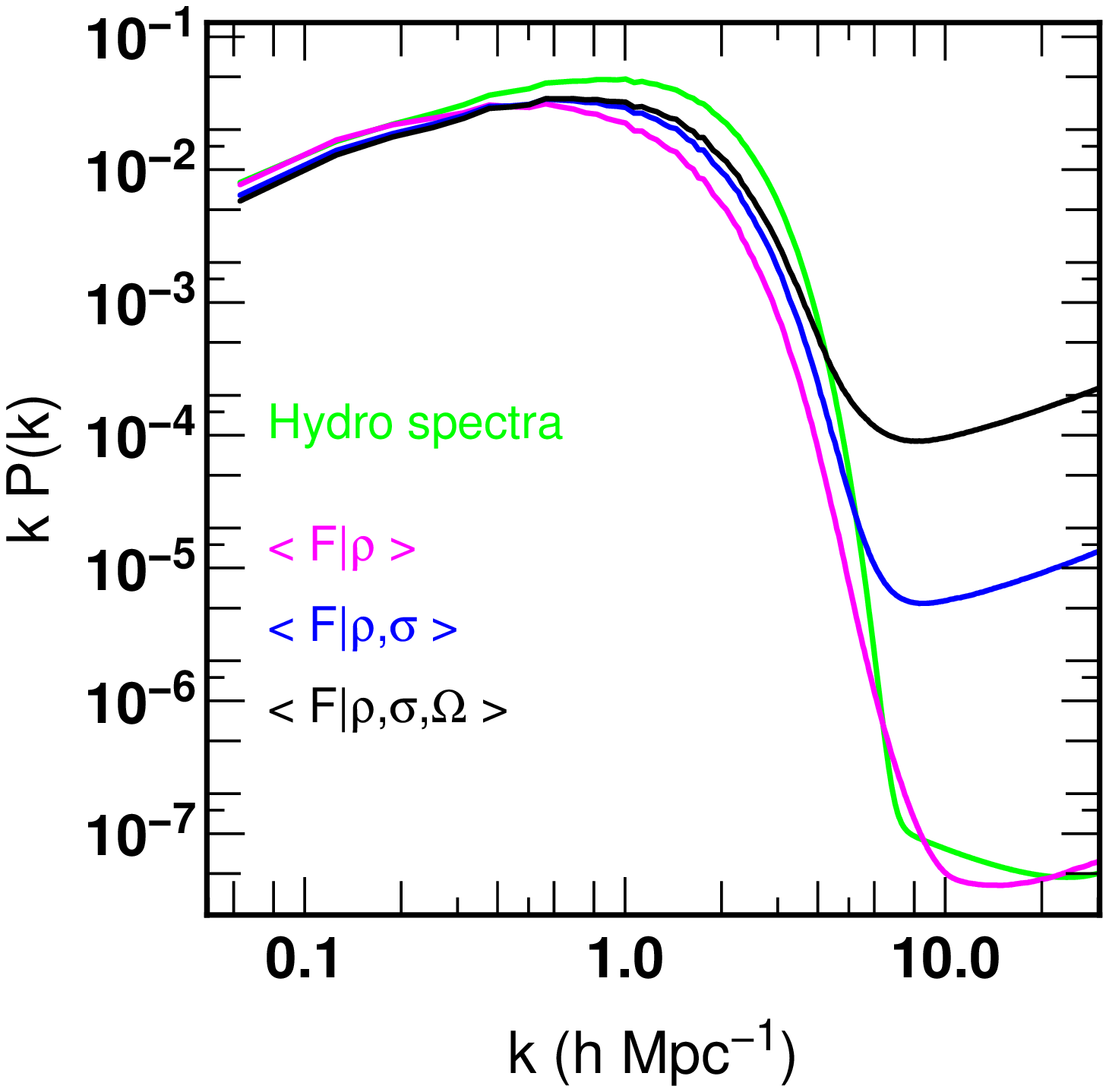}}
\rotatebox{0}{\includegraphics[width=\columnwidth]{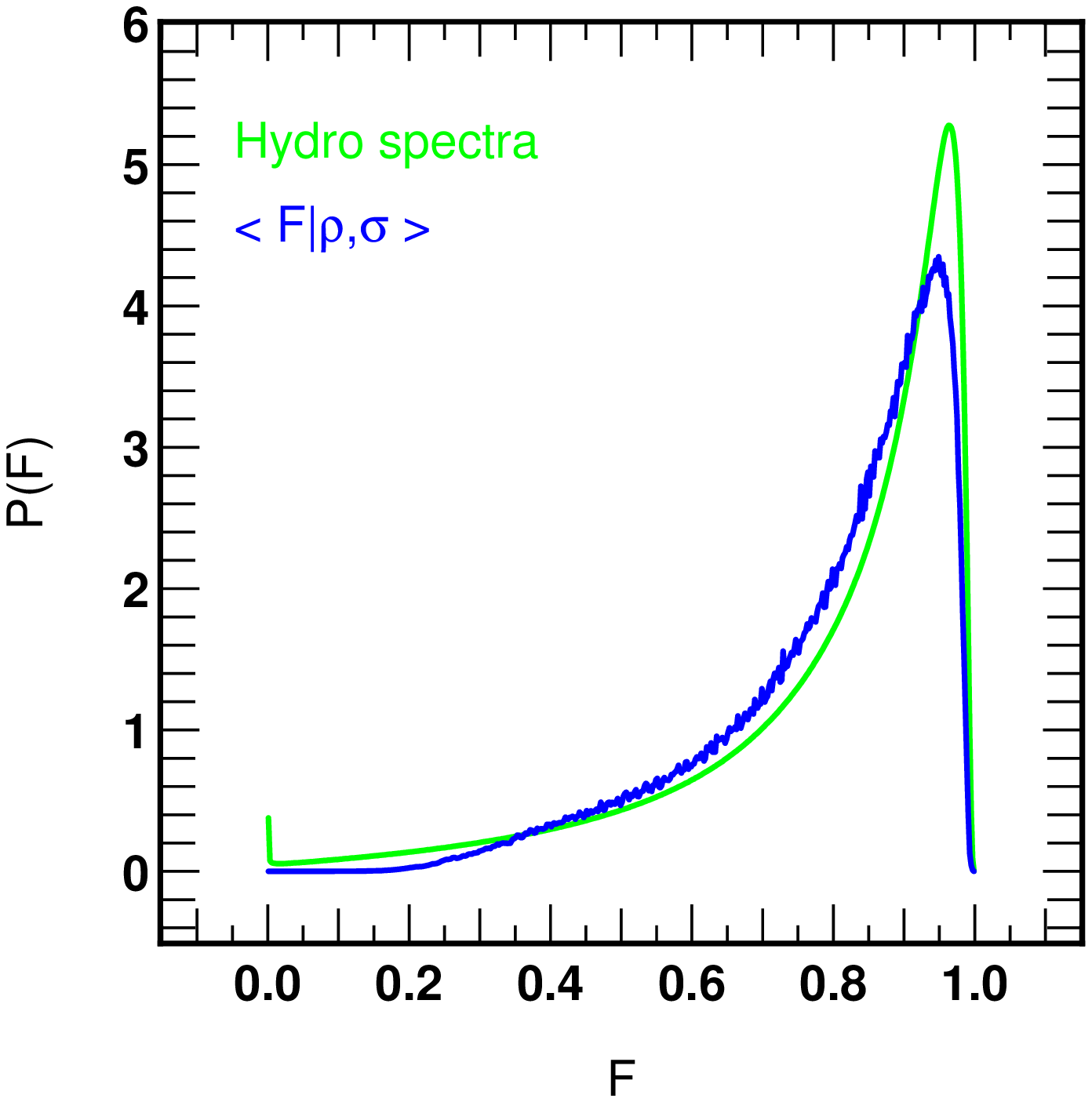}}
\caption{Top panel: the 1d power spectrum of pseudo spectra generate
from \hnoagnn DM fields (smoothed at 0.5 Mpc/h; $z=2.5$) and using
deterministic relations described in the text. 
The discrepancies with the hydro 1d-Pk are quite pronounced especially 
at small scales.
Bottom panel: an example of PDF of pseudo-spectra compare to hydro spectra, showing
again a noticeable disagreement.
All results are presented without a full iteration in the scheme (flux 1d-Pk and PDF rescaling).
}
\label{fig_appendix_5}
\end{center}
 \end{figure}

\begin{figure}
\begin{center}
\rotatebox{0}{\includegraphics[width=\columnwidth]{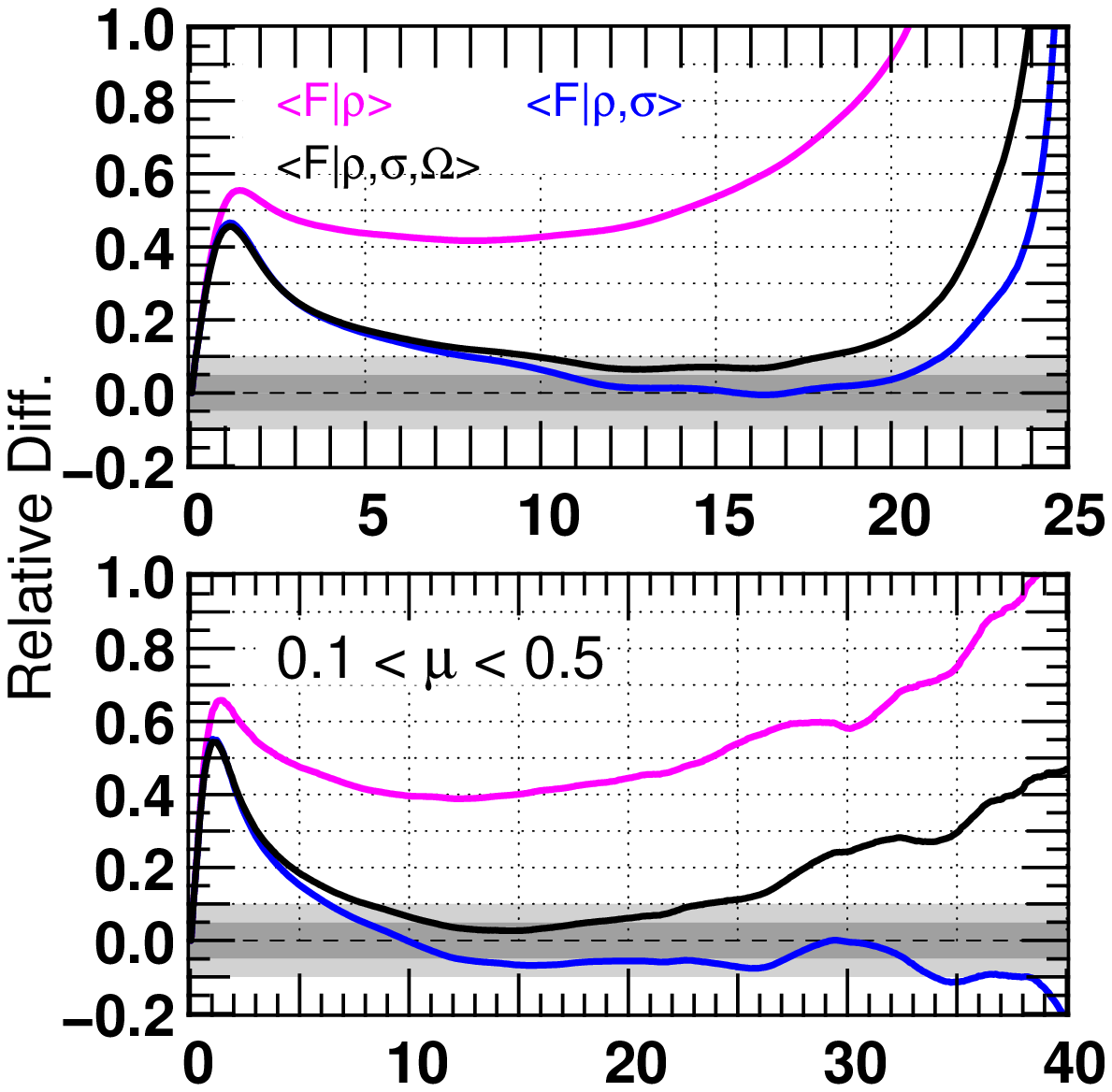}}
\rotatebox{0}{\includegraphics[width=\columnwidth]{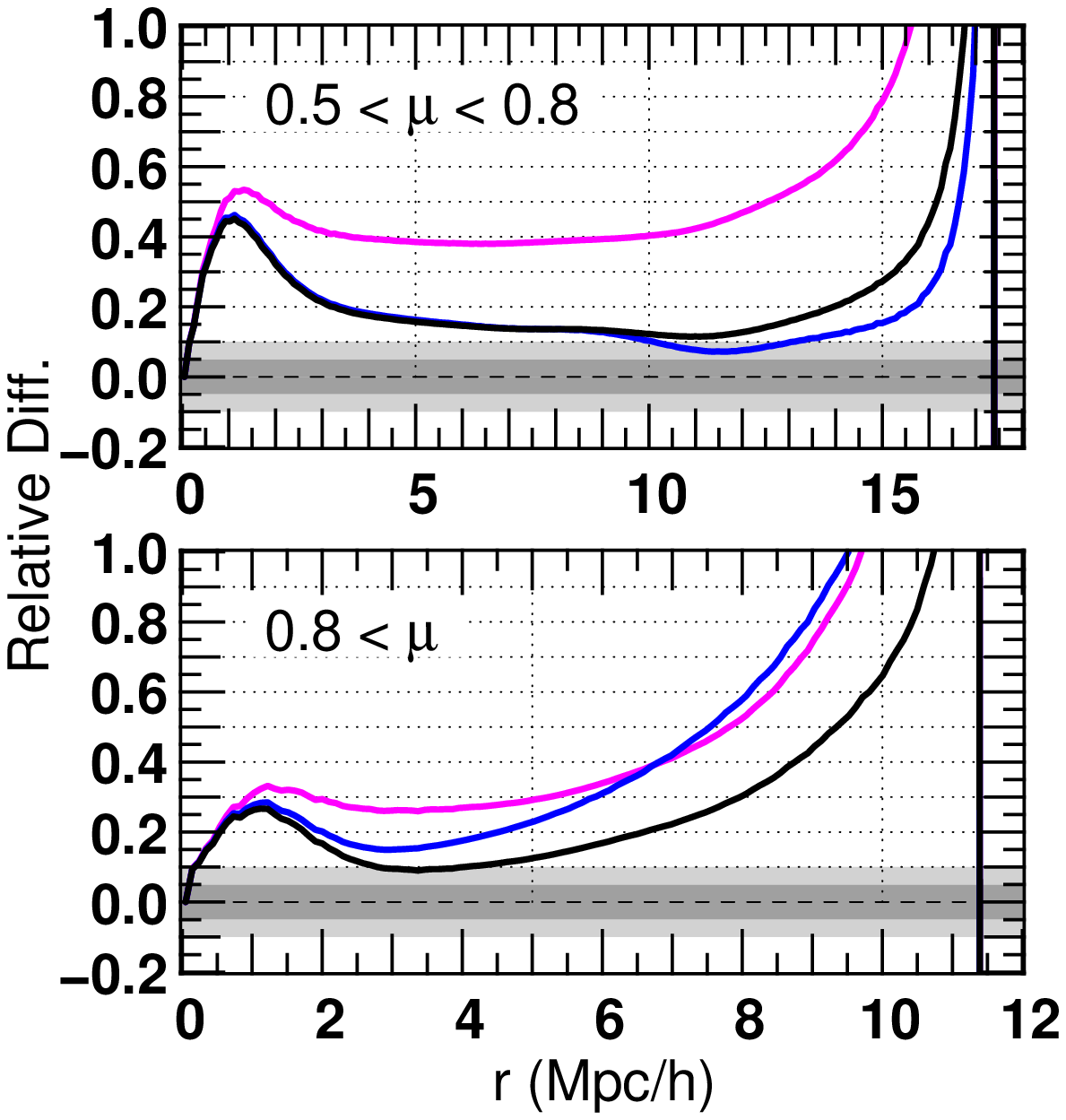}}
\caption{The relative difference of the two-point correlation functions of pseudo-spectra
produced with deterministic sampling (and a full iteration in the scheme).
It appears clearly that the different deterministic sampling do no reach the level
of accuracy of LyMAS2 especially for high angles where the error are quite high 
($>10\%$). The light and dark grey shade represent error lower that (absolute)
10 and 5\% respectively.}
\label{fig_appendix_6}
\end{center}
 \end{figure}

We have then produced grids of pseudo-spectra from DM fields extracted
from the \hnoagnn simulation (see section~\ref{sec:simu}), smoothed at 0.5 Mpc/h and
using three different deterministic relations. The first one considers
the DM overdensity field only (1-field), the second one both overdensity and
velocity dispersion fields (2-fields) while the last one associates
the DM density field to the velocity dispersion and vorticity fields (3-fields).
To estimate the mean value of the flux from a given value of $\rho$
or a given set of ($\rho$,$\sigma$) or ($\rho$,$\sigma$,$\Omega$),
we use respectively interpolations, bilinear interpolations and trilinear interpolations
depending of the number of input DM fields.
First, Fig.~\ref{fig_appendix_5} shows the 1d power spectrum and PDF of pseudo spectra 
(without iteration) for the 2-fields case.
We notice that the 1-point PDF is in general not well recovered. 
The predictions of the 1d-$P_k$ are also not as accurate than LyMAS and things especially
for the 2-fields and 3-fields cases. Indeed, the power spectra at small scales are
considerably overestimated. 
Although a full iteration can improved these trends,
the predicted 2 point correlation functions, shown in Fig.~\ref{fig_appendix_6},
 present errors that are generally quite high especially 
when an angle $\mu$ is considered. For instance the errors are much higher
than 10\% when $\mu>0.8$. 

Also, one main drawback of this approach when creating large mocks ($>$1 Gpc/h)
is to have enough statistics from the hydrodynamics simulations 
to cover most of the parameter space of the large cosmological simulation.
This should not be an issue for the 1-field case since simple interpolations and extrapolations
can be done (e.g. from Fig.~\ref{fig_appendix_3}  $\langle F\rangle\sim$1 and $\langle F\rangle\sim$0 for $1+\delta<-1.5$ and
$1+\delta>1.5$ respectively).
For the 2-fields and the 3-fields cases, efficient interpolations and
extrapolations could be obviously much more complicated to realized.

\section{Adaptive smoothing}
\label{appendix3}

In this section we provide some detail on how the dark matter density field,
velocity field and velocity dispersion are interpolated adaptively on a mesh from
the output of a cosmological dark matter $N$-body simulation, before further treatment by
LyMAS2, in particular additional smoothing with a Gaussian
window. 

Note that, while the notations below assume standard
configuration space, the calculation naturally extend to redshift space, by just
modifying particle coordinates accordingly with the local peculiar
velocity component contribution. In LyMAS2, we use the
infinitely remote observer approximation by just accounting for
redshift distortion along $z$ axis.

 For a smooth phase-space distribution function $f(\bmath{x},\bmath{v})$,
the projected density is given by
\begin{equation}
  \rho_{\rm E}(\bmath{x}) = \int {\rm d}^3 v\, f(\bmath{x},\bmath{v}),
\end{equation}
the Eulerian mean velocity field by
\begin{equation}
  \bmath{v}_{\rm E}(\bmath{x})=\langle \bmath{v} \rangle_{\rm E}= \frac{1}{\rho_{\rm E}(\bmath{x})} \int {\rm d}^3 v\,
  \bmath{v}\, f(\bmath{x},\bmath{v}), \label{eq:vEdef}
\end{equation}
and the local mean square velocity reads
\begin{equation}
 \langle v^2 \rangle_{\rm E}(\bmath{x})=\frac{1}{\rho_{\rm E}(\bmath{x})} \int {\rm d}^3 v\,
 v^2\, f(\bmath{x},\bmath{v}).
 \label{eq:v2E}
\end{equation}
Obviously, the last two equations stand for points of space where
$\rho(\bmath{x}) > 0$. 
From equation (\ref{eq:v2E}),  we can derive the local velocity dispersion
\begin{equation}
  \sigma_{v,{\rm E}}^2 \equiv \langle v^2 \rangle_{\rm E} - v_{\rm
    E}^2.
  \label{eq:sigvE}
  \end{equation}
We notice that the local velocity field can be considered
as a statistical average, this is why we used the $\langle \cdots
\rangle_{\rm E}$ notation above,
\begin{equation}
  \langle \bmath{v} \rangle_{\rm E}(\bmath{x})=\int {\rm d}^3 v\, \bmath{v}\, 
  f_{\rm E}(\bmath{x},\bmath{v}),
\end{equation}
with the local density probability
\begin{equation}
  f_{\rm E}(\bmath{x},\bmath{v}) \equiv \frac{1}{\rho_{\rm
      E}(\bmath{x})} f(\bmath{x},\bmath{v}).
  \label{eq:probability}
  \end{equation}
  In this probabilistic approach, the mean square velocity is given by
  equation (\ref{eq:v2E}), since
\begin{equation}
 \langle v^2 \rangle_{\rm E}(\bmath{x})=\int {\rm d}^3 v\,
 v^2\, f_{\rm E}(\bmath{x},\bmath{v}).
\end{equation}  
 and, likewise, its local variance by equation (\ref{eq:sigvE}). 

 What we have actually access to is not a smooth distribution function,
unfortunately, but a distribution of $N$ simulation particles of individual masses
$m_i$, positions $\bmath{x}_i$ and velocities $\bmath{v}_i$. This
means that the phase-space distribution function has the following
form
\begin{equation}
  f(\bmath{x},\bmath{v}) = \sum_i m_i \,\delta_{\rm D}
  (\bmath{v}-\bmath{v}_i)\, \delta_{\rm D} (\bmath{x}-\bmath{x}_i),
  \label{eq:fdisc}
\end{equation}
where $\delta_{\rm D}$ is the Dirac distribution function. 
From eq.~(\ref{eq:fdisc}), one can compute the projected Eulerian density
\begin{equation}
  \rho_{\rm E}(\bmath{x})=\sum_i m_i \,\delta_{\rm
    D}(\bmath{x}-\bmath{x}_i),
\end{equation}
but the Eulerian velocity field $\bmath{v}_{\rm E}(\bmath{x})$ and
mean square velocity are ill-defined.

However, the underlying distribution of true dark matter particles
is much smoother than its crude numerical representation in terms of
macroparticles of the $N$-body simulation. While, strictly speaking,  the phase-space
density is still of the form (\ref{eq:fdisc}) at the microscopic level, it can be considered as
a smooth function at the macroscopic level, at least in terms of
probability density.

In order to recover a good approximation of the
continuum, \citet{dens_smooth} proposed a locally adaptive smoothing
algorithm, SmoothDens, inspired from smooth particle hydrodynamics
(hereafter SPH),
using, to compute various fiels, an
interpolation window $F_{\bmath{x}}$, of which the shape parameters, in
  particular the typical size  $\ell(\bmath{x})$,
  depend on position $\bmath{x}$. This window function is normalized to unity, i.e. $\int {\rm
  d}^3 x'\, F_{\bmath{x}}(\bmath{x}')=1$.

In principle, for a given function $h(\bmath{x},\bmath{v})$, the smoothed counterpart is given by
\begin{equation}
[F_{\bmath{x}}*h](\bmath{x},\bmath{v})=\int {\rm
  d}^3x'\, F_{\bmath{x}}(\bmath{x}-\bmath{x}')\,
h(\bmath{x}',\bmath{v}).
\end{equation}
Setting $h = \rho_{\rm E}$, after simple algebraic calculations exploiting the properties of the Dirac
distribution function, we obtain the simple expression for the
adaptively smoothed density:
\begin{equation}
  \rho_{\rm F}(\bmath{x})=\sum_i m_i\, F_{\bmath{x}}(\bmath{x}-\bmath{x}_i).
  \label{eq:rhopart}
\end{equation}
From there, we can formally define the analogous of equation (\ref{eq:probability}) but
with adaptive smoothing performed in the spatial position,  
\begin{equation}
  f_{\rm F}(\bmath{x},\bmath{v})\equiv \frac{1}{\rho_{\rm F}(\bmath{x})} [F_{\bmath{x}}*f](\bmath{x},\bmath{v}),
\end{equation}
from which one can derive estimates in the mean field limit of velocity related quantities:
\begin{equation}
  \bmath{v}_{\rm F}(\bmath{x})=\frac{\sum_i m_i \,\bmath{v}_i
   \, F_{\bmath{x}}(\bmath{x}-\bmath{x}_i)}{\sum_i m_i\, F_{\bmath{x}}(\bmath{x}-\bmath{x}_i)},
    \label{eq:vpart}
  \end{equation}
  \begin{equation}
  {\langle v^2 \rangle}_{\rm G}=\frac{ \sum_i m_i\, v^2_i\,
F_{\bmath{x}}(\bmath{x}-\bmath{x}_i)}{\sum_i m_i\,
         F_{\bmath{x}}(\bmath{x}-\bmath{x}_i)}.
     \label{eq:v2part}
   \end{equation}   
In the algorithm SmoothDens, the adaptive
procedure is used to compute various fields for a particular set of positions
$\bmath{x}=\bmath{x}_j$ on a cubical grid of size $n_{\rm g}$.
Function $F$ is a  compact \citet[][]{1985A&A...149..135M} spline of size
$\ell(\bmath{x})$, with $\ell(\bmath{x})$ the distance of the
$N_{\rm SPH}^{\rm th}$ closest simulation particle to position $\bmath{x}$. The
value of $N_{\rm SPH}$ we adopt here is $N_{\rm SPH}=32$.
As an additional recipe, adaptive
smoothing is locally replaced with nearest grid point (NGP) interpolation
\citep[][]{1988csup.book.....H}
when $\ell(\bmath{x})$ is smaller or of the order of the grid cell size $L/n_{\rm
  g}$, where $L$ is the simulation box size. Also, a local weight is
given to each particle $i$ so that at the end, its total contribution sums
up to the particle mass $m_i$. Note that, due to the finite
extension of the spline function, some particles
belonging to dense clusters or close to dense clusters may not
contribute at all. In the latest implementation of the algorithm, 
  SmoothDens5, which we use in LyMAS2, an option allows one to affect these particles to the grid with NGP
interpolation in order to conserve total mass. The effect of not doing
so is however generally small. 

The outcome of SmoothDens mainly depends on two
parameters, the resolution $n_{\rm g}$ of the grid and the value used
for the number of neighbors, $N_{\rm SPH}$. Changing both these
parameters can have drastic impact on the results, especially
the local velocity dispersion and the velocity derivatives
estimates. Additional Gaussian smoothing performed in LyMAS2 is
however expected to reduce considerably the dependence on these two
parameters provided that the smoothing scale $R_{\rm G}$ associated to the
smoothing window is large enough compared
to $L/n_{\rm g}$. Yet one
has to bear in mind that the influence of $N_{\rm SPH}$ cannot be
negligible in underdense regions as long as it can influence scales 
larger than $R_{\rm G}$, which is unfortunately very likely. Despite
these non trivial issues, the
reason why LyMAS2 still works so accurately is that it is calibrated
relying on probability distributions mappings, which naturally corrects for
intrinsic biases introduced by local adaptive smoothing. 
\end{document}